\documentclass{article}

\usepackage{arxiv}

\usepackage[utf8]{inputenc} 
\usepackage[T1]{fontenc}    
\usepackage{lmodern}        
\usepackage{hyperref}       
\usepackage{url}            
\usepackage{booktabs}       
\usepackage{amsfonts}       
\usepackage{nicefrac}       
\usepackage{microtype}      
\usepackage{lipsum}
\usepackage{graphicx}

\title{In a Nutshell---The Sequential Parameter Optimization Toolbox}

\author{
    Thomas Bartz-Beielstein
   \\
    Institute of Data Science, Engineering, and Analytics \\
    Technische Hochschule Koeln \\
  51643 Gummersbach \\
  \texttt{\href{mailto:thomas.bartz-beielstein@th-koeln.de}{\nolinkurl{thomas.bartz-beielstein@th-koeln.de}}} \\
   \And
    Martin Zaefferer
   \\
    Institute of Data Science, Engineering, and Analytics \\
    Technische Hochschule Koeln \\
  51643 Gummersbach \\
  \texttt{\href{mailto:martin.zaefferer@th-koeln.de}{\nolinkurl{martin.zaefferer@th-koeln.de}}} \\
   \And
    Frederik Rehbach
   \\
    Institute of Data Science, Engineering, and Analytics \\
    Technische Hochschule Koeln \\
  51643 Gummersbach \\
  \texttt{\href{mailto:frederik.rehbach@th-koeln.de}{\nolinkurl{frederik.rehbach@th-koeln.de}}} \\
  }

\usepackage{color}
\usepackage{fancyvrb}

\DefineVerbatimEnvironment{Highlighting}{Verbatim}{commandchars=\\\{\}}
\usepackage{framed}
\definecolor{shadecolor}{RGB}{248,248,248}
\newenvironment{Shaded}{\begin{snugshade}}{\end{snugshade}}

\newcommand{\AttributeTok}[1]{\textcolor[rgb]{0.77,0.63,0.00}{#1}}

\newcommand{\CommentTok}[1]{\textcolor[rgb]{0.56,0.35,0.01}{\textit{#1}}}

\newcommand{\ConstantTok}[1]{\textcolor[rgb]{0.00,0.00,0.00}{#1}}
\newcommand{\ControlFlowTok}[1]{\textcolor[rgb]{0.13,0.29,0.53}{\textbf{#1}}}

\newcommand{\DecValTok}[1]{\textcolor[rgb]{0.00,0.00,0.81}{#1}}
\newcommand{\DocumentationTok}[1]{\textcolor[rgb]{0.56,0.35,0.01}{\textbf{\textit{#1}}}}

\newcommand{\FloatTok}[1]{\textcolor[rgb]{0.00,0.00,0.81}{#1}}
\newcommand{\FunctionTok}[1]{\textcolor[rgb]{0.00,0.00,0.00}{#1}}

\newcommand{\NormalTok}[1]{#1}

\newcommand{\OtherTok}[1]{\textcolor[rgb]{0.56,0.35,0.01}{#1}}

\newcommand{\SpecialCharTok}[1]{\textcolor[rgb]{0.00,0.00,0.00}{#1}}

\newcommand{\StringTok}[1]{\textcolor[rgb]{0.31,0.60,0.02}{#1}}

\usepackage{longtable}

\begin{document}
\maketitle

\def\tightlist{}

\begin{abstract}
The performance of optimization algorithms relies crucially on their
parameterizations. Finding good parameter settings is called algorithm
tuning. The sequential parameter optimization (SPOT) package for R is a
toolbox for tuning and understanding simulation and optimization
algorithms. Model-based investigations are common approaches in
simulation and optimization. Sequential parameter optimization has been
developed, because there is a strong need for sound statistical analysis
of simulation and optimization algorithms. SPOT includes methods for
tuning based on classical regression and analysis of variance
techniques; tree-based models such as CART and random forest; Gaussian
process models (Kriging), and combinations of different meta-modeling
approaches. Using a simple simulated annealing algorithm, we will
demonstrate how optimization algorithms can be tuned using SPOT. The
underling concepts of the SPOT approach are explained. This includes key
techniques such as exploratory fitness landscape analysis and
sensititvity analysis. Many examples illustrate how SPOT can be used for
understanding the performance of algorithms and gaining insight into
algorithm's behavior. Furthermore, we demonstrate how SPOT can be used
as an optimizer and how a sophisticated ensemble approach is able to
combine several meta models via stacking. This article exemplifies how
SPOT can be used for automatic and interactive tuning.
\end{abstract}


\hypertarget{introduction}{%
\section{Introduction}\label{introduction}}

The performance of search heuristics such as evolution strategies
(\texttt{ES}), differential evolution (\texttt{DE}), or simulated
annealing (\texttt{SANN}) relies crucially on their
parameterizations---or, statistically speaking, on their factor
settings.\\
Finding good parameter settings for an optimization algorithm will be
referred to as \emph{tuning}. We will illustrate how an existing search
heuristic can be tuned using the sequential parameter optimization
toolbox (\texttt{SPOT}), which is one possible implementation of the
sequential parameter optimization (\texttt{SPO}) framework introduced in
\cite{Bart06a}.

The version of \texttt{SPOT} presented in this article is implemented in
\texttt{R}. \texttt{R} is a freely available language and environment
for statistical computing and graphics which provides a wide variety of
statistical and graphical techniques: linear and nonlinear modelling,
statistical tests, time series analysis, classification, clustering,
etc. \cite{rcor21a}. R can be downloaded from
\href{https://cran.r-project.org}{CRAN}. The \texttt{SPOT} package can
be installed from within \texttt{R} using the \texttt{installPackages}
command.

\begin{Shaded}
\begin{Highlighting}[]
\FunctionTok{install.packages}\NormalTok{(}\StringTok{"SPOT"}\NormalTok{)}
\end{Highlighting}
\end{Shaded}

Note, that the \texttt{SPOT} package only has to be installed once
(unless an update to a more recent version is needed). Unlike
installation, the package has to be loaded to the \texttt{R} work space
every time a new \texttt{R} session is started. \texttt{SPOT} can be
loaded to the work space with \texttt{R}'s \texttt{library} command.
Package version \texttt{SPOT} should be greater equal than
\texttt{2.2.24}, which can be checked with \texttt{R}'s
\texttt{packageVersion()} command.

\begin{Shaded}
\begin{Highlighting}[]
\FunctionTok{library}\NormalTok{(}\StringTok{"SPOT"}\NormalTok{)}
\end{Highlighting}
\end{Shaded}

\begin{verbatim}
## Registered S3 method overwritten by 'sensitivity':
##   method    from 
##   print.src dplyr
\end{verbatim}

\begin{Shaded}
\begin{Highlighting}[]
\FunctionTok{packageVersion}\NormalTok{(}\StringTok{"SPOT"}\NormalTok{)}
\end{Highlighting}
\end{Shaded}

\begin{verbatim}
## [1] '2.2.24'
\end{verbatim}

In order to keep the setup as simple as possible, we will use a standard
optimizer that is part of every base \texttt{R} installation for
illustrating the tuning procedure: simulated annealing \cite{Kirk83a}.
This implementation of the simulated annealing heuristic will be
referred to as \texttt{SANN} in the following.

This report is structured as follows. First, the three levels of the
algorithm tuning methodology are described in Section
\ref{sec:algtunesmbo}, which are useful for distinguishing several
problem domains.

\begin{enumerate}
\def\labelenumi{\arabic{enumi}.}
\tightlist
\item
  The first level describes the objective function.
\item
  The second level presents the optimization algorithm.
\item
  The third level implements the tuning strategy.
\end{enumerate}

Next, details of the \texttt{SPOT} configuration are described. Because
visual inspection of the results plays an important role in the
\texttt{SPOT} approach, plotting functions that are provided in the
\texttt{SPOT} toolbox are described ins Section \ref{sec:plot}. These
enable an exploratory fitness landscape analysis as described in Section
\ref{sec:exploratory}. The response surface methodology is introduced in
Section \ref{sec:rsm}. Furthermore, tools for an statistical analysis
are described in Section \ref{sec:stats}. Applying \texttt{SPOT} to
deterministic problems is briefly explained in Section \ref{sec:determ}.
Ensemble models via stacking are presented in Section \ref{sec:stack}.
Hybrid approached are discussed in Section \ref{sec:hybrid}. Section
\ref{sec:summary} gives a summary. The Appendix \ref{sec:appendix} lists
many examples.

\hypertarget{sec:algtunesmbo}{%
\section{Algorithm Tuning and SMBO}\label{sec:algtunesmbo}}

\hypertarget{levels-during-tuning-and-optimization}{%
\subsection{\texorpdfstring{Levels During Tuning and Optimization
}{Levels During Tuning and Optimization }}\label{levels-during-tuning-and-optimization}}

We will consider \emph{algorithm tuning} and \emph{surrogate model based
optimization} in this article.

\hypertarget{algorithm-tuning}{%
\subsubsection{Algorithm Tuning}\label{algorithm-tuning}}

\begin{figure}

{\centering \includegraphics[width=0.5\linewidth]{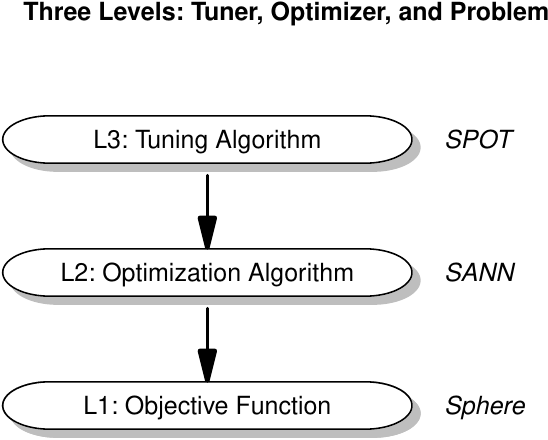} 

}

\caption{Levels: If SPOT is used as a tuner, three levels occur.}\label{fig:threeLevels}
\end{figure}

Algorithm tuning is also referred to as off-line parameter tuning in
contrast to on-line parameter tuning \cite{Eibe03a}. Algorithm tuning
involves the three levels, which describe the experimental setup.
Consider Figure \ref{fig:threeLevels}: the three levels that occur when
tuning an optimization algorithm with \texttt{SPOT}.

\begin{itemize}
\tightlist
\item
  (L1) The real-world system. This system allows the specification of an
  objective function, say \(f\). As an example, we will use the sphere
  function in the following.
\item
  (L2) The optimization algorithm, in our example \texttt{SANN}. It
  requires the specification of algorithm parameters, e.g., the initial
  temperature or the mutation rate of evolution strategies. These
  parameters determine the performance of the optimization algorithm.
  Therefore, they should be tuned. The algorithm is in turn used to
  determine optimal values of the objective function \(f\) from level
  (L1).
\item
  (L3) The tuning algorithm, here \texttt{SPOT}.
\end{itemize}

This setting will be referred to as \emph{algorithm tuning}.

\hypertarget{surrogate-model-based-optimization}{%
\subsection{Surrogate Model Based
Optimization}\label{surrogate-model-based-optimization}}

\begin{figure}

{\centering \includegraphics[width=0.5\linewidth]{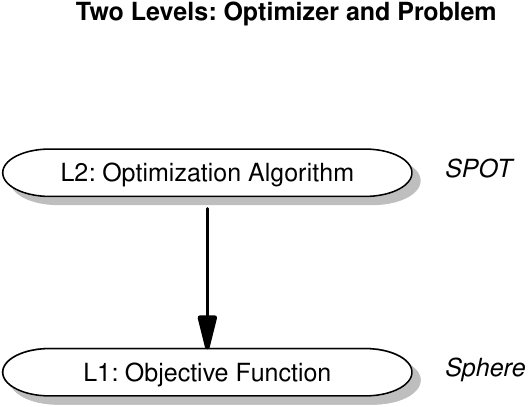} 

}

\caption{Levels: SPOT as an Optimizer}\label{fig:twoLevels}
\end{figure}

Alternatively, \texttt{SPOT} can be applied as an optimizer. This
situation is shown in Figure \ref{fig:twoLevels}. In this case,
\texttt{SPOT} tries to find arguments of the objective function that
result in an optimal function value. Following the taxonomy introduced
in \cite{Bart16n}, this setting will be referred to as
\emph{surrogate-model based optimization} (SMBO), because \texttt{SPOT}
uses an internal surrogate model.

\hypertarget{algorithm-and-problem-designs}{%
\subsubsection{Algorithm and Problem
Designs}\label{algorithm-and-problem-designs}}

The term \emph{algorithm design} summarizes factors that influence the
behavior (performance) of an algorithm, whereas \emph{problem design}
refers to factors from the optimization (simulation) problem. The
initial temperature in \texttt{SANN} is one typical factor which belongs
to the algorithm design, the search space dimension belongs to the
problem design.

\hypertarget{surrogate-model-based-optimization-1}{%
\subsubsection{Surrogate Model Based
Optimization}\label{surrogate-model-based-optimization-1}}

Instead of tuning an optimization algorithm, \texttt{SPOT} itself can be
used as a surrogate model based optimization algorithm. Then,
\texttt{SPOT} has the same role as \texttt{SANN} in the algorithm tuning
scenario. In this the surrogate model based optimization setting, only
two levels exist:

\begin{enumerate}
\def\labelenumi{\arabic{enumi}.}
\tightlist
\item
  the objective function (L1) and
\item
  the optimization algorithm (L2).
\end{enumerate}

This situation can be seen in Figure \ref{fig:twoLevels}.

\hypertarget{the-spot-algorithm}{%
\subsubsection{The SPOT Algorithm}\label{the-spot-algorithm}}

\texttt{SPOT} finds improved solutions in the following way (see the
following pseudo code):

\begin{itemize}
\tightlist
\item
  Initially, a population of (random) solutions is created.\\
  The initialization step is shown in State 1 in the following pseudo
  code of the \texttt{SPOT} algorithm.
\item
  A set of surrogate models is specified (Step 2).
\item
  Then, the solutions are evaluated on the objective function (level
  L1). This is State 3.
\item
  Next, surrogate models are built (State 4).
\item
  A global search is performed to generate new candidate solutions
  (State 5).
\item
  The new solutions are evaluated on the objective function (level L1)
  (State 6).
\end{itemize}

These steps are repeated, until a satisfying solution has been found.

\hypertarget{sequential-parameter-optimization}{%
\paragraph{Sequential parameter
optimization}\label{sequential-parameter-optimization}}

\begin{itemize}
\tightlist
\item
  State 1: \(t=0\). \(P(t) =\) SetInitialPopulation().
\item
  State 2: Select one or several surrogate models \(\mathfrak{M}\).
\item
  State 3: Evaluate(\(P(t)\)) on \(f\).
\item
  While\{not TerminationCriterion()\}

  \begin{itemize}
  \tightlist
  \item
    State 4: Use \(P(t)\) to build a model \(M(t)\) using
    \(\mathfrak{M}\).
  \item
    State 5: \(P'(t+1)=\) results from GlobalSearch(\(M(t)\)).
  \item
    State 6: Evaluate(\(P'(t+1)\)) on \(f\).
  \item
    State 7: \(P(t+1) = P(t) \cup P'(t+1)\).
  \item
    State 8: \(t = t+1\).
  \end{itemize}
\item
  EndWhile
\end{itemize}

\hypertarget{level-l1-objective-function}{%
\subsection{\texorpdfstring{Level L1: Objective Function
}{Level L1: Objective Function }}\label{level-l1-objective-function}}

Before \texttt{SANN} can be started, the user has to specify an
objective function \(f\). To keep things as simple as possible, the
\texttt{sphere} function will be used: \[
f(x) = \sum_{i=1}^n x_i^2
\]

\begin{Shaded}
\begin{Highlighting}[]
\NormalTok{sphere }\OtherTok{\textless{}{-}} \ControlFlowTok{function}\NormalTok{ (x)\{}
  \FunctionTok{sum}\NormalTok{(x}\SpecialCharTok{\^{}}\DecValTok{2}\NormalTok{)}
\NormalTok{  \}}
\FunctionTok{sphere}\NormalTok{( }\FunctionTok{c}\NormalTok{(}\DecValTok{1}\NormalTok{,}\DecValTok{2}\NormalTok{) )}
\end{Highlighting}
\end{Shaded}

\begin{verbatim}
## [1] 5
\end{verbatim}

The \texttt{sphere} function shown here uses vector inputs. A
matrix-based implementation is defined as \texttt{funSphere} in the
\texttt{SPOT} package.

\begin{Shaded}
\begin{Highlighting}[]
\NormalTok{funSphere}
\end{Highlighting}
\end{Shaded}

\begin{verbatim}
## function (x) {
##   matrix(apply(x, # matrix
##                1, # margin (apply over rows)
##                function(x) {
##                  sum(x ^ 2)  # objective function
##                }),
##          , 1) # number of columns
## }
## <bytecode: 0x7fb857a85088>
## <environment: namespace:SPOT>
\end{verbatim}

A surface plot of this function in the interval \([0;1] \times [0;1]\)
can be generated using the following command:

\begin{Shaded}
\begin{Highlighting}[]
\FunctionTok{plotFunction}\NormalTok{(funSphere)}
\end{Highlighting}
\end{Shaded}

\begin{center}\includegraphics[width=0.7\linewidth]{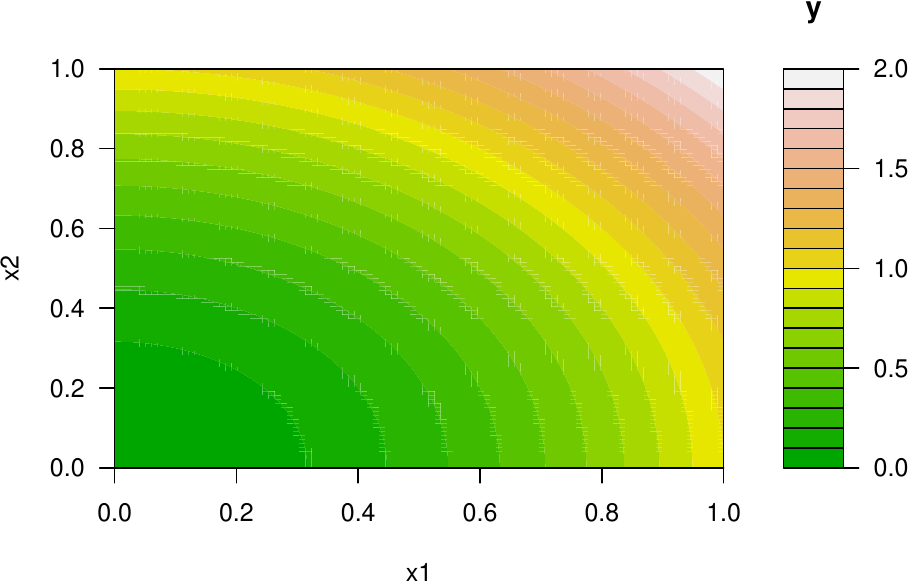} \end{center}

\hypertarget{level-l2-simulated-annealing-sann}{%
\subsection{\texorpdfstring{Level L2: Simulated Annealing
\texttt{SANN}}{Level L2: Simulated Annealing SANN}}\label{level-l2-simulated-annealing-sann}}

\emph{Simulated annealing} is a generic probabilistic heuristic
algorithm for global optimization \cite{Kirk83a}. The name comes from
annealing in metallurgy and describes controlled heating and cooling of
a material, which reduces defects. Heating enables atoms to leave their
initial positions (which are local minima of their internal energy), and
controlled cooling improves the probability to find positions with lower
states of internal energy than the initial positions. The \texttt{SANN}
algorithm replaces the current solution with a randomly generated new
solution.

Better solutions are accepted deterministically, where worse solutions
are accepted with a probability that depends on the difference between
the corresponding function values and on a global parameter, which is
commonly referred to as the \texttt{temperature}. The algorithm
parameter \texttt{temp} specifies the initial temperature of the
\texttt{SANN} algorithm. The temperature is gradually decreased during
the optimization. A second parameter, \texttt{tmax}, is used to model
this cooling scheme.

We consider the \texttt{R} implementation of \texttt{SANN}, which is
available via the general-purpose optimization function \texttt{optim()}
from the \texttt{R} package \texttt{stats}, which is part of every
\texttt{R} installation. The function \texttt{optim()} is parametrized
as follows

\texttt{optim(par,\ fn,\ gr\ =\ NULL,\ ...,\ method\ =\ c("Nelder-Mead",\ "BFGS",\ "CG",\ "L-BFGS-B",\ "SANN",\ "Brent"),\ lower\ =\ -Inf,\ upper\ =\ Inf,\ control\ =\ list(),\ hessian\ =\ FALSE)}

\begin{itemize}
\tightlist
\item
  Here, \texttt{par} denotes initial values for the parameters to be
  optimized over. Note, the problem dimension is specified by the length
  of this vector, so \texttt{par=c(1,1,1,1)} denotes a four-dimensional
  optimization problem.
\item
  \texttt{fn} is a function to be minimized (or maximized), with first
  argument the vector of parameters over which minimization is to take
  place.
\item
  \texttt{gr} defines a function to return the gradient for the
  \texttt{BFGS}, \texttt{CG} and \texttt{L-BFGS-B} methods.

  \begin{itemize}
  \tightlist
  \item
    If it is \texttt{NULL}, a finite-difference approximation will be
    used.
  \item
    For the \texttt{SANN} method it specifies a function to generate a
    new candidate point.
  \item
    If it is \texttt{NULL}, a default Gaussian Markov kernel is used.
  \end{itemize}
\item
  The symbol \texttt{...} represents further arguments (optional) that
  can be be passed to \texttt{fn} and \texttt{gr}.
\item
  The parameter \texttt{method} denotes the optimization method to be
  used. Here, we will use the parameter value \texttt{SANN}.
\end{itemize}

The parameters \texttt{lower}, \texttt{upper} specify bounds on the
variables for the ``L-BFGS-B'' method, or bounds in which to search for
method \texttt{Brent}. So, we will not use these variables in our
examples.

The argument \texttt{control} defines a relatively long list of control
parameters. We will use the following parameters from this list:

\begin{enumerate}
\def\labelenumi{\arabic{enumi}.}
\tightlist
\item
  \texttt{maxit}, i.e., the maximum number of iterations, which is for
  \texttt{SANN} the maximum number of function valuations. This is the
  stopping criterion.
\item
  \texttt{temp} controls the \texttt{SANN} algorithm. It is the starting
  temperature for the cooling schedule with a default value of 10.
\item
  Finally, we will use \texttt{tmax}, which is the number of function
  evaluations at each temperature for the \texttt{SANN} method. Its
  default value is also 10.
\end{enumerate}

To obtain reproducible results, we will set the random number generator
(RNG) seed. Using a two-dimensional objective function (\texttt{sphere})
and the starting point (initial values for the parameters to be
optimized over) \((10,10)\), we can execute the optimization runs as
follows:

\begin{Shaded}
\begin{Highlighting}[]
\FunctionTok{set.seed}\NormalTok{(}\DecValTok{123}\NormalTok{)}
\NormalTok{resSANN }\OtherTok{\textless{}{-}} \FunctionTok{optim}\NormalTok{(}\FunctionTok{c}\NormalTok{(}\DecValTok{10}\NormalTok{,}\DecValTok{10}\NormalTok{), sphere, }\AttributeTok{method=}\StringTok{"SANN"}\NormalTok{,}
                \AttributeTok{control=}\FunctionTok{list}\NormalTok{(}\AttributeTok{maxit=}\DecValTok{100}\NormalTok{, }\AttributeTok{temp=}\DecValTok{10}\NormalTok{, }\AttributeTok{tmax =} \DecValTok{10}\NormalTok{))}
\NormalTok{resSANN}
\end{Highlighting}
\end{Shaded}

\begin{verbatim}
## $par
## [1] 4.835178 4.664964
## 
## $value
## [1] 45.14084
## 
## $counts
## function gradient 
##      100       NA 
## 
## $convergence
## [1] 0
## 
## $message
## NULL
\end{verbatim}

The best, i.e., smallest, function value, which was found by
\texttt{SANN}, reads 45.14084. The corresponding point in the search
space is approximately (4.835178, 4.664964). No gradient information was
used and one hundred function evaluations were performed. The variable
\texttt{convergence} is an integer code, and its value \texttt{0}
indicates successful completion of the \texttt{SANN} run. No additional
\texttt{message} is returned.

Now that we have performed a first run of the \texttt{SANN} algorithm on
our simple test function, we are interested in improving \texttt{SANN}'s
performance. The \texttt{SANN} heuristic requires some parameter
settings, namely \texttt{temp} and \texttt{tmax}. If these values are
omitted, a default value of ten is used. The questions is:

\begin{itemize}
\tightlist
\item
  Are the default algorithm parameter settings, namely \texttt{temp}=10
  and \texttt{tmax}=10, adequate for \texttt{SANN} or can these values
  be improved?
\end{itemize}

That is, we are trying to tune the \texttt{SANN} optimization algorithm.
A typical beginner in algorithm tuning would try to improve the
algorithm's performance by manually increasing or decreasing the
algorithm parameter values, e.g., choosing \texttt{temp} = 20 and
\texttt{tmax} = 5. This is shown the following piece of \texttt{R} code:

\begin{Shaded}
\begin{Highlighting}[]
\FunctionTok{set.seed}\NormalTok{(}\DecValTok{123}\NormalTok{)}
\NormalTok{resSANN }\OtherTok{\textless{}{-}} \FunctionTok{optim}\NormalTok{(}\AttributeTok{par =} \FunctionTok{c}\NormalTok{(}\DecValTok{10}\NormalTok{,}\DecValTok{10}\NormalTok{), }\AttributeTok{fn =}\NormalTok{ sphere, }\AttributeTok{method=}\StringTok{"SANN"}\NormalTok{,}
                \AttributeTok{control =} \FunctionTok{list}\NormalTok{(}\AttributeTok{maxit =} \DecValTok{100}\NormalTok{, }\AttributeTok{temp =} \DecValTok{20}\NormalTok{, }\AttributeTok{tmax =} \DecValTok{5}\NormalTok{))}
\NormalTok{resSANN}
\end{Highlighting}
\end{Shaded}

\begin{verbatim}
## $par
## [1] 6.163905 6.657100
## 
## $value
## [1] 82.3107
## 
## $counts
## function gradient 
##      100       NA 
## 
## $convergence
## [1] 0
## 
## $message
## NULL
\end{verbatim}

Obviously, the manual tuning step worsened the result. And, this
procedure is very time consuming and does not allow efficient
statistical conclusions. Therefore, we will present a different
approach, which uses \texttt{SPOT}. Although the setup for the tuning
procedure with \texttt{SPOT} is very similar to the setup discussed in
this section, it enables deeper insights into the algorithm's
performance.

\hypertarget{level-l3-spot}{%
\subsection{Level L3: SPOT}\label{level-l3-spot}}

This section presents an example, which demonstrates, how \texttt{SPOT}
can be used to tune the \texttt{SANN} algorithm defined at level L2. The
goal of this tuning procedure is to determine improved parameter
settings for the \texttt{SANN} algorithm. As a level L1 function, which
will be optimized by \texttt{SANN}, the sphere function was chosen.

\hypertarget{example-using-random-forest}{%
\subsubsection{Example: Using Random
Forest}\label{example-using-random-forest}}

First, the problem setup for level L1 has to be specified. To keep the
situation as simple as possible, we will use the \texttt{sphere()} test
function, which was introduced above. The problem design requires the
specification of the starting point \texttt{x0} for the search.

\begin{Shaded}
\begin{Highlighting}[]
\NormalTok{x0 }\OtherTok{=}  \FunctionTok{c}\NormalTok{(}\SpecialCharTok{{-}}\DecValTok{1}\NormalTok{,}\DecValTok{1}\NormalTok{,}\SpecialCharTok{{-}}\DecValTok{1}\NormalTok{) }
\end{Highlighting}
\end{Shaded}

Because \texttt{x0} has three elements, we are facing a three
dimensional optimization problem. \texttt{SANN} will be used to
determine its minimum function value. Secondly, the problem setup for
level L2 has to be defined. Again, several settings have to be specified
for the \texttt{SANN} algorithm to be tuned. The budget, i.e., the
maximum number of function evaluations that can be used by \texttt{SANN}
is specified via \texttt{maxit}:

\begin{Shaded}
\begin{Highlighting}[]
\NormalTok{maxit }\OtherTok{=} \DecValTok{100} 
\end{Highlighting}
\end{Shaded}

As above, the \texttt{R} implementation of \texttt{SANN} will be used
via the \texttt{optim()} function. We will consider two parameters:

\begin{enumerate}
\def\labelenumi{\arabic{enumi}.}
\tightlist
\item
  the initial temperature (\texttt{temp}) and
\item
  the number of function evaluations at each temperature
  (\texttt{tmax}).
\end{enumerate}

Both are integer values. All parameters and settings of \texttt{SANN},
which were used for this simple example are summarized in the following
table. The first two parameters belong to the algorithm design, whereas
the remaining parameters are from the problem design. Note, the starting
point defines the problem dimension, i.e., by specifying a three
dimensional starting point the problem dimension is set to three. The
initial seed is the value that the \emph{random number generator} (RNG)
is initialized with.

\begin{longtable}[]{@{}lll@{}}
\toprule
Name & Symbol & Factor name\tabularnewline
\midrule
\endhead
Initial temperature & \(t\) & \texttt{temp}\tabularnewline
Number of function evaluations at each temperature & \(t_{\max}\) &
\texttt{tmax}\tabularnewline
Starting point & \(\vec{x_0} = (-1,1,-1)\) & \texttt{x0}\tabularnewline
Problem dimension & \(n=3\) &\tabularnewline
Objective function & sphere & \texttt{sphere()}\tabularnewline
Quality measure & Expected performance, e.g., \(E(y)\) &
\texttt{y}\tabularnewline
Initial seed & \(s\) & \texttt{1}\tabularnewline
Budget & \(\textrm{maxit} = 100\) & \texttt{maxit}\tabularnewline
\bottomrule
\end{longtable}

Thirdly, the tuning procedure at level L3 has to be specified.

\hypertarget{an-interface-to-sannsphere-for-spot}{%
\subsubsection{An interface to SANN+Sphere for
SPOT}\label{an-interface-to-sannsphere-for-spot}}

To interface \texttt{SANN} with \texttt{SPOT}, the wrapper function
\texttt{sann2spot()} is used. Note, \texttt{SPOT} uses matrices as the
basic data structure. The matrix format was chosen as a compromise
between speed and flexibility. The \texttt{matrix()} command can be used
as follows:

\texttt{matrix(data\ =\ NA,\ nrow\ =\ 1,\ ncol\ =\ 1,\ byrow\ =\ FALSE,\ dimnames\ =\ NULL)}

The interface function receives a matrix where each row is proposed
parameter setting (\texttt{temp}, \texttt{tmax}), and each column
specifies the parameters. It generates a \((n,1)\)-matrix as output,
where \(n\) is the number of (\texttt{temp}, \texttt{tmax}) parameter
settings.

\begin{Shaded}
\begin{Highlighting}[]
\NormalTok{sann2spot }\OtherTok{\textless{}{-}} \ControlFlowTok{function}\NormalTok{(algpar)\{}
\NormalTok{  performance }\OtherTok{\textless{}{-}} \ConstantTok{NULL}
  \ControlFlowTok{for}\NormalTok{ (i }\ControlFlowTok{in} \DecValTok{1}\SpecialCharTok{:}\FunctionTok{nrow}\NormalTok{(algpar))\{}
\NormalTok{    resultList }\OtherTok{\textless{}{-}} \FunctionTok{optim}\NormalTok{(}\AttributeTok{par =} \FunctionTok{c}\NormalTok{(}\DecValTok{10}\NormalTok{,}\DecValTok{10}\NormalTok{),}
                    \AttributeTok{fn =}\NormalTok{ sphere,}
                    \AttributeTok{method =} \StringTok{"SANN"}\NormalTok{,}
                    \AttributeTok{control =} \FunctionTok{list}\NormalTok{(}\AttributeTok{maxit =} \DecValTok{100}\NormalTok{,}
                                  \AttributeTok{temp =}\NormalTok{ algpar[i,}\DecValTok{1}\NormalTok{],}
                                  \AttributeTok{tmax =}\NormalTok{ algpar[i,}\DecValTok{2}\NormalTok{]))}
\NormalTok{    performance }\OtherTok{\textless{}{-}} \FunctionTok{c}\NormalTok{(performance,resultList}\SpecialCharTok{$}\NormalTok{value)}
\NormalTok{    \}}
\FunctionTok{return}\NormalTok{(}\FunctionTok{matrix}\NormalTok{(performance,}\AttributeTok{ncol=}\DecValTok{1}\NormalTok{))}
\NormalTok{\}}
\end{Highlighting}
\end{Shaded}

Now we can test the interface. First, we run \texttt{SANN} with
\texttt{temp} = 10 and \texttt{tmax} = 10. A second \texttt{SANN} run is
performed using \texttt{temp} = 20 and \texttt{tmax} = 5.

\begin{Shaded}
\begin{Highlighting}[]
\FunctionTok{set.seed}\NormalTok{(}\DecValTok{123}\NormalTok{)}
\FunctionTok{sann2spot}\NormalTok{(}\AttributeTok{algpar =} \FunctionTok{matrix}\NormalTok{(}\FunctionTok{c}\NormalTok{(}\DecValTok{10}\NormalTok{,}\DecValTok{10}\NormalTok{),}\DecValTok{1}\NormalTok{))}
\end{Highlighting}
\end{Shaded}

\begin{verbatim}
##          [,1]
## [1,] 45.14084
\end{verbatim}

\begin{Shaded}
\begin{Highlighting}[]
\FunctionTok{set.seed}\NormalTok{(}\DecValTok{123}\NormalTok{)}
\FunctionTok{sann2spot}\NormalTok{(}\AttributeTok{algpar=}\FunctionTok{matrix}\NormalTok{(}\FunctionTok{c}\NormalTok{(}\DecValTok{5}\NormalTok{,}\DecValTok{20}\NormalTok{),}\DecValTok{1}\NormalTok{))}
\end{Highlighting}
\end{Shaded}

\begin{verbatim}
##          [,1]
## [1,] 4.469163
\end{verbatim}

\hypertarget{the-configuration-list}{%
\subsubsection{The Configuration List}\label{the-configuration-list}}

\texttt{SPOT} itself has various parameters that need to be configured
so that it can solve the tuning problem efficiently. A configuration or
\texttt{control} list can be defined for \texttt{SPOT}. If no
configuration list is specified, the default configuration that works
fine for many problems is used. In the following, some elements of the
configuration list that are important for our example (tuning
\texttt{SANN} + \texttt{sphere()}) will be explained.

\hypertarget{types}{%
\paragraph{types:}\label{types}}

Since \texttt{SANN}'s parameters \texttt{temp} and \texttt{tmax} are
integers, we provide this type information via
\texttt{types\ =\ c("integer",\ "integer")}.

\hypertarget{funevals}{%
\paragraph{\texorpdfstring{\texttt{funEvals}:}{funEvals:}}\label{funevals}}

The number of algorithm runs, i.e., runs of the \texttt{SANN} algorithm,
is specified via \texttt{funEvals}.

\hypertarget{noise}{%
\paragraph{\texorpdfstring{\texttt{noise}:}{noise:}}\label{noise}}

\texttt{SANN} is a stochastic optimizer, so we specify
\texttt{noise\ =\ TRUE}.

\hypertarget{seedfun}{%
\paragraph{\texorpdfstring{\texttt{seedFun}:}{seedFun:}}\label{seedfun}}

Also due to the stochasticity of the optimizer, a random number
generator seed has to be specified, \texttt{seedFun\ =\ 1}. Every time
an algorithm configuration is tested, the RNG seed will be set. For the
first evaluation, the seed will be \texttt{seedFun}, subsequent
evaluations will increment this value.

\hypertarget{replicates}{%
\paragraph{\texorpdfstring{\texttt{replicates}:}{replicates:}}\label{replicates}}

Since our evaluations are subject to noise, we can make replicates to
mitigate it. In our example, each algorithm configuration will be
evaluated twice, so the setting \texttt{replicates\ =\ 2} is used.

\hypertarget{seedspot}{%
\paragraph{\texorpdfstring{\texttt{seedSPOT}:}{seedSPOT:}}\label{seedspot}}

An additional seed for \texttt{SPOT} can be specified using
\texttt{seedSPOT\ =\ 1}. This second seed is only set once in the
beginning. This ensures that the \texttt{SPOT} run is reproducible,
since \texttt{SPOT} itself may also be of a stochastic nature (depending
on the configuration).

\hypertarget{design}{%
\paragraph{\texorpdfstring{\texttt{design}:}{design:}}\label{design}}

The \texttt{design} parameter defines the method to be used to generate
an initial design, i.e., a set of initial algorithm settings (here: a
number of pairs of \texttt{temp} and \texttt{tmax}). A Latin hyper cube
design (LHD) is specified via \texttt{design\ =\ designLHD}.

\hypertarget{model}{%
\paragraph{\texorpdfstring{\texttt{model}:}{model:}}\label{model}}

Based on the initial design and subsequent evaluations, a \texttt{model}
can be trained to learn the relation between algorithm parameters
(\texttt{temp},\texttt{tmax}) and algorithm performance. To generate the
meta \texttt{model}, we use a random forest implementation
\cite{Brei01a}. This can be specified via
\texttt{model\ =\ buildRandomForest}. Random forest was chosen, because
it is a robust method which can handle categorical and numerical
variables.

\hypertarget{optimizer}{%
\paragraph{\texorpdfstring{\texttt{optimizer}:}{optimizer:}}\label{optimizer}}

Once a meta \texttt{model}is trained, we need an \texttt{optimizer} to
find the best potential algorithm configuration, based on the
\texttt{model}. Here, we choose a very simple optimizer, which creates a
large LHD and evaluates it on the \texttt{model}:
\texttt{optimizer\ =\ optimLHD}.

\hypertarget{optimizercontrol}{%
\paragraph{\texorpdfstring{\texttt{optimizerControl}:}{optimizerControl:}}\label{optimizercontrol}}

The specified \texttt{optimizer} may have options that need to be set.
Here, we only specify the number of \texttt{model} evaluations to be
performed by the \texttt{optimizer} with
\texttt{optimizerControl\ =\ list(funEvals=100)}. Overall, we obtain the
following configuration:

\begin{Shaded}
\begin{Highlighting}[]
\NormalTok{spotConfig }\OtherTok{\textless{}{-}} \FunctionTok{list}\NormalTok{(}
  \AttributeTok{types =} \FunctionTok{c}\NormalTok{(}\StringTok{"integer"}\NormalTok{, }\StringTok{"integer"}\NormalTok{), }\CommentTok{\#data type of tuned parameters}
  \AttributeTok{funEvals =} \DecValTok{50}\NormalTok{, }\CommentTok{\#maximum number of SANN runs}
  \AttributeTok{noise =} \ConstantTok{TRUE}\NormalTok{, }\CommentTok{\#problem is noisy (SANN is non{-}deterministic)}
  \AttributeTok{seedFun =} \DecValTok{1}\NormalTok{, }\CommentTok{\#RNG start seed for algorithm calls (iterated)}
  \AttributeTok{replicates =} \DecValTok{2}\NormalTok{, }\CommentTok{\#2 replicates for each SANN parameterization}
  \AttributeTok{seedSPOT =} \DecValTok{1}\NormalTok{, }\CommentTok{\#main RNG}
  \AttributeTok{design =}\NormalTok{ designLHD, }\CommentTok{\#initial design: Latin Hypercube}
  \AttributeTok{model =}\NormalTok{ buildRandomForest, }\CommentTok{\# model = buildKriging Kriging surrogate model}
  \AttributeTok{optimizer =}\NormalTok{ optimLHD, }\CommentTok{\#Use LHD to optimize on model }
  \AttributeTok{optimizerControl =} \FunctionTok{list}\NormalTok{(}\AttributeTok{funEvals=}\DecValTok{100}\NormalTok{) }\CommentTok{\#100 model evals in each iteration  }
\NormalTok{)}
\end{Highlighting}
\end{Shaded}

\hypertarget{the-region-of-interest}{%
\subsection{The Region of Interest}\label{the-region-of-interest}}

The region of interest (ROI) specifies \texttt{SPOT}'s search intervals
for the \texttt{SANN} parameters, i.e., for \texttt{tmax} and
\texttt{temp}. Here, both parameters \texttt{temp} and \texttt{tmax}
will be tuned in the region between one and 100.

\begin{Shaded}
\begin{Highlighting}[]
\NormalTok{tempLo }\OtherTok{=} \DecValTok{1}
\NormalTok{tempHi }\OtherTok{=} \DecValTok{100}
\NormalTok{tmaxLo }\OtherTok{=} \DecValTok{1}
\NormalTok{tmaxHi }\OtherTok{=} \DecValTok{100}
\NormalTok{lower}\OtherTok{=}\FunctionTok{c}\NormalTok{(tempLo,tmaxLo)}
\NormalTok{upper}\OtherTok{=}\FunctionTok{c}\NormalTok{(tempHi,tmaxHi)}
\end{Highlighting}
\end{Shaded}

The order (first \texttt{temp} then \texttt{tmax}) has to be the same as
in the \texttt{sann2spot()} interface.

\hypertarget{the-first-spot-run-calling-spot}{%
\subsection{\texorpdfstring{The First \texttt{SPOT} Run: Calling
\texttt{spot()}}{The First SPOT Run: Calling spot()}}\label{the-first-spot-run-calling-spot}}

Now we are ready to perform our first \texttt{SPOT} experiment. We will
start \texttt{SPOT} via \texttt{spot()}. The result is stored in the
list \texttt{resRf}:

\begin{Shaded}
\begin{Highlighting}[]
\CommentTok{\# library(SPOT)}
\CommentTok{\# source(\textquotesingle{}\textasciitilde{}/workspace/SPOT/R/spot.R\textquotesingle{})}
\CommentTok{\# source(\textquotesingle{}\textasciitilde{}/workspace/SPOT/R/initialInputCheck.R\textquotesingle{})}
\NormalTok{resRf }\OtherTok{\textless{}{-}} \FunctionTok{spot}\NormalTok{(}\AttributeTok{x=}\ConstantTok{NULL}\NormalTok{,}
              \AttributeTok{fun=}\NormalTok{sann2spot,}
              \AttributeTok{lower=}\NormalTok{lower,}
              \AttributeTok{upper=}\NormalTok{upper,}
              \AttributeTok{control=}\NormalTok{spotConfig)}
\FunctionTok{is.null}\NormalTok{(spotConfig}\SpecialCharTok{$}\NormalTok{optimizerControl}\SpecialCharTok{$}\NormalTok{eval\_g\_ineq)}
\end{Highlighting}
\end{Shaded}

\begin{verbatim}
## [1] TRUE
\end{verbatim}

We are able to take a look at the results from the tuning procedure.
Output from the \texttt{SPOT} run, which is stored in the list
\texttt{resRf}, has the following structure:

\begin{Shaded}
\begin{Highlighting}[]
\FunctionTok{str}\NormalTok{(resRf)}
\end{Highlighting}
\end{Shaded}

\begin{verbatim}
## List of 9
##  $ xbest   : num [1, 1:2] 3 82
##  $ ybest   : num [1, 1] 0.00123
##  $ x       : num [1:50, 1:2] 9 72 45 15 26 57 70 93 33 88 ...
##  $ y       : num [1:50, 1] 0.226 28.891 59.309 0.192 2.864 ...
##  $ logInfo : logi NA
##  $ count   : int 50
##  $ msg     : chr "budget exhausted"
##  $ modelFit:List of 3
##   ..$ rfFit:List of 17
##   .. ..$ call           : language randomForest(x = x, y = y)
##   .. ..$ type           : chr "regression"
##   .. ..$ predicted      : num [1:49] 0.458 29.661 24.833 31.003 23.679 ...
##   .. ..$ mse            : num [1:500] 360 410 327 188 133 ...
##   .. ..$ rsq            : num [1:500] -0.00606 -0.14405 0.08703 0.47442 0.62732 ...
##   .. ..$ oob.times      : int [1:49] 182 193 179 191 188 201 188 176 174 182 ...
##   .. ..$ importance     : num [1:2, 1] 7925 7835
##   .. .. ..- attr(*, "dimnames")=List of 2
##   .. .. .. ..$ : chr [1:2] "1" "2"
##   .. .. .. ..$ : chr "IncNodePurity"
##   .. ..$ importanceSD   : NULL
##   .. ..$ localImportance: NULL
##   .. ..$ proximity      : NULL
##   .. ..$ ntree          : num 500
##   .. ..$ mtry           : num 1
##   .. ..$ forest         :List of 11
##   .. .. ..$ ndbigtree    : int [1:500] 7 11 9 9 9 11 7 7 7 5 ...
##   .. .. ..$ nodestatus   : int [1:17, 1:500] -3 -3 -1 -3 -1 -1 -1 0 0 0 ...
##   .. .. ..$ leftDaughter : int [1:17, 1:500] 2 4 0 6 0 0 0 0 0 0 ...
##   .. .. ..$ rightDaughter: int [1:17, 1:500] 3 5 0 7 0 0 0 0 0 0 ...
##   .. .. ..$ nodepred     : num [1:17, 1:500] 5.288 0.241 49.701 0.18 2.864 ...
##   .. .. ..$ bestvar      : int [1:17, 1:500] 1 1 0 2 0 0 0 0 0 0 ...
##   .. .. ..$ xbestsplit   : num [1:17, 1:500] 48 19 0 82.5 0 0 0 0 0 0 ...
##   .. .. ..$ ncat         : num [1:2] 1 1
##   .. .. ..$ nrnodes      : int 17
##   .. .. ..$ ntree        : num 500
##   .. .. ..$ xlevels      :List of 2
##   .. .. .. ..$ : num 0
##   .. .. .. ..$ : num 0
##   .. ..$ coefs          : NULL
##   .. ..$ y              : num [1:49, 1] 0.226 28.891 59.309 0.192 2.864 ...
##   .. ..$ test           : NULL
##   .. ..$ inbag          : NULL
##   .. ..- attr(*, "class")= chr "randomForest"
##   ..$ x    : num [1:49, 1:2] 9 72 45 15 26 57 70 93 33 88 ...
##   ..$ y    : num [1:49, 1] 0.226 28.891 59.309 0.192 2.864 ...
##   ..- attr(*, "class")= chr "spotRandomForest"
##  $ ybestVec: num [1:48] 0.192 0.192 0.192 0.192 0.192 ...
\end{verbatim}

\texttt{SPOT} generates many information which can be used for a
statistical analysis. For example, the best configuration found can be
displayed as follows:

\begin{Shaded}
\begin{Highlighting}[]
\FunctionTok{cbind}\NormalTok{(resRf}\SpecialCharTok{$}\NormalTok{xbest, resRf}\SpecialCharTok{$}\NormalTok{ybest)}
\end{Highlighting}
\end{Shaded}

\begin{verbatim}
##      [,1] [,2]        [,3]
## [1,]    3   82 0.001232935
\end{verbatim}

\texttt{SPOT} recommends using \texttt{temp} =

\begin{Shaded}
\begin{Highlighting}[]
\NormalTok{resRf}\SpecialCharTok{$}\NormalTok{xbest[}\DecValTok{1}\NormalTok{]}
\end{Highlighting}
\end{Shaded}

\begin{verbatim}
## [1] 3
\end{verbatim}

and \texttt{tmax} =

\begin{Shaded}
\begin{Highlighting}[]
\NormalTok{resRf}\SpecialCharTok{$}\NormalTok{xbest[}\DecValTok{2}\NormalTok{] }
\end{Highlighting}
\end{Shaded}

\begin{verbatim}
## [1] 82
\end{verbatim}

These parameter settings differ significantly from the default values.
These values also make sense: the starting temperature \texttt{temp}
should be low and the number of evaluations at each temperature
\texttt{tmax} should be high. Hence, worse solutions will rarely be
accepted by \texttt{SANN}, which leads to a very localized search. This
is a good configuration for this example, since the sphere function is
unimodal.

\hypertarget{details-1-the-spot-interface}{%
\subsection{\texorpdfstring{Details 1: The \texttt{spot()}
Interface}{Details 1: The spot() Interface}}\label{details-1-the-spot-interface}}

\hypertarget{arguments}{%
\subsubsection{Arguments}\label{arguments}}

\texttt{SPOT} uses the same interface as \texttt{R}'s standard
\texttt{optim()} function, i.e., arguments reported in the following
Table can be used:

\begin{longtable}[]{@{}ll@{}}
\toprule
name & description\tabularnewline
\midrule
\endhead
\texttt{x} & Optional start point (or set of start points), specified as
a matrix.\tabularnewline
& One row for each point, and one column for each optimized
parameter.\tabularnewline
\texttt{fun} & Objective function. It should receive a matrix \texttt{x}
and return a matrix \texttt{y}.\tabularnewline
& In case the function uses external code and is noisy, an additional
seed parameter may be used,\tabularnewline
& see the \texttt{control\$seedFun} argument in the function
documentation for details.\tabularnewline
\texttt{lower} & Vector that defines the lower boundary of search
space\tabularnewline
\texttt{upper} & Vector that defines the upper boundary of search
space\tabularnewline
\texttt{control} & List of additional settings\tabularnewline
\bottomrule
\end{longtable}

Note, take care of consistency of \texttt{upper}, \texttt{lower} and, if
specified, \texttt{x}. In cases of inconsistency, the dimension of the
variable \texttt{lower} will be taken into account to establish the
dimension of the problem.

\hypertarget{return-values}{%
\subsubsection{Return Values}\label{return-values}}

The function \texttt{spot()} returns a list with the values shown in the
following Table.

\begin{longtable}[]{@{}lll@{}}
\toprule
name & type & description\tabularnewline
\midrule
\endhead
\texttt{xbest} & \texttt{matrix} & Parameters of the best found
solution\tabularnewline
\texttt{ybest} & \texttt{matrix} & Objective function value of the best
found solution\tabularnewline
\texttt{x} & \texttt{matrix} & Archive of all evaluation
parameters\tabularnewline
\texttt{y} & \texttt{matrix} & Archive of the respective objective
function values\tabularnewline
\texttt{count} & \texttt{integer} & Number of performed objective
function evaluations\tabularnewline
\texttt{msg} & \texttt{character} & Message specifying the reason of
termination\tabularnewline
\texttt{modelFit} & \texttt{list} & The fit of the model from the last
\texttt{SPOT} iteration,\tabularnewline
& & i.e., an object returned by the last call to the
function\tabularnewline
& & specified by \texttt{control\$model}\tabularnewline
\bottomrule
\end{longtable}

\hypertarget{details-2-spot-configuration}{%
\subsection{Details 2: SPOT
Configuration}\label{details-2-spot-configuration}}

\texttt{SPOT} configuration settings are listed in following Table.

\begin{longtable}[]{@{}lll@{}}
\toprule
name & default & description\tabularnewline
\midrule
\endhead
\texttt{funEvals} & \texttt{20} & Budget of function evaluations.
\texttt{spot}\tabularnewline
& & uses no more than funEvals evaluations of fun.\tabularnewline
\texttt{types} & \texttt{numeric} & Vector of data type of each variable
as a string.\tabularnewline
\texttt{design} & \texttt{designLHD} & A function that creates an
initial design of experiment.\tabularnewline
& & Functions that accept the same parameters, and return a matrix
like\tabularnewline
& & \texttt{designLHD} or \texttt{designUniformRandom} can be
used.\tabularnewline
\texttt{designControl} & \texttt{list()} & List of controls passed to
the \texttt{control} list of the \texttt{design}
function.\tabularnewline
\texttt{model} & \texttt{buildKriging} & Function that builds a model of
the observed data. Functions that accept\tabularnewline
& & the same parameters, and return a matrix like \texttt{buildKriging}
or\tabularnewline
& & \texttt{buildRandomForest} can be used.\tabularnewline
\texttt{modelControl} & \texttt{list()} & List of controls passed to the
\texttt{control} list of the \texttt{model} function.\tabularnewline
\texttt{optimizer} & \texttt{optimLHD} & Function that is used to
optimize the \texttt{model}, finding the most promising\tabularnewline
& & candidate solutions. Functions that accept the same
parameters,\tabularnewline
& & and return a matrix like \texttt{optimLHD} or \texttt{optimLBFGSB}
can be used.\tabularnewline
\texttt{optimizerControl} & \texttt{list()} & List of controls passed to
the \texttt{control} list of the \texttt{optimizer}
function.\tabularnewline
\texttt{noise} & \texttt{FALSE} & Logical, whether the objective
function has noise.\tabularnewline
\texttt{OCBA} & \texttt{FALSE} & Logical, indicating whether OCBA should
be used in case of a noisy\tabularnewline
& & objective function.\tabularnewline
& & \texttt{replicates} should be larger than one, the initial
experimental\tabularnewline
& & design (see \texttt{design}) should have replicates larger than
one.\tabularnewline
\texttt{OCBAbudget} & \texttt{3} & Number of objective function
evaluations that OCBA can distribute\tabularnewline
& & in each iteration.\tabularnewline
\texttt{replicates} & \texttt{1} & Number of times a candidate solution
is initially evaluated, i.e.,\tabularnewline
& & in the initial design, or when created by the
optimizer.\tabularnewline
\texttt{seedFun} & \texttt{NA} & Initial seed for the objective function
in case of noise.\tabularnewline
& & The default means that no seed is set.\tabularnewline
& & This seed is by default set prior to each evaluation.\tabularnewline
& & A replicated evaluation will receive an incremented value of the
seed.\tabularnewline
\texttt{seedSPOT} & \texttt{1} & Value used to initialize the random
number generator.\tabularnewline
& & It ensures that experiments are reproducible.\tabularnewline
\texttt{duplicate} & \texttt{"EXPLORE"} & In case of a deterministic
(non-noisy) objective function,\tabularnewline
& & this handles duplicated candidate\tabularnewline
& & solutions. By default
(\texttt{duplicate\ =\ "EXPLORE"}),\tabularnewline
& & duplicates are replaced by new candidate solutions,\tabularnewline
& & generated by random sampling with uniform
distribution.\tabularnewline
\texttt{plots} & \texttt{FALSE} & Logical. Should the progress be
tracked by a line plot?\tabularnewline
\bottomrule
\end{longtable}

Note:

\begin{itemize}
\tightlist
\item
  \texttt{seedFun}: Sometimes, the user may want to call external code
  using random numbers. To allow for that case, the user can specify an
  objective function (\texttt{fun}), which has a second parameter
  \texttt{seed}, in addition to first parameter (matrix \texttt{x}).
  This seed can then be passed to the external code, for random number
  generator initialization. See end of examples section in the
  documentation of \texttt{SPOT} for a demonstration.
\item
  \texttt{duplicate}: If desired, the user can set this to
  \texttt{"STOP"}, which means that the optimization stops and results
  are returned to the user (with a warning). This may be desirable, as
  duplicates can be a indicator of convergence, or problems with the
  configuration. In case of noise, duplicates are allowed regardless of
  this parameter.
\end{itemize}

The configuration list stores information about \texttt{SPOT} specific
settings. All settings not specified in the list will be set to their
default values.

\hypertarget{details-3-initial-designs}{%
\subsection{Details 3: Initial
Designs}\label{details-3-initial-designs}}

The problem design comprehends the information about the optimization
problem, e.g., the starting point \texttt{x0\ =\ (-1,1,-1)} belongs to
the problem design. A \emph{region of interest} (ROI) specifies
algorithm parameters and associated lower and upper bounds for the
algorithm parameters.

\begin{itemize}
\tightlist
\item
  Values for \texttt{temp} are chosen from the interval \([1; 100]\).
\item
  Values for \texttt{tmax} are chosen from the interval \([1; 100]\).
\end{itemize}

\texttt{SPOT} implements a sequential approach, i.e., the available
budget is not used in one step. Rather, sequential steps are made,
comprised of model training, optimization, and evaluation. To initialize
this procedure, some first data set is required to train the first,
coarse-grained meta model.

\hypertarget{example-modifying-the-number-of-function-evaluations}{%
\subsubsection{Example: Modifying the number of function
evaluations}\label{example-modifying-the-number-of-function-evaluations}}

For example:

\begin{itemize}
\tightlist
\item
  the total number of \texttt{SANN} algorithm runs, i.e., the available
  budget, can be set to 10 using \texttt{funEvals\ =\ 10},
\item
  the size of the initial design can be modified via
  \texttt{designControl\$size},
\item
  the number of repeated evaluations of the initial design can be
  modified via \texttt{designControl\$replicates}, and
\item
  the number of replicates used for each new algorithm design point
  suggested by \texttt{SPOT} can be modified via \texttt{replicates}.
\end{itemize}

\begin{Shaded}
\begin{Highlighting}[]
\NormalTok{spotConfig10 }\OtherTok{\textless{}{-}} \FunctionTok{list}\NormalTok{(}
  \AttributeTok{funEvals =} \DecValTok{10}\NormalTok{,}
  \AttributeTok{designControl =} \FunctionTok{list}\NormalTok{(}
    \AttributeTok{size =} \DecValTok{6}\NormalTok{,}
    \AttributeTok{replicates =} \DecValTok{1}
\NormalTok{  ),}
  \AttributeTok{noise =} \ConstantTok{TRUE}\NormalTok{,}
  \AttributeTok{seedFun =} \DecValTok{1}\NormalTok{,}
  \AttributeTok{seedSPOT =} \DecValTok{1}\NormalTok{,}
  \AttributeTok{replicates =} \DecValTok{2}\NormalTok{,}
  \AttributeTok{model =}\NormalTok{ buildRandomForest }
\NormalTok{)}
\end{Highlighting}
\end{Shaded}

Using this configuration, the budget will be spent as follows:

\begin{itemize}
\tightlist
\item
  Six initial algorithm design points (parameter sets) will be created,
  and each will be evaluated just once.
\item
  Afterwards, \texttt{SPOT} will have a remaining budget of four
  evaluations. These can be spent on sequentially testing two additional
  design points. Each of those will be evaluated twice.
\end{itemize}

\begin{Shaded}
\begin{Highlighting}[]
\NormalTok{res10 }\OtherTok{\textless{}{-}} \FunctionTok{spot}\NormalTok{(}\AttributeTok{x =} \ConstantTok{NULL}\NormalTok{,}
              \AttributeTok{fun=}\NormalTok{sann2spot,}
              \AttributeTok{lower=}\NormalTok{lower,}
              \AttributeTok{upper=}\NormalTok{upper,}
              \AttributeTok{control=}\NormalTok{spotConfig10)}
\end{Highlighting}
\end{Shaded}

Results from these runs can be displayed as follows. The first two
columns show the settings of the algorithm parameters \texttt{temp} and
\texttt{tmax}, respectively. The third row shows the corresponding
function values:

\begin{Shaded}
\begin{Highlighting}[]
\FunctionTok{cbind}\NormalTok{(res10}\SpecialCharTok{$}\NormalTok{x, res10}\SpecialCharTok{$}\NormalTok{y)}
\end{Highlighting}
\end{Shaded}

\begin{verbatim}
##            [,1]      [,2]         [,3]
##  [1,]  8.954322 63.418391 2.935315e-01
##  [2,] 93.392836 43.125099 1.073285e+02
##  [3,] 25.643432 26.240373 1.614929e+01
##  [4,] 70.072590 96.524378 1.129012e+01
##  [5,] 47.651660 67.384965 3.933353e+00
##  [6,] 61.529701  8.874296 7.312226e+01
##  [7,] 12.121097 80.404416 8.496108e-03
##  [8,] 12.121097 80.404416 4.645919e-01
##  [9,]  8.156284 79.499694 3.392544e-01
## [10,]  8.156284 79.499694 2.126833e-02
\end{verbatim}

The results show the desired outcome: the first six configurations are
unique, the following four contain replications (two identical settings,
but with different stochastic result).

\hypertarget{initial-designs}{%
\subsubsection{Initial Designs}\label{initial-designs}}

The function \texttt{designLHD()} is the default setting to generate
designs. A simple one-dimensional design with values from the interval
\([-1, 1]\) can be generated as follows:

\begin{Shaded}
\begin{Highlighting}[]
\FunctionTok{designLHD}\NormalTok{(}\AttributeTok{x=}\ConstantTok{NULL}\NormalTok{, }\SpecialCharTok{{-}}\DecValTok{1}\NormalTok{, }\DecValTok{1}\NormalTok{) }
\end{Highlighting}
\end{Shaded}

\begin{verbatim}
##              [,1]
##  [1,]  0.33411001
##  [2,]  0.46001819
##  [3,] -0.75700111
##  [4,]  0.69957356
##  [5,] -0.06232477
##  [6,]  0.92854939
##  [7,] -0.32472347
##  [8,] -0.97695432
##  [9,]  0.09826695
## [10,] -0.54409864
\end{verbatim}

A more complicated design, which consists of numeric values for the
first variable in the range from -1 to +1, of integer values for the
second variable in the range from -2 to 4, of integer values for the
third variable in the range from 1 to 9, and with two factors 0 and 1
for the fourth variable, can be generated as follows:

\begin{Shaded}
\begin{Highlighting}[]
\FunctionTok{designLHD}\NormalTok{(}\AttributeTok{x=}\ConstantTok{NULL}\NormalTok{, }\FunctionTok{c}\NormalTok{(}\SpecialCharTok{{-}}\DecValTok{1}\NormalTok{,}\SpecialCharTok{{-}}\DecValTok{2}\NormalTok{,}\DecValTok{1}\NormalTok{,}\DecValTok{0}\NormalTok{),}\FunctionTok{c}\NormalTok{(}\DecValTok{1}\NormalTok{,}\DecValTok{4}\NormalTok{,}\DecValTok{9}\NormalTok{,}\DecValTok{1}\NormalTok{)}
\NormalTok{          , }\AttributeTok{control=}\FunctionTok{list}\NormalTok{(}\AttributeTok{size=}\DecValTok{5}\NormalTok{, }\AttributeTok{retries=}\DecValTok{100}\NormalTok{, }\AttributeTok{types=}\FunctionTok{c}\NormalTok{(}\StringTok{"numeric"}\NormalTok{,}\StringTok{"integer"}\NormalTok{,}\StringTok{"factor"}\NormalTok{,}\StringTok{"factor"}\NormalTok{)))}
\end{Highlighting}
\end{Shaded}

\begin{verbatim}
##             [,1] [,2] [,3] [,4]
## [1,]  0.99895818    0    4    1
## [2,] -0.83535083   -1    7    1
## [3,] -0.30061528    4    2    1
## [4,]  0.56058677    2    9    0
## [5,] -0.05172332    0    3    0
\end{verbatim}

Designs can also be combined as illustrated in the following example.
The corresponding plot shows the resulting design.

\begin{Shaded}
\begin{Highlighting}[]
\FunctionTok{set.seed}\NormalTok{(}\DecValTok{123}\NormalTok{)}
\NormalTok{x1 }\OtherTok{\textless{}{-}} \FunctionTok{designLHD}\NormalTok{(}\AttributeTok{x=}\ConstantTok{NULL}\NormalTok{,}\FunctionTok{c}\NormalTok{(}\SpecialCharTok{{-}}\DecValTok{1}\NormalTok{,}\SpecialCharTok{{-}}\DecValTok{1}\NormalTok{),}\FunctionTok{c}\NormalTok{(}\DecValTok{1}\NormalTok{,}\DecValTok{1}\NormalTok{),}\AttributeTok{control=}\FunctionTok{list}\NormalTok{(}\AttributeTok{size=}\DecValTok{50}\NormalTok{,}\AttributeTok{retries=}\DecValTok{100}\NormalTok{))}
\NormalTok{x2 }\OtherTok{\textless{}{-}} \FunctionTok{designLHD}\NormalTok{(x1,}\FunctionTok{c}\NormalTok{(}\SpecialCharTok{{-}}\DecValTok{2}\NormalTok{,}\SpecialCharTok{{-}}\DecValTok{2}\NormalTok{),}\FunctionTok{c}\NormalTok{(}\DecValTok{2}\NormalTok{,}\DecValTok{2}\NormalTok{),}\AttributeTok{control=}\FunctionTok{list}\NormalTok{(}\AttributeTok{size=}\DecValTok{50}\NormalTok{,}\AttributeTok{retries=}\DecValTok{100}\NormalTok{))}
\FunctionTok{plot}\NormalTok{(x2,}\AttributeTok{pch=}\DecValTok{1}\NormalTok{)}
\FunctionTok{points}\NormalTok{(x1, }\AttributeTok{pch=}\DecValTok{4}\NormalTok{)}
\end{Highlighting}
\end{Shaded}

\begin{figure}
\centering
\includegraphics{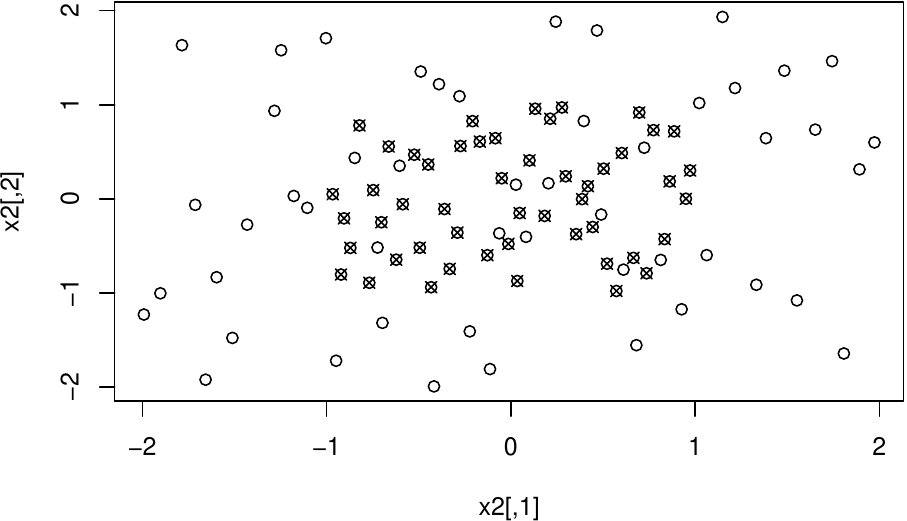}
\caption{Combined designs}
\end{figure}

\hypertarget{uniform-random-sampling-designs}{%
\subsubsection{Uniform Random Sampling
Designs}\label{uniform-random-sampling-designs}}

Designs based on uniform random sampling can be generated with the
function \texttt{designUniformRandom()} as follows:

\begin{Shaded}
\begin{Highlighting}[]
\FunctionTok{designUniformRandom}\NormalTok{(}\AttributeTok{x=}\ConstantTok{NULL}\NormalTok{,}\FunctionTok{c}\NormalTok{(}\SpecialCharTok{{-}}\DecValTok{1}\NormalTok{,}\DecValTok{0}\NormalTok{),}\FunctionTok{c}\NormalTok{(}\DecValTok{1}\NormalTok{,}\DecValTok{10}\NormalTok{),}\AttributeTok{control=}\FunctionTok{list}\NormalTok{(}\AttributeTok{size=}\DecValTok{5}\NormalTok{))}
\end{Highlighting}
\end{Shaded}

\begin{verbatim}
##            [,1]     [,2]
## [1,]  0.3480606 3.447254
## [2,] -0.3917176 2.413147
## [3,]  0.7143919 8.768452
## [4,] -0.4602907 1.681719
## [5,] -0.5792664 9.799514
\end{verbatim}

\hypertarget{details-4-models}{%
\subsection{Details 4: Models}\label{details-4-models}}

In \texttt{SPOT}, a meta model or surrogate model is used to determine
promising algorithm design points. To that end, it aims to learn the
relation between algorithm parameters and the corresponding algorithm
performance. The different models and their options are described next.

\hypertarget{kriging-models}{%
\subsubsection{Kriging Models}\label{kriging-models}}

As default, the \texttt{buildKriging()} function is used for modeling.
This function builds a Kriging model (also known as Gaussian process
regression) loosely based on code by \cite{Forr08a}. Kriging models are
based on measures of similarity, or kernels. Here, a Gaussian kernel is
used: \[
k(x,x') = \exp \left(-\sum^n_{i=1} \theta_i |x_i-x'_i|^{p_i} \right).
\] By default exponents are fixed at a value of two, \(p_i = 2\),
(\(i=1,\ldots,n\)), and the nugget effect (or regularization constant)
is used. To correct the uncertainty estimates in case of using the
nugget effect, re-interpolation is also by default turned on (see
\cite{Forr08a} for more details on these features of the model).

\hypertarget{example-spot-as-a-surrogate-model-based-algorithm}{%
\subsubsection{\texorpdfstring{Example: \texttt{SPOT} as a surrogate
model based
algorithm}{Example: SPOT as a surrogate model based algorithm}}\label{example-spot-as-a-surrogate-model-based-algorithm}}

The following code exemplifies how a Kriging model is built with an
artificial data set. Note, this example exemplifies how \texttt{SPOT}
can be used as a surrogate model based optimization algorithm, i.e., no
algorithm is tuned. The Branin function, which is implemented in
\texttt{SPOT} as \texttt{funBranin()} is used in our example. \[
f(x_1,x_2) = 
\left(x_2  - 5.1x_1^2/(4 \pi^2)  + 5 x_1 / \pi  - 6 \right)^2 + 10  (1 - 1/8 \pi)  \cos(x_1 ) + 10
\]

We proceed as follows:

\begin{enumerate}
\def\labelenumi{\arabic{enumi}.}
\tightlist
\item
  Create 20 design points
\item
  Compute observations at design points (for Branin function)
\item
  Create model with default settings
\item
  Print model parameters
\item
  Define a new location
\item
  Predict at new location
\item
  And compare result with true value at location
\end{enumerate}

\begin{Shaded}
\begin{Highlighting}[]
\FunctionTok{set.seed}\NormalTok{(}\DecValTok{1}\NormalTok{)}
\NormalTok{x }\OtherTok{\textless{}{-}} \FunctionTok{cbind}\NormalTok{(}\FunctionTok{runif}\NormalTok{(}\DecValTok{20}\NormalTok{)}\SpecialCharTok{*}\DecValTok{15{-}5}\NormalTok{, }\FunctionTok{runif}\NormalTok{(}\DecValTok{20}\NormalTok{)}\SpecialCharTok{*}\DecValTok{15}\NormalTok{)}
\NormalTok{y }\OtherTok{\textless{}{-}} \FunctionTok{funBranin}\NormalTok{(x)}
\NormalTok{fit }\OtherTok{\textless{}{-}} \FunctionTok{buildKriging}\NormalTok{(x,y,}\AttributeTok{control =} \FunctionTok{list}\NormalTok{(}\AttributeTok{algTheta=}\NormalTok{optimLHD))}
\FunctionTok{print}\NormalTok{(fit)}
\end{Highlighting}
\end{Shaded}

\begin{verbatim}
## ------------------------
## Forrester Kriging model.
## ------------------------
## Estimated activity parameters (theta) sorted 
## from most to least important variable 
## x1  x2 
## 7.575502 1.361329
##  
## exponent(s) p:
## 2
##  
## Estimated regularization constant (or nugget) lambda:
## 5.239774e-06
##  
## Number of Likelihood evaluations during MLE:
## 600
## ------------------------
\end{verbatim}

\begin{Shaded}
\begin{Highlighting}[]
\NormalTok{newloc }\OtherTok{\textless{}{-}} \FunctionTok{matrix}\NormalTok{(}\FunctionTok{c}\NormalTok{(}\DecValTok{1}\NormalTok{,}\DecValTok{2}\NormalTok{),}\AttributeTok{nrow =}\DecValTok{1}\NormalTok{ )}
\FunctionTok{predict}\NormalTok{(fit,newloc)}
\end{Highlighting}
\end{Shaded}

\begin{verbatim}
## $y
## [1] 21.08753
\end{verbatim}

\begin{Shaded}
\begin{Highlighting}[]
\FunctionTok{funBranin}\NormalTok{(newloc)}
\end{Highlighting}
\end{Shaded}

\begin{verbatim}
##          [,1]
## [1,] 21.62764
\end{verbatim}

\hypertarget{handling-factor-variables-in-the-kriging-model}{%
\subsection{Handling Factor Variables in the Kriging
Model}\label{handling-factor-variables-in-the-kriging-model}}

Sometimes, parameters optimized or modeled by \texttt{SPOT} functions
will not be numerical, but rather categorical. This may, e.g., occur if
an evolutionary algorithm is tuned: while some parameters like mutation
rates may be real valued, the selection between different mutation
operators may be a categorical parameter.

Hence, if \(x_i\), the \(i\)th dimension of a parameter configuration
\(\vec{x}\), is a factor variable (see parameter \texttt{types}),
Hamming distance, that determines the number of positions at which the
corresponding values are different, will be used instead of
\(|x_i-x'_i|\). To illustrate how factor variables can be handled, we
create a test function that uses a factor variable. Here, the third
dimension \(x_3\) is categorical.

\begin{Shaded}
\begin{Highlighting}[]
\NormalTok{braninFunctionFactor }\OtherTok{\textless{}{-}} \ControlFlowTok{function}\NormalTok{ (x) \{  }
\NormalTok{y }\OtherTok{\textless{}{-}}\NormalTok{ (x[}\DecValTok{2}\NormalTok{] }\SpecialCharTok{{-}} \FloatTok{5.1} \SpecialCharTok{/}\NormalTok{ (}\DecValTok{4} \SpecialCharTok{*}\NormalTok{ pi}\SpecialCharTok{\^{}}\DecValTok{2}\NormalTok{) }\SpecialCharTok{*}\NormalTok{ (x[}\DecValTok{1}\NormalTok{]}\SpecialCharTok{\^{}}\DecValTok{2}\NormalTok{) }
      \SpecialCharTok{+} \DecValTok{5} \SpecialCharTok{/}\NormalTok{ pi }\SpecialCharTok{*}\NormalTok{ x[}\DecValTok{1}\NormalTok{] }\SpecialCharTok{{-}} \DecValTok{6}\NormalTok{)}\SpecialCharTok{\^{}}\DecValTok{2} 
      \SpecialCharTok{+} \DecValTok{10} \SpecialCharTok{*}\NormalTok{ (}\DecValTok{1} \SpecialCharTok{{-}} \DecValTok{1} \SpecialCharTok{/}\NormalTok{ (}\DecValTok{8} \SpecialCharTok{*}\NormalTok{ pi)) }\SpecialCharTok{*} \FunctionTok{cos}\NormalTok{(x[}\DecValTok{1}\NormalTok{]) }\SpecialCharTok{+} \DecValTok{10}
    \ControlFlowTok{if}\NormalTok{(x[}\DecValTok{3}\NormalTok{] }\SpecialCharTok{==} \DecValTok{1}\NormalTok{)}
\NormalTok{        y }\OtherTok{\textless{}{-}}\NormalTok{ y }\SpecialCharTok{+} \DecValTok{1}
    \ControlFlowTok{else} \ControlFlowTok{if}\NormalTok{(x[}\DecValTok{3}\NormalTok{]}\SpecialCharTok{==}\DecValTok{2}\NormalTok{)}
\NormalTok{        y }\OtherTok{\textless{}{-}}\NormalTok{ y }\SpecialCharTok{{-}} \DecValTok{1}
\NormalTok{    y  }
\NormalTok{\}}
\end{Highlighting}
\end{Shaded}

To test how this affects our model, we first generate some training data
and fit the model with default settings, which ignores factor
information and uses the standard kernel.

\begin{Shaded}
\begin{Highlighting}[]
\FunctionTok{set.seed}\NormalTok{(}\DecValTok{1}\NormalTok{)}
\DocumentationTok{\#\# Replace x with new data}
\NormalTok{x }\OtherTok{\textless{}{-}} \FunctionTok{cbind}\NormalTok{(}\FunctionTok{runif}\NormalTok{(}\DecValTok{50}\NormalTok{)}\SpecialCharTok{*}\DecValTok{15{-}5}\NormalTok{,}\FunctionTok{runif}\NormalTok{(}\DecValTok{50}\NormalTok{)}\SpecialCharTok{*}\DecValTok{15}\NormalTok{,}\FunctionTok{sample}\NormalTok{(}\DecValTok{1}\SpecialCharTok{:}\DecValTok{3}\NormalTok{,}\DecValTok{50}\NormalTok{,}\AttributeTok{replace=}\ConstantTok{TRUE}\NormalTok{))}
\DocumentationTok{\#\#}
\NormalTok{y }\OtherTok{\textless{}{-}} \FunctionTok{as.matrix}\NormalTok{(}\FunctionTok{apply}\NormalTok{(x,}\DecValTok{1}\NormalTok{,braninFunctionFactor))}
\NormalTok{fitDefault }\OtherTok{\textless{}{-}} \FunctionTok{buildKriging}\NormalTok{(x,y,}\AttributeTok{control =} \FunctionTok{list}\NormalTok{(}\AttributeTok{algTheta=}\NormalTok{optimLBFGSB))}
\end{Highlighting}
\end{Shaded}

Afterwards we fit the model, which includes information about the factor
variable.

\begin{Shaded}
\begin{Highlighting}[]
\NormalTok{fitFactor }\OtherTok{\textless{}{-}} \FunctionTok{buildKriging}\NormalTok{(x,y,}\AttributeTok{control =} \FunctionTok{list}\NormalTok{(}\AttributeTok{algTheta=}\NormalTok{optimLBFGSB,}\AttributeTok{types=}\FunctionTok{c}\NormalTok{(}\StringTok{"numeric"}\NormalTok{,}\StringTok{"numeric"}\NormalTok{,}\StringTok{"factor"}\NormalTok{)))}
\end{Highlighting}
\end{Shaded}

We generate some new, unseen data for testing and perform some
predictions with both models.

\begin{Shaded}
\begin{Highlighting}[]
\DocumentationTok{\#\#Replace xtest with new data}
\NormalTok{xtest }\OtherTok{\textless{}{-}} \FunctionTok{cbind}\NormalTok{(}\FunctionTok{runif}\NormalTok{(}\DecValTok{200}\NormalTok{)}\SpecialCharTok{*}\DecValTok{15{-}5}\NormalTok{,}\FunctionTok{runif}\NormalTok{(}\DecValTok{200}\NormalTok{)}\SpecialCharTok{*}\DecValTok{15}\NormalTok{,}\FunctionTok{sample}\NormalTok{(}\DecValTok{1}\SpecialCharTok{:}\DecValTok{3}\NormalTok{,}\DecValTok{200}\NormalTok{,}\AttributeTok{replace=}\ConstantTok{TRUE}\NormalTok{))}
\DocumentationTok{\#\#}
\NormalTok{ytest }\OtherTok{\textless{}{-}} \FunctionTok{as.matrix}\NormalTok{(}\FunctionTok{apply}\NormalTok{(xtest,}\DecValTok{1}\NormalTok{,braninFunctionFactor))}
\DocumentationTok{\#\# Predict test data with both models, and compute error}
\NormalTok{ypredDef }\OtherTok{\textless{}{-}} \FunctionTok{predict}\NormalTok{(fitDefault,xtest)}\SpecialCharTok{$}\NormalTok{y}
\NormalTok{ypredFact }\OtherTok{\textless{}{-}} \FunctionTok{predict}\NormalTok{(fitFactor,xtest)}\SpecialCharTok{$}\NormalTok{y}
\FunctionTok{mean}\NormalTok{((ypredDef}\SpecialCharTok{{-}}\NormalTok{ytest)}\SpecialCharTok{\^{}}\DecValTok{2}\NormalTok{)}
\end{Highlighting}
\end{Shaded}

\begin{verbatim}
## [1] 0.153288
\end{verbatim}

\begin{Shaded}
\begin{Highlighting}[]
\FunctionTok{mean}\NormalTok{((ypredFact}\SpecialCharTok{{-}}\NormalTok{ytest)}\SpecialCharTok{\^{}}\DecValTok{2}\NormalTok{)}
\end{Highlighting}
\end{Shaded}

\begin{verbatim}
## [1] 0.1482254
\end{verbatim}

The error of the factor-aware model is lower. This demonstrates that
users should make sure to declare the nature of the modeled variables
correctly, via the \texttt{types} variable.

\hypertarget{details-5-optimization-on-the-meta-model}{%
\subsection{Details 5: Optimization on the Meta
Model}\label{details-5-optimization-on-the-meta-model}}

Minimization by \emph{Latin Hyper cube Sampling} (LHS) is the default
optimizer used for finding the next algorithm design parameters on the
meta model. The LHS procedure generates a set of LHD points. The
function \texttt{optimLHD()} uses LHS to optimize a specified target
function as follows:

\begin{itemize}
\tightlist
\item
  A Latin hyper cube Design (LHD) is created with \texttt{designLHD()},
  then evaluated by the objective function.
\item
  All results are reported, including the best (minimal) objective
  value, and corresponding design point.
\end{itemize}

A standalone optimization run using \texttt{optimLHD()} can be
implemented as follows. It uses 100 design points as a default value.

\begin{Shaded}
\begin{Highlighting}[]
\NormalTok{resOptimumLHD }\OtherTok{\textless{}{-}} \FunctionTok{optimLHD}\NormalTok{(}\AttributeTok{x=}\ConstantTok{NULL}\NormalTok{,}\AttributeTok{fun =}\NormalTok{ funSphere,}\AttributeTok{lower =} \FunctionTok{c}\NormalTok{(}\SpecialCharTok{{-}}\DecValTok{10}\NormalTok{,}\SpecialCharTok{{-}}\DecValTok{20}\NormalTok{),}\AttributeTok{upper=}\FunctionTok{c}\NormalTok{(}\DecValTok{20}\NormalTok{,}\DecValTok{8}\NormalTok{))}
\FunctionTok{str}\NormalTok{(resOptimumLHD)}
\end{Highlighting}
\end{Shaded}

\begin{verbatim}
## List of 6
##  $ x    : num [1:100, 1:2] -5.528 -6.509 0.718 -4.327 15.608 ...
##  $ y    : num [1:100, 1] 48.8 62.5 295.3 260.7 366.2 ...
##  $ xbest: num [1, 1:2] 0.136 -0.558
##  $ ybest: num [1, 1] 0.33
##  $ count: num 100
##  $ msg  : chr "success"
\end{verbatim}

\begin{Shaded}
\begin{Highlighting}[]
\NormalTok{resOptimumLHD}\SpecialCharTok{$}\NormalTok{ybest}
\end{Highlighting}
\end{Shaded}

\begin{verbatim}
##         [,1]
## [1,] 0.32992
\end{verbatim}

Using more sophisticated algorithms, e.g., the variable metric algorithm
(L-BFGS-B), might lead to better results. However, they are not as
robust as the simple \texttt{optimLHD()} search. Also, \texttt{L-BFGS-B}
is a pure local search, which may not be ideal to solve potentially
multi-modal tuning problems.

\begin{Shaded}
\begin{Highlighting}[]
\NormalTok{resOptimBFGS }\OtherTok{\textless{}{-}} \FunctionTok{optimLBFGSB}\NormalTok{(}\AttributeTok{x=}\ConstantTok{NULL}\NormalTok{,}\AttributeTok{fun =}\NormalTok{ funSphere,}\AttributeTok{lower =} \FunctionTok{c}\NormalTok{(}\SpecialCharTok{{-}}\DecValTok{10}\NormalTok{,}\SpecialCharTok{{-}}\DecValTok{20}\NormalTok{),}\AttributeTok{upper=}\FunctionTok{c}\NormalTok{(}\DecValTok{20}\NormalTok{,}\DecValTok{8}\NormalTok{))}
\NormalTok{resOptimBFGS}\SpecialCharTok{$}\NormalTok{ybest}
\end{Highlighting}
\end{Shaded}

\begin{verbatim}
## [1] 2.098584e-40
\end{verbatim}

Hence, \texttt{SPOT} also includes interfaces to more sophisticated
algorithms, such as differential evolution from \texttt{DEoptim} package
or various methods included in the \texttt{nloptr} package. For further
details on these, see e.g., the help of the corresponding functions.

\hypertarget{details-6-spot-loop-continued-evaluation}{%
\subsection{Details 6: SPOT Loop: Continued
Evaluation}\label{details-6-spot-loop-continued-evaluation}}

Sometimes, users may desire to continue a previously finished
\texttt{SPOT} run. We will demonstrate, how \texttt{SPOT} can be
restarted, reusing the existing data.

\hypertarget{example-spot-with-continued-evaluation}{%
\subsubsection{\texorpdfstring{Example: \texttt{SPOT} with continued
evaluation}{Example: SPOT with continued evaluation}}\label{example-spot-with-continued-evaluation}}

The surrogate model based optimization setting will be used to exemplify
the continued evaluation. The two dimensional sphere function is used as
an objective function (level L1) and \texttt{SPOT} will be used at level
L2. \texttt{SPOT} uses 5 function evaluations.

\begin{Shaded}
\begin{Highlighting}[]
\NormalTok{control01 }\OtherTok{\textless{}{-}} \FunctionTok{list}\NormalTok{(}
    \AttributeTok{designControl =} \FunctionTok{list}\NormalTok{(}\AttributeTok{size =} \DecValTok{5}\NormalTok{, }
                       \AttributeTok{replicates =} \DecValTok{1}\NormalTok{),    }
    \AttributeTok{funEvals =} \DecValTok{5}\NormalTok{)}
\NormalTok{res1 }\OtherTok{\textless{}{-}} \FunctionTok{spot}\NormalTok{(}\AttributeTok{x=}\ConstantTok{NULL}\NormalTok{,funSphere,}
             \AttributeTok{lower =} \FunctionTok{c}\NormalTok{(}\SpecialCharTok{{-}}\DecValTok{2}\NormalTok{,}\SpecialCharTok{{-}}\DecValTok{3}\NormalTok{),}
             \AttributeTok{upper =} \FunctionTok{c}\NormalTok{(}\DecValTok{1}\NormalTok{,}\DecValTok{2}\NormalTok{),}
\NormalTok{             control01)}
\FunctionTok{cbind}\NormalTok{(res1}\SpecialCharTok{$}\NormalTok{x, res1}\SpecialCharTok{$}\NormalTok{y)}
\end{Highlighting}
\end{Shaded}

\begin{verbatim}
##            [,1]       [,2]     [,3]
## [1,] -0.8963358  0.2026923 0.844502
## [2,] -1.7919899 -1.2888788 4.872436
## [3,]  0.6002650 -0.8783081 1.131743
## [4,] -0.5141893 -2.7545115 7.851724
## [5,]  0.3353190  1.1433044 1.419584
\end{verbatim}

Now, we continue with a larger budget. If we would like to add 3
function evaluations, the total number of function evaluations is
\(5+3=8\). To continue a \texttt{SPOT} run, the command
\texttt{spotLoop()} can be used as follows:

\begin{itemize}
\tightlist
\item
  \texttt{spotLoop(x,\ y,\ fun,\ lower,\ upper,\ control,\ ...)}.
\end{itemize}

The arguments are:

\begin{itemize}
\tightlist
\item
  \texttt{x}: the known candidate solutions that the \texttt{SPOT} loop
  is started with, specified as a matrix. One row for each point, and
  one column for each optimized parameter.
\item
  \texttt{y}: the corresponding observations for each solution in
  \texttt{x}, specified as a matrix. One row for each point.
\item
  \texttt{fun}: is the objective function. It should receive a matrix
  \texttt{x} and should return a matrix \texttt{y}.
\item
  \texttt{lower}: is the vector that defines the lower boundary of
  search space. This determines also the dimensionality of the problem.
\item
  \texttt{upper}: is the vector that defines the upper boundary of
  search space.
\item
  \texttt{control}: is the list with control settings for spot.
\item
  \texttt{...}: additional parameters passed to fun.
\end{itemize}

\begin{Shaded}
\begin{Highlighting}[]
\NormalTok{control01}\SpecialCharTok{$}\NormalTok{funEvals }\OtherTok{\textless{}{-}} \DecValTok{8}
\NormalTok{res2 }\OtherTok{\textless{}{-}} \FunctionTok{spotLoop}\NormalTok{(res1}\SpecialCharTok{$}\NormalTok{x,}
\NormalTok{                 res1}\SpecialCharTok{$}\NormalTok{y,}
\NormalTok{                 funSphere,}
                 \AttributeTok{lower =} \FunctionTok{c}\NormalTok{(}\SpecialCharTok{{-}}\DecValTok{2}\NormalTok{,}\SpecialCharTok{{-}}\DecValTok{3}\NormalTok{),}
                 \AttributeTok{upper =} \FunctionTok{c}\NormalTok{(}\DecValTok{1}\NormalTok{,}\DecValTok{2}\NormalTok{),}
\NormalTok{                 control01)}
\FunctionTok{cbind}\NormalTok{(res2}\SpecialCharTok{$}\NormalTok{x, res2}\SpecialCharTok{$}\NormalTok{y)}
\end{Highlighting}
\end{Shaded}

\begin{verbatim}
##            [,1]        [,2]       [,3]
## [1,] -0.8963358  0.20269226 0.84450200
## [2,] -1.7919899 -1.28887878 4.87243633
## [3,]  0.6002650 -0.87830808 1.13174310
## [4,] -0.5141893 -2.75451149 7.85172411
## [5,]  0.3353190  1.14330438 1.41958374
## [6,]  0.6450130  0.20991043 0.46010412
## [7,]  0.7041126 -0.08754435 0.50343851
## [8,] -0.2917148  0.07870320 0.09129173
\end{verbatim}

\hypertarget{sec:plot}{%
\section{Introduction to SPOT's Plot Functions}\label{sec:plot}}

The \texttt{SPOT} package offers three plot functions that can be used
to visualize data or evaluate a model's performance creating 2D and 3D
surface plots.

\begin{itemize}
\tightlist
\item
  \texttt{plotFunction()} plots function objects
\item
  \texttt{plotData()} plots data
\item
  \texttt{plotModel()} plots model objects, created by \texttt{build*}
  functions from the \texttt{SPOT} package.
\end{itemize}

\hypertarget{plotfunction}{%
\subsection{plotFunction}\label{plotfunction}}

The function \texttt{plotFunction()} visualizes the fitness landscape.
It generates a (filled) contour plot or perspective / surface plot of a
function \texttt{f()}. It has several options for changing plotting
styles, colors etc., but default values should be fine in many cases.
The basic arguments are:
\texttt{plotFunction(f\ ,\ lower\ ,\ upper\ ,\ type)}

\hypertarget{example-plot-of-the-sphere-function}{%
\subsubsection{Example: Plot of the sphere
function}\label{example-plot-of-the-sphere-function}}

The following code generates a 2D filled contour plot of the sphere
function. Note that \texttt{plotFunction()} requires a function that
handles matrix objects, so \texttt{funSphere()} is used.

\begin{Shaded}
\begin{Highlighting}[]
\FunctionTok{plotFunction}\NormalTok{(funSphere, }\FunctionTok{rep}\NormalTok{(}\SpecialCharTok{{-}}\DecValTok{1}\NormalTok{,}\DecValTok{2}\NormalTok{), }\FunctionTok{rep}\NormalTok{(}\DecValTok{1}\NormalTok{,}\DecValTok{2}\NormalTok{)) }
\end{Highlighting}
\end{Shaded}

\begin{center}\includegraphics[width=0.7\linewidth]{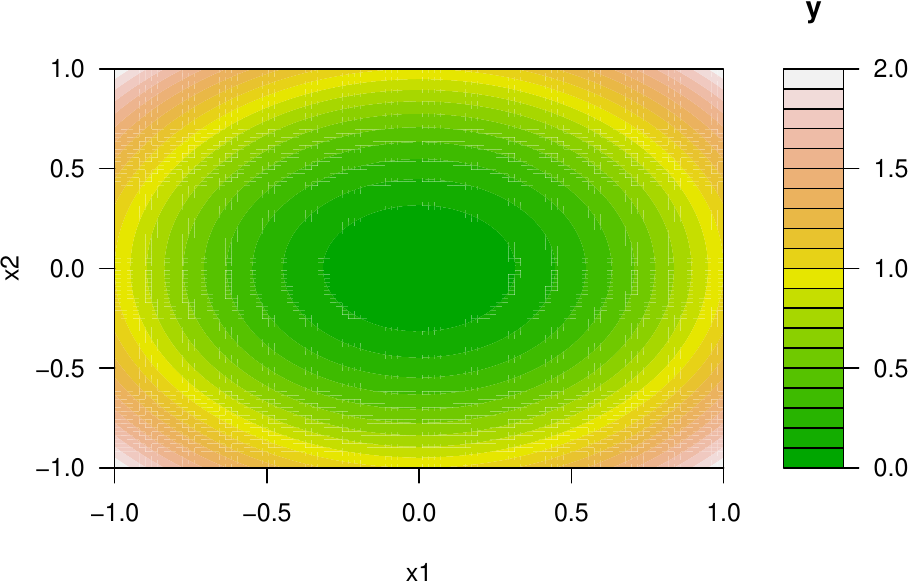} \end{center}

\hypertarget{example-plotting-a-user-defined-function}{%
\subsubsection{Example: Plotting a user defined
function}\label{example-plotting-a-user-defined-function}}

The following code examples show how user defined functions can be
plotted. It also illustrates how the color scheme can be modified.
Consider the following function: \[
f(\vec{x}) = \sum_{i=1}^n (x_i^3 -1).
\]

\begin{Shaded}
\begin{Highlighting}[]
\NormalTok{myFunction }\OtherTok{\textless{}{-}} \ControlFlowTok{function}\NormalTok{ (x)\{}
  \FunctionTok{matrix}\NormalTok{(}\FunctionTok{apply}\NormalTok{(x, }\CommentTok{\# matrix}
               \DecValTok{1}\NormalTok{, }\CommentTok{\# margin (apply over rows)}
               \ControlFlowTok{function}\NormalTok{(x) }\FunctionTok{sum}\NormalTok{(x}\SpecialCharTok{\^{}}\DecValTok{3{-}1}\NormalTok{) }\CommentTok{\# objective function}
\NormalTok{               ),}
\NormalTok{         , }\DecValTok{1}\NormalTok{) }\CommentTok{\# number of columns}
\NormalTok{  \}}
\FunctionTok{plotFunction}\NormalTok{(myFunction, }
             \FunctionTok{rep}\NormalTok{(}\SpecialCharTok{{-}}\DecValTok{1}\NormalTok{,}\DecValTok{2}\NormalTok{), }
             \FunctionTok{rep}\NormalTok{(}\DecValTok{1}\NormalTok{,}\DecValTok{2}\NormalTok{), }
             \AttributeTok{color.palette =}\NormalTok{ rainbow) }
\end{Highlighting}
\end{Shaded}

\begin{center}\includegraphics[width=0.7\linewidth]{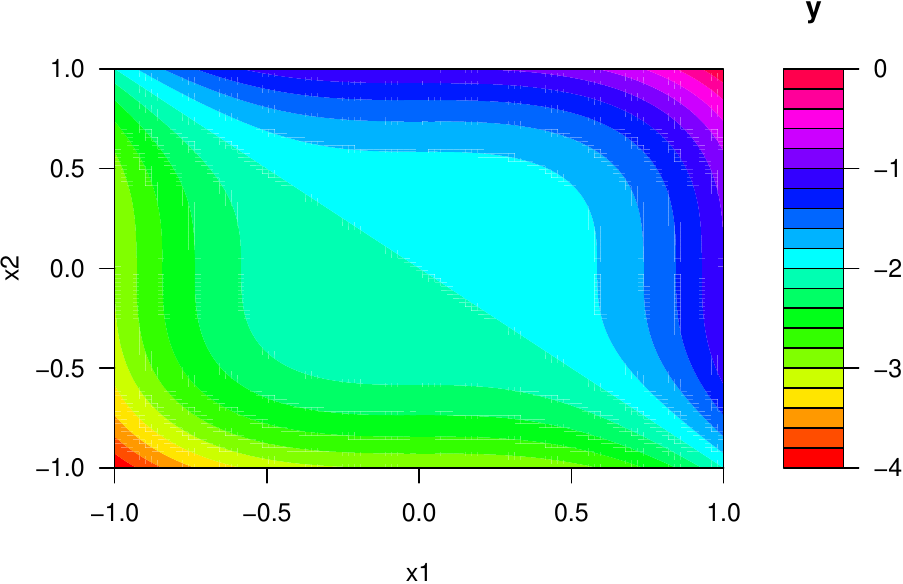} \end{center}

We can also generate a perspective plot of the user defined function as
shown in the following code.

\begin{Shaded}
\begin{Highlighting}[]
\FunctionTok{plotFunction}\NormalTok{(myFunction,}
             \FunctionTok{rep}\NormalTok{(}\SpecialCharTok{{-}}\DecValTok{1}\NormalTok{,}\DecValTok{2}\NormalTok{), }
             \FunctionTok{rep}\NormalTok{(}\DecValTok{1}\NormalTok{,}\DecValTok{2}\NormalTok{), }
             \AttributeTok{type=}\StringTok{"persp"}\NormalTok{, }
             \AttributeTok{theta=}\DecValTok{10}\NormalTok{, }
             \AttributeTok{phi=}\DecValTok{25}\NormalTok{, }
             \AttributeTok{border =} \ConstantTok{NA}\NormalTok{)}
\end{Highlighting}
\end{Shaded}

\begin{center}\includegraphics[width=0.4\linewidth]{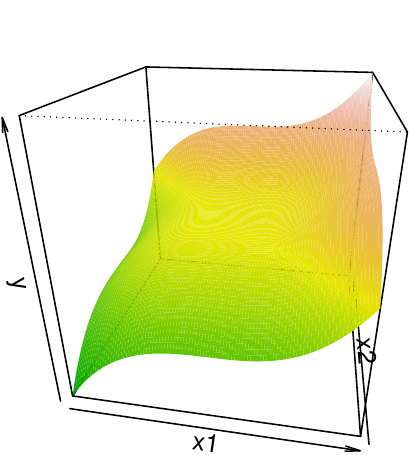} \end{center}

\hypertarget{plotmodel}{%
\subsection{plotModel()}\label{plotmodel}}

Furthermore, \texttt{plotModel()} offers the possibility to visualize
(suurogate) models that have already been trained, e.g., during a
\texttt{SPOT} run. Some simple examples are given below.

\hypertarget{example-plotting-a-trained-model}{%
\subsubsection{Example: Plotting a trained
model}\label{example-plotting-a-trained-model}}

First, we generate some training data. To demonstrate how plots from
data with more than two input dimensions can be generated, we generate a
three-dimensional input design, i.e., we consider a functional
relationship of the type \[
f: \mathbb{R}^3 \to \mathbb{R}, \qquad y = f(x_1,x_2,x_3).
\] * The output is one dimensional.

\begin{Shaded}
\begin{Highlighting}[]
\FunctionTok{set.seed}\NormalTok{(}\DecValTok{123}\NormalTok{)}
\NormalTok{k }\OtherTok{\textless{}{-}} \DecValTok{30}
\NormalTok{x.test }\OtherTok{\textless{}{-}} \FunctionTok{designLHD}\NormalTok{(}\AttributeTok{x=}\ConstantTok{NULL}\NormalTok{,}\FunctionTok{rep}\NormalTok{(}\SpecialCharTok{{-}}\DecValTok{1}\NormalTok{,}\DecValTok{3}\NormalTok{),}\FunctionTok{rep}\NormalTok{(}\DecValTok{1}\NormalTok{,}\DecValTok{3}\NormalTok{), }\AttributeTok{control =} \FunctionTok{list}\NormalTok{(}\AttributeTok{size =}\NormalTok{ k)) }
\NormalTok{y.test }\OtherTok{\textless{}{-}} \FunctionTok{funSphere}\NormalTok{(x.test)}
\FunctionTok{head}\NormalTok{( }\FunctionTok{cbind}\NormalTok{(x.test, y.test))}
\end{Highlighting}
\end{Shaded}

\begin{verbatim}
##            [,1]        [,2]        [,3]      [,4]
## [1,]  0.6721562 -0.67693655 -0.06068276 0.9137194
## [2,] -0.0813375  0.74944336  0.59109728 0.9176771
## [3,]  0.5462156 -0.39765660 -0.09992404 0.4664671
## [4,]  0.2437057 -0.29365065  0.21488375 0.1917982
## [5,] -0.3043235  0.25574608 -0.30583872 0.2515562
## [6,]  0.4131432  0.02414995  0.12575375 0.1870845
\end{verbatim}

Then, we train a standard response surface using \texttt{SPOT}'s
\texttt{buildRSM()} function. We generate the default contour plot using
\texttt{plotModel()}.

\begin{Shaded}
\begin{Highlighting}[]
\NormalTok{fit.test }\OtherTok{\textless{}{-}} \FunctionTok{buildRSM}\NormalTok{(x.test,y.test)}
\FunctionTok{plotModel}\NormalTok{(fit.test)}
\end{Highlighting}
\end{Shaded}

\begin{center}\includegraphics[width=0.7\linewidth]{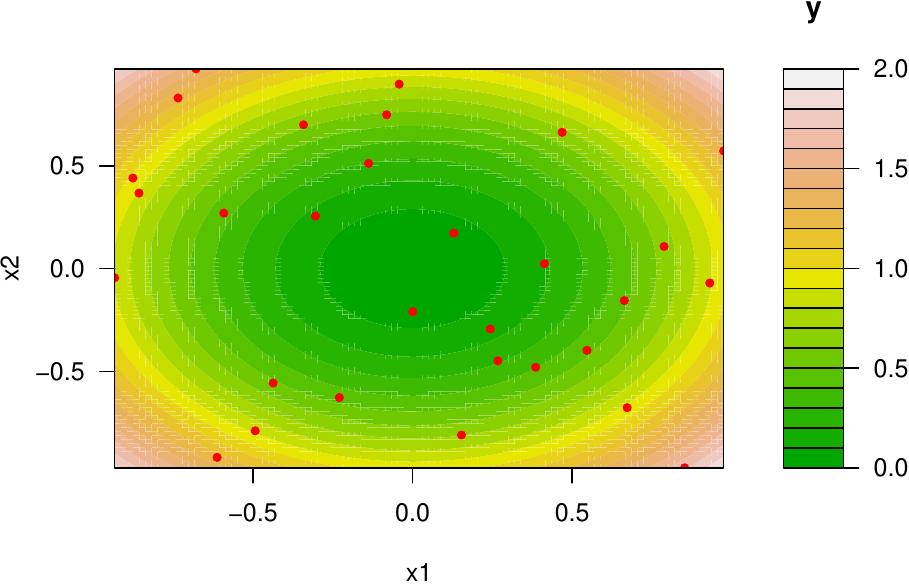} \end{center}

Passing the argument \texttt{type="contour"} to the \texttt{plotModel()}
function, a 2D contour plot can be generated as shown in the Figure. By
default, the dependent variable \(y\) is plotted against the first two
\(x_i\) variables. Note, that the argument \texttt{which} specifies the
independent variables \(x_i\) that are plotted. To plot \(y\) against
\(x_1\) and \(x_3\), the argument \texttt{which=c(1,3)} can be used.

\begin{Shaded}
\begin{Highlighting}[]
\FunctionTok{plotModel}\NormalTok{(fit.test,}\AttributeTok{which=}\FunctionTok{c}\NormalTok{(}\DecValTok{1}\NormalTok{,}\DecValTok{3}\NormalTok{),}\AttributeTok{type=}\StringTok{"contour"}\NormalTok{,}\AttributeTok{pch1=}\DecValTok{24}\NormalTok{,}\AttributeTok{col1=}\StringTok{"blue"}\NormalTok{)}
\end{Highlighting}
\end{Shaded}

\begin{center}\includegraphics[width=0.7\linewidth]{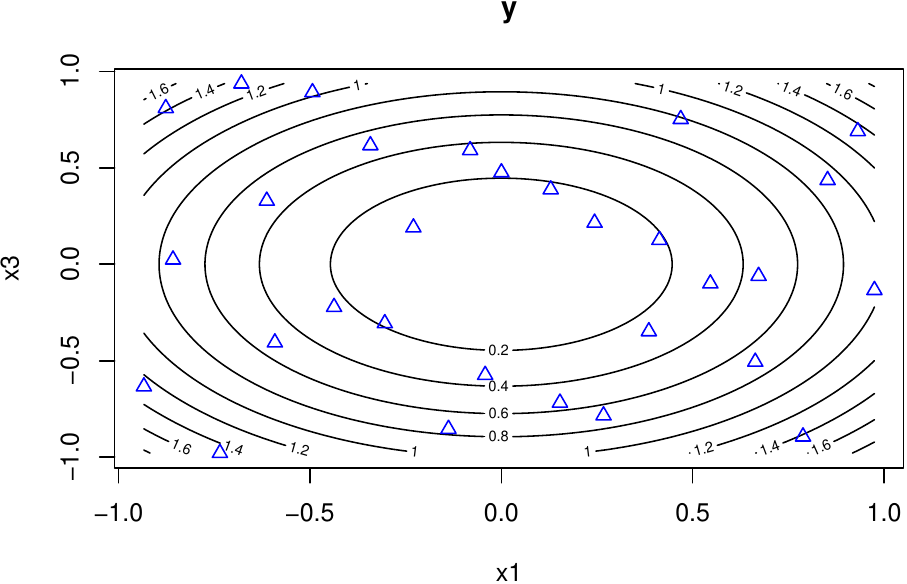} \end{center}

Perspective plots of the same model can be generated as follows. The
arguments \texttt{theta} and \texttt{phi} can be used to modify the view
point. The figure can be generated with the following code:

\begin{Shaded}
\begin{Highlighting}[]
\FunctionTok{plotModel}\NormalTok{(fit.test,}\AttributeTok{which=}\FunctionTok{c}\NormalTok{(}\DecValTok{1}\NormalTok{,}\DecValTok{3}\NormalTok{),}\AttributeTok{type=}\StringTok{"persp"}\NormalTok{,}\AttributeTok{border=}\StringTok{"NA"}\NormalTok{,}\AttributeTok{theta=}\DecValTok{255}\NormalTok{,}\AttributeTok{phi=}\DecValTok{20}\NormalTok{)}
\end{Highlighting}
\end{Shaded}

\begin{figure}

{\centering \includegraphics[width=0.4\linewidth]{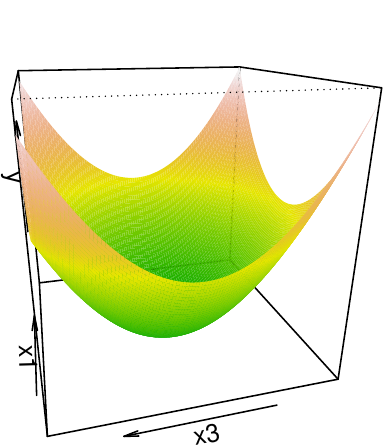} 

}

\caption{Perspective plot}\label{fig:plotModel2}
\end{figure}

\hypertarget{plotdata}{%
\subsection{\texorpdfstring{\texttt{plotData()}}{plotData()}}\label{plotdata}}

Finally, using \texttt{plotData()}, different models built on provided
data can be compared. The \texttt{plotData()} function generates a
(filled) contour or perspective plot of a data set with two independent
and one dependent variable. The plot is generated by some interpolation
or regression model. By default, the \texttt{LOESS} function is used.
Some simple examples are given below.

\hypertarget{example-plotting-data-using-loess-and-random-forest}{%
\paragraph{\texorpdfstring{Example: Plotting data using \texttt{LOESS}
and
\texttt{random\ forest}}{Example: Plotting data using LOESS and random forest}}\label{example-plotting-data-using-loess-and-random-forest}}

The following code shows a comparison of two different models fitting
the same data. First, the default \texttt{LOESS} model is used for
interpolation.

\begin{Shaded}
\begin{Highlighting}[]
\FunctionTok{plotData}\NormalTok{(x.test,y.test)}
\end{Highlighting}
\end{Shaded}

\begin{center}\includegraphics[width=0.7\linewidth]{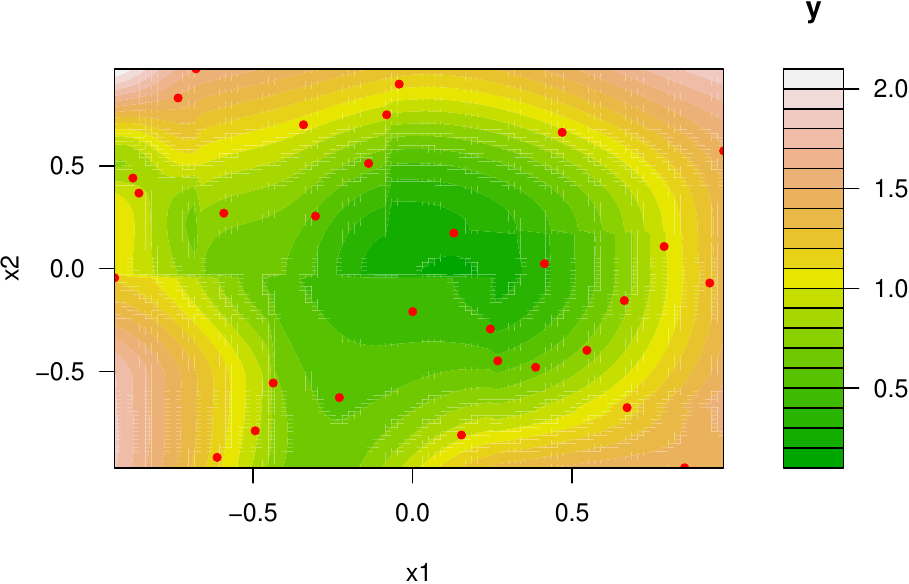} \end{center}

Then, the random forest was used for interpolation.

\begin{Shaded}
\begin{Highlighting}[]
\FunctionTok{plotData}\NormalTok{(x.test,y.test,}\AttributeTok{type=}\StringTok{"filled.contour"}\NormalTok{,}\AttributeTok{cex1=}\DecValTok{1}\NormalTok{,}\AttributeTok{col1=}\StringTok{"red"}\NormalTok{,}\AttributeTok{pch1=}\DecValTok{21}\NormalTok{,}\AttributeTok{model=}\NormalTok{buildRandomForest)}
\end{Highlighting}
\end{Shaded}

\begin{center}\includegraphics[width=0.7\linewidth]{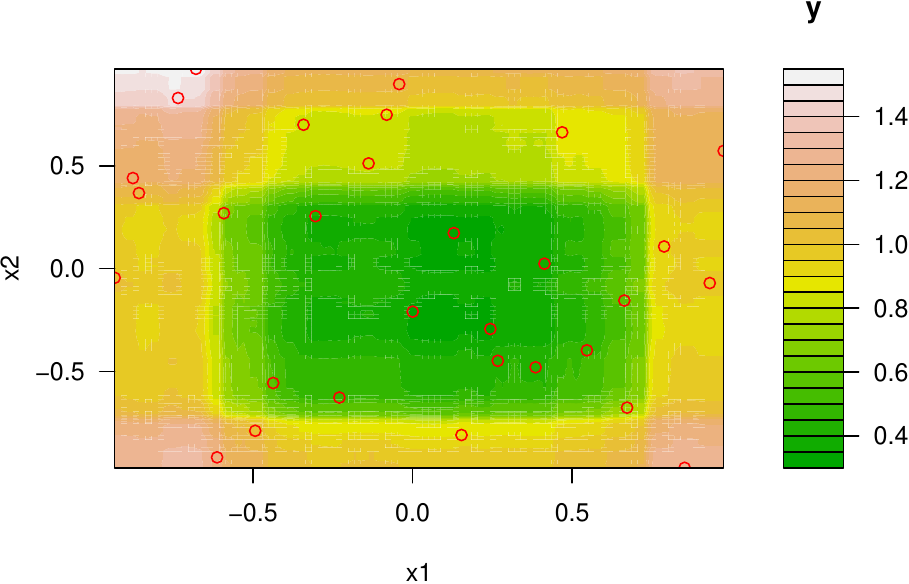} \end{center}

\hypertarget{example-perspective-plot}{%
\paragraph{Example: Perspective plot}\label{example-perspective-plot}}

\begin{figure}

{\centering \includegraphics[width=0.4\linewidth]{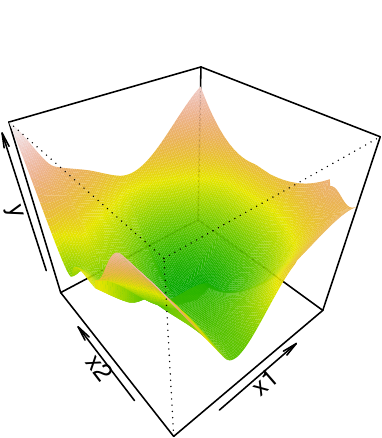} 

}

\caption{Perspective plot}\label{fig:plotPersp1}
\end{figure}

Figure \ref{fig:plotPersp1} shows a perspective plot, which is based on
the same data that were used in the previous Example.

\begin{Shaded}
\begin{Highlighting}[]
\FunctionTok{plotData}\NormalTok{(x.test,y.test,}\AttributeTok{type=}\StringTok{"persp"}\NormalTok{,}\AttributeTok{border=}\ConstantTok{NA}\NormalTok{,}\AttributeTok{model=}\NormalTok{buildLOESS)}
\end{Highlighting}
\end{Shaded}

\hypertarget{sec:exploratory}{%
\section{Exploratory Fitness Landscape Analysis Using
SPOT}\label{sec:exploratory}}

This section demonstrates how functions from the \texttt{SPOT} package
can be used to perform a visual inspection of the fitness landscape
during an interactive \texttt{SPOT} run. These results can also be used
to illustrate the algorithm's performance. In this section, reference
will be made to the application, in which \texttt{SPOT} is used for the
tuning procedure of two \texttt{SANN} design parameters:

\begin{itemize}
\tightlist
\item
  starting temperature \texttt{temp} and the
\item
  number of function evaluations at each temperature \texttt{tmax}.
\end{itemize}

The fitness landscape can be visualized in two different ways:

\begin{enumerate}
\def\labelenumi{\arabic{enumi}.}
\tightlist
\item
  Because \texttt{SPOT} builds a surrogate model during the sequential
  optimization, this model can be used to visualize the fitness
  landscape. In this case, the \texttt{plotModel()} function will be
  used.
\item
  Using standard interpolation or local regression functions. In this
  case, the \texttt{plotData()} function will be used. Note, that the
  \texttt{plotData()} function allows the specification of several
  interpolation functions (\texttt{LOESS} is default).
\end{enumerate}

Using \texttt{plotFunction()} is usually not applicable, because the
underlying (true) analytical function is not known. We consider the
\texttt{SPOT} model based approach first.

\hypertarget{plotting-the-final-model-with-plotmodel}{%
\subsection{\texorpdfstring{Plotting the Final Model with
\texttt{plotModel()}}{Plotting the Final Model with plotModel()}}\label{plotting-the-final-model-with-plotmodel}}

Plotting the final model from the \texttt{SPOT} run might be the most
generic way of visualizing the results, because during the optimization,
the optimizer trusted this model. So, why should it be considered
unreliable after the optimization is finished? Based on the tuning
results from the previoius example (using the \texttt{resRf} data), we
will demonstrate how the final model, which was built during the
\texttt{SPOT} run, can be plotted. Since the model is stored in the
result list from the \texttt{SPOT} run, i.e., in \texttt{resRf}, the
parameter \texttt{resRf\$modelFit()} can be passed as an argument to the
\texttt{plotModel()} function. The result is shown in the following
Figure. It shows a surface plot from the \texttt{SPOT} run with random
forest.

\begin{Shaded}
\begin{Highlighting}[]
\FunctionTok{plotModel}\NormalTok{(resRf}\SpecialCharTok{$}\NormalTok{modelFit)}
\end{Highlighting}
\end{Shaded}

\begin{center}\includegraphics[width=0.7\linewidth]{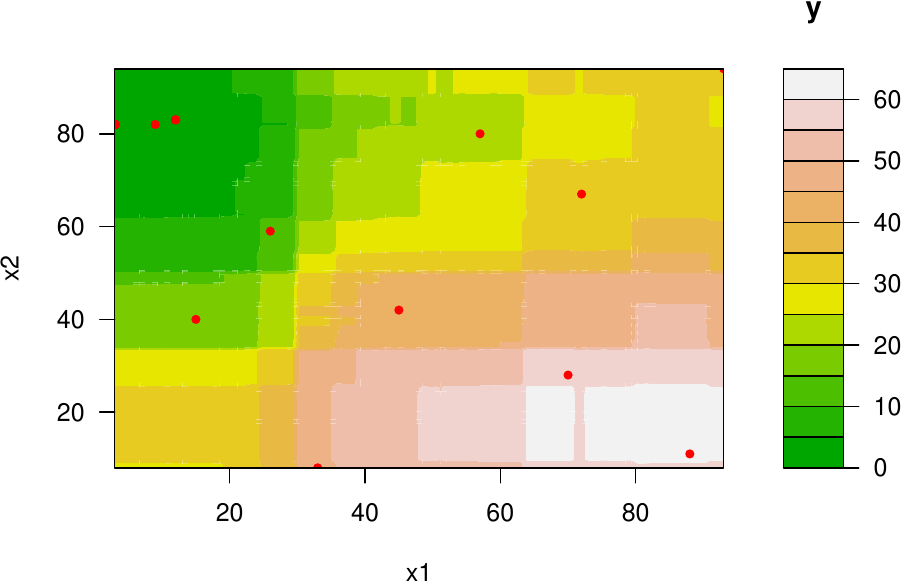} \end{center}

\hypertarget{plotting-the-data-with-plotdata}{%
\subsubsection{\texorpdfstring{Plotting the Data with
\texttt{plotData()}}{Plotting the Data with plotData()}}\label{plotting-the-data-with-plotdata}}

Results from Example with \texttt{resRf} result data were obtained with
the random forest model. But, the data can be fitted to a different
model, e.g., a locally (weighted) scatter plot smoothing
(\texttt{LOESS}) or Kriging model as follows.

\hypertarget{example-plotting-data-using-loess}{%
\paragraph{\texorpdfstring{Example: Plotting data using
\texttt{LOESS}}{Example: Plotting data using LOESS}}\label{example-plotting-data-using-loess}}

The following code illustrates the \texttt{LOESS} model fit, which uses
the data generated with a random forest model.

\begin{Shaded}
\begin{Highlighting}[]
\FunctionTok{suppressWarnings}\NormalTok{(}\FunctionTok{suppressMessages}\NormalTok{(}\FunctionTok{plotData}\NormalTok{(resRf}\SpecialCharTok{$}\NormalTok{x,resRf}\SpecialCharTok{$}\NormalTok{y,}\AttributeTok{model=}\NormalTok{buildLOESS)))}
\end{Highlighting}
\end{Shaded}

\begin{center}\includegraphics[width=0.7\linewidth]{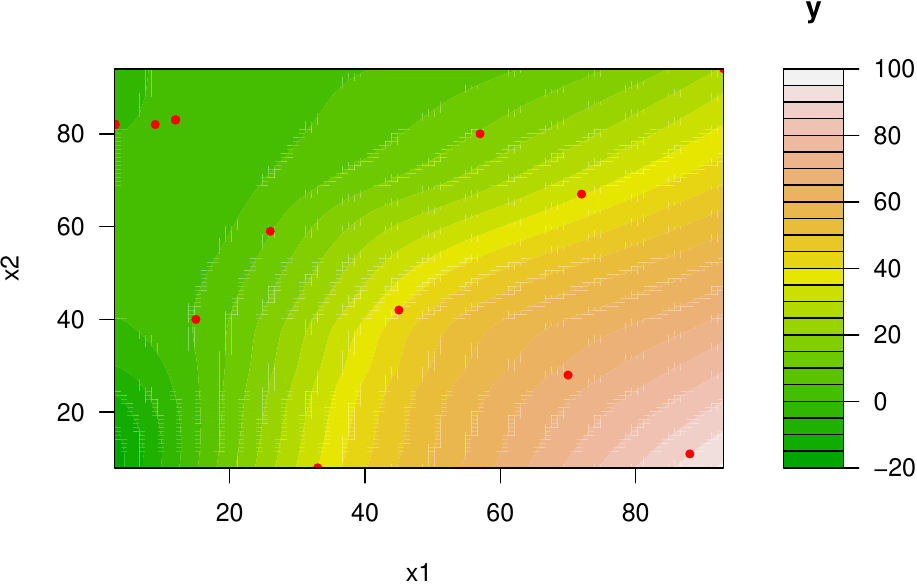} \end{center}

\hypertarget{example-plotting-data-using-kriging}{%
\paragraph{Example: Plotting data using
Kriging}\label{example-plotting-data-using-kriging}}

Plotting the same data as in the previous Figure with a Kriging model
can easily be done. The result can be generated as follows.

\begin{Shaded}
\begin{Highlighting}[]
\FunctionTok{plotData}\NormalTok{(resRf}\SpecialCharTok{$}\NormalTok{x,resRf}\SpecialCharTok{$}\NormalTok{y,}\AttributeTok{model=}\NormalTok{buildKriging)}
\end{Highlighting}
\end{Shaded}

\begin{center}\includegraphics[width=0.7\linewidth]{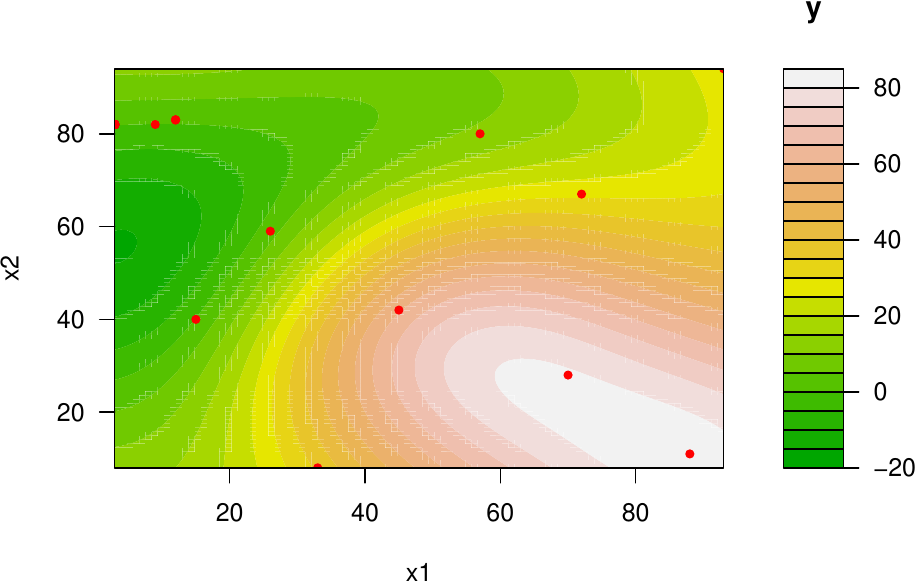} \end{center}

This result can be compared to a case in which Kriging models were used
during the \texttt{SPOT} run and for the final illustration. The same
setting as in the random forest based optimization will be used. Only
the meta model will be changed. The \texttt{SPOT} configuration
parameters can be changed as follows:

\begin{Shaded}
\begin{Highlighting}[]
\NormalTok{spotConfig}\SpecialCharTok{$}\NormalTok{model }\OtherTok{=}\NormalTok{ buildKriging}
\NormalTok{spotConfig}\SpecialCharTok{$}\NormalTok{optimizer }\OtherTok{=}\NormalTok{ optimLBFGSB}
\NormalTok{spotConfig}\SpecialCharTok{$}\NormalTok{modelControl }\OtherTok{=} \FunctionTok{list}\NormalTok{(}\AttributeTok{algTheta=}\NormalTok{optimLBFGSB)}
\NormalTok{resK }\OtherTok{\textless{}{-}} \FunctionTok{spot}\NormalTok{(}\AttributeTok{x=}\ConstantTok{NULL}\NormalTok{,}
             \AttributeTok{fun=}\NormalTok{sann2spot,}
             \AttributeTok{lower=}\NormalTok{lower,}
             \AttributeTok{upper=}\NormalTok{upper,}
             \AttributeTok{control=}\NormalTok{spotConfig)}
\end{Highlighting}
\end{Shaded}

Now, the Kriging model used during the \texttt{SPOT} run will be used to
visualize the fitness landscape. The following code can be used for
visualization:

\begin{Shaded}
\begin{Highlighting}[]
\FunctionTok{plotModel}\NormalTok{(resK}\SpecialCharTok{$}\NormalTok{modelFit)}
\end{Highlighting}
\end{Shaded}

\begin{center}\includegraphics[width=0.7\linewidth]{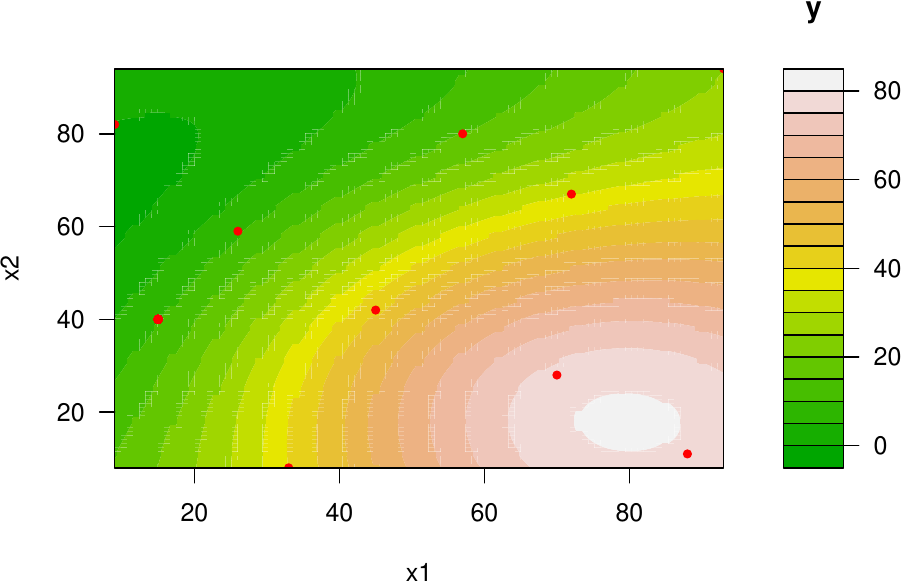} \end{center}

The landscape generated with the random forest based \texttt{SPOT} run
and the landscape from the Example, which used the Kriging-based
\texttt{SPOT} run, differ. We can check whether longer runs increase the
similarities.

\hypertarget{continued-runs}{%
\subsubsection{Continued Runs}\label{continued-runs}}

The optimization procedure from Example which generated the
\texttt{resRf} data, used 50 runs of the \texttt{SANN} algorithm. Now
this number is increased to 100 and the procedure is continued. This
means, an additional 50 evaluations are necessary since the results from
the first runs are reused.

\hypertarget{example-continued-with-50-additional-runs}{%
\paragraph{Example: Continued with 50 additional
runs}\label{example-continued-with-50-additional-runs}}

The fitness landscape of the random forest meta model with 100 function
evaluations is shown at the top in the Figure.

\begin{Shaded}
\begin{Highlighting}[]
\NormalTok{spotConfig}\SpecialCharTok{$}\NormalTok{funEvals }\OtherTok{\textless{}{-}} \DecValTok{100}
\NormalTok{spotConfig}\SpecialCharTok{$}\NormalTok{model }\OtherTok{\textless{}{-}}\NormalTok{ buildRandomForest}
\NormalTok{res100Rf }\OtherTok{\textless{}{-}} \FunctionTok{spotLoop}\NormalTok{(resRf}\SpecialCharTok{$}\NormalTok{x,}
\NormalTok{                     resRf}\SpecialCharTok{$}\NormalTok{y,}
                     \AttributeTok{fun=}\NormalTok{sann2spot,}
                     \AttributeTok{lower=}\NormalTok{lower,}
                     \AttributeTok{upper=}\NormalTok{upper,}
                     \AttributeTok{control=}\NormalTok{spotConfig)}
\end{Highlighting}
\end{Shaded}

\hypertarget{example-resk-continued-with-50-additional-runs}{%
\paragraph{\texorpdfstring{Example: \texttt{resK} continued with 50
additional
runs}{Example: resK continued with 50 additional runs}}\label{example-resk-continued-with-50-additional-runs}}

In a similar manner, new results can be added to the Kriging based
optimization runs from Example \texttt{resK}.

\begin{Shaded}
\begin{Highlighting}[]
\NormalTok{spotConfig}\SpecialCharTok{$}\NormalTok{model }\OtherTok{=}\NormalTok{ buildKriging}
\NormalTok{spotConfig}\SpecialCharTok{$}\NormalTok{optimizer }\OtherTok{=}\NormalTok{ optimLBFGSB}
\NormalTok{spotConfig}\SpecialCharTok{$}\NormalTok{modelControl }\OtherTok{=} \FunctionTok{list}\NormalTok{(}\AttributeTok{algTheta=}\NormalTok{optimLBFGSB)}
\NormalTok{res100K }\OtherTok{\textless{}{-}} \FunctionTok{spotLoop}\NormalTok{(resK}\SpecialCharTok{$}\NormalTok{x,}
\NormalTok{                    resK}\SpecialCharTok{$}\NormalTok{y,}
                    \AttributeTok{fun=}\NormalTok{sann2spot,}
                    \AttributeTok{lower=}\NormalTok{lower,}
                    \AttributeTok{upper=}\NormalTok{upper,}
                    \AttributeTok{control=}\NormalTok{spotConfig)}
\end{Highlighting}
\end{Shaded}

A comparison of the models resulting from the two long runs (Example
\texttt{resRf100} and \texttt{res100K}) is shown in the following
Figures. A visual inspection indicates that the landscapes are
qualitatively similar, e.g., poor algorithm settings can be found in the
lower right corner. Comparison of the long runs with 100 function
evaluations and two different surrogate models.

First: Example \texttt{resRf100}. Long run using Random forest model.

\begin{Shaded}
\begin{Highlighting}[]
\FunctionTok{plotModel}\NormalTok{(res100Rf}\SpecialCharTok{$}\NormalTok{modelFit)}
\end{Highlighting}
\end{Shaded}

\begin{center}\includegraphics[width=0.7\linewidth]{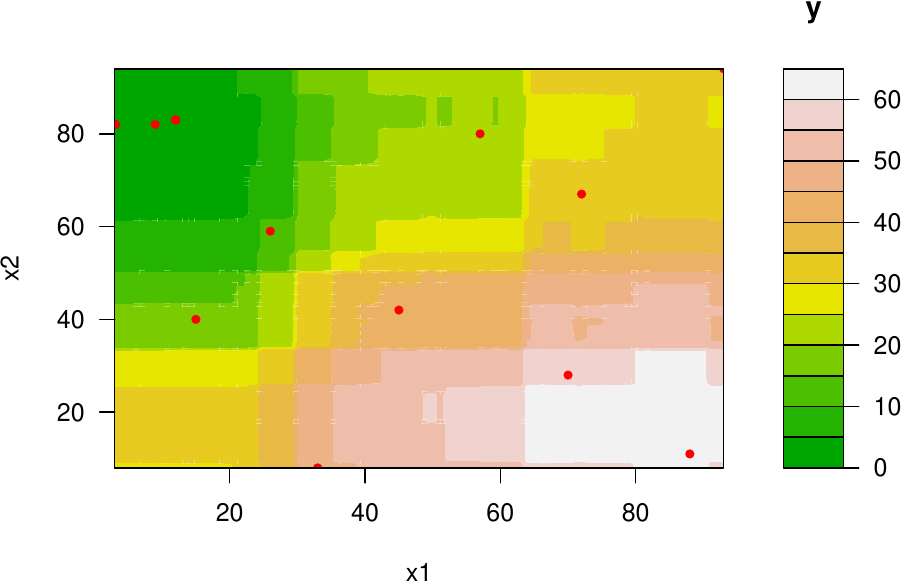} \end{center}

Second: Example \texttt{res100K}. Long run with 100 function evaluations
using a Kriging model.

\begin{Shaded}
\begin{Highlighting}[]
\FunctionTok{plotModel}\NormalTok{(res100K}\SpecialCharTok{$}\NormalTok{modelFit)}
\end{Highlighting}
\end{Shaded}

\begin{center}\includegraphics[width=0.7\linewidth]{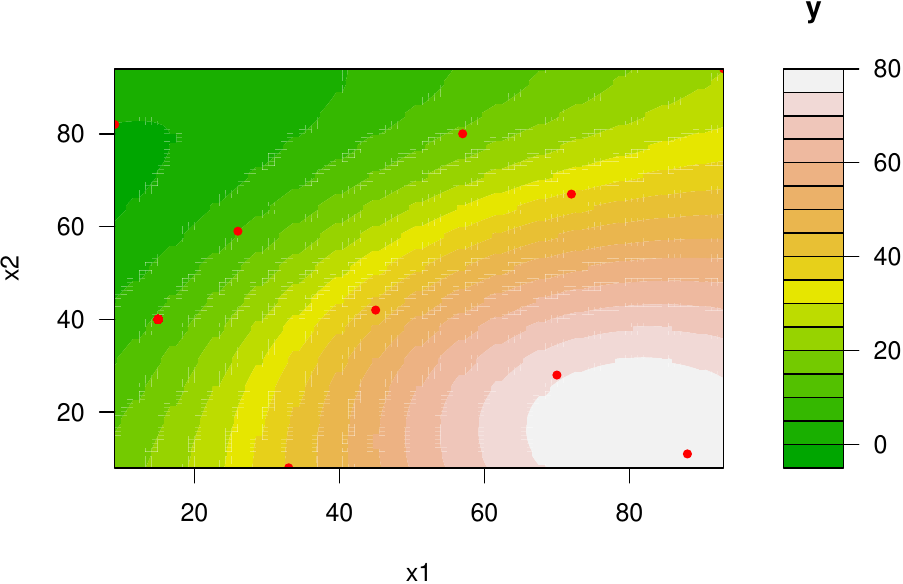} \end{center}

\hypertarget{sec:rsm}{%
\section{Response Surface Methodology (RSM)}\label{sec:rsm}}

\emph{Response surface methodology} (RSM) can be seen as a set of
statistical methods for empirical model building \cite{Box87a}.

Using design of experiments, a response (dependent variable, output
variable, or fitness value, \(y\)) that depends on one or several input
variables (independent variables or solutions, \(\vec{x}\)) is
optimized. The underlying model can be formulated as \[
y = f(\vec{x}) + \epsilon,
\] where \(\epsilon\) represents some noise (uncertainty, error
observed) in the response \(y\).

The term \emph{response surface} refers to the surface represented by
\(f(\vec{x})\). In order to estimate the quality of a solution, the term
\texttt{fitness} is used in evolutionary optimization. In physics, the
concept of a potential or energy function is used. Since we are dealing
with minimization, a low fitness value \(f(\vec{x})\) implies that
\(\vec{x}\) is a good solution.

Using the \texttt{rsm} package, which is maintained by \cite{Lent12a},
the \texttt{buildRSM()} function builds a linear response surface model.
The arguments of \texttt{buildRSM(x,\ y,\ control\ =\ list())} are as
follows:

\begin{itemize}
\tightlist
\item
  \texttt{x}: design matrix (sample locations), rows for each sample,
  columns for each variable.
\item
  \texttt{y}: vector of observations at \texttt{x}
\item
  \texttt{control}: list, with the options for the model building
  procedure:

  \begin{itemize}
  \tightlist
  \item
    \texttt{mainEffectsOnly}: logical, defaults to \texttt{FALSE}. Set
    to \texttt{TRUE} if a model with main effects only is desired (no
    interactions, second order effects).
  \item
    \texttt{canonical}: logical, defaults to \texttt{FALSE}. If this is
    \texttt{TRUE}, use the canonical path to descent from saddle points.
    Else, simply use steepest descent
  \end{itemize}
\end{itemize}

\hypertarget{example-path-of-the-steepest-descent}{%
\subsection{Example: Path of the steepest
descent}\label{example-path-of-the-steepest-descent}}

First, we create some design points and compute observations at design
points. Then, using \texttt{buildRSM()}, the response surface model is
build. The function \texttt{descentSpotRSM()} returns the path of the
steepest descent.

\begin{Shaded}
\begin{Highlighting}[]
\NormalTok{x }\OtherTok{\textless{}{-}} \FunctionTok{designUniformRandom}\NormalTok{(}\AttributeTok{lower=}\FunctionTok{rep}\NormalTok{(}\SpecialCharTok{{-}}\DecValTok{5}\NormalTok{,}\DecValTok{2}\NormalTok{), }
                         \AttributeTok{upper=}\FunctionTok{rep}\NormalTok{(}\DecValTok{15}\NormalTok{,}\DecValTok{2}\NormalTok{), }
                         \AttributeTok{control=}\FunctionTok{list}\NormalTok{(}\AttributeTok{size=}\DecValTok{20}\NormalTok{))}
\NormalTok{y }\OtherTok{\textless{}{-}} \FunctionTok{funSphere}\NormalTok{(x)}
\end{Highlighting}
\end{Shaded}

Create model with default settings.

\begin{Shaded}
\begin{Highlighting}[]
\NormalTok{fit }\OtherTok{\textless{}{-}} \FunctionTok{buildRSM}\NormalTok{(x,y)}
\end{Highlighting}
\end{Shaded}

Predict new point.

\begin{Shaded}
\begin{Highlighting}[]
\FunctionTok{predict}\NormalTok{(fit,}\FunctionTok{cbind}\NormalTok{(}\DecValTok{1}\NormalTok{,}\DecValTok{2}\NormalTok{))}
\end{Highlighting}
\end{Shaded}

\begin{verbatim}
## $y
##      [,1]
## [1,]    5
\end{verbatim}

True value at location:

\begin{Shaded}
\begin{Highlighting}[]
\FunctionTok{sphere}\NormalTok{(}\FunctionTok{c}\NormalTok{(}\DecValTok{1}\NormalTok{,}\DecValTok{2}\NormalTok{))}
\end{Highlighting}
\end{Shaded}

\begin{verbatim}
## [1] 5
\end{verbatim}

\begin{Shaded}
\begin{Highlighting}[]
\FunctionTok{descentSpotRSM}\NormalTok{(fit)}
\end{Highlighting}
\end{Shaded}

\begin{verbatim}
## Path of steepest descent from ridge analysis:
\end{verbatim}

\begin{verbatim}
## $x
##           V1      V2
## 1   4.470955  4.7340
## 2   3.849775  4.0405
## 3   3.219460  3.3470
## 4   2.589145  2.6630
## 5   1.949695  1.9790
## 6   1.301110  1.3140
## 7   0.643390  0.6490
## 8  -0.014330 -0.0160
## 9  -0.681185 -0.6620
## 10 -1.348040 -1.3080
## 
## $y
##               [,1]
##  [1,] 4.240019e+01
##  [2,] 3.114641e+01
##  [3,] 2.156733e+01
##  [4,] 1.379524e+01
##  [5,] 7.717752e+00
##  [6,] 3.419483e+00
##  [7,] 8.351517e-01
##  [8,] 4.613489e-04
##  [9,] 9.022570e-01
## [10,] 3.528076e+00
\end{verbatim}

This situation is illustrated in the following Figure, which illustrates
the response surface, which fits the sphere function.

\begin{Shaded}
\begin{Highlighting}[]
\FunctionTok{plot}\NormalTok{(fit)}
\end{Highlighting}
\end{Shaded}

\begin{center}\includegraphics[width=0.7\linewidth]{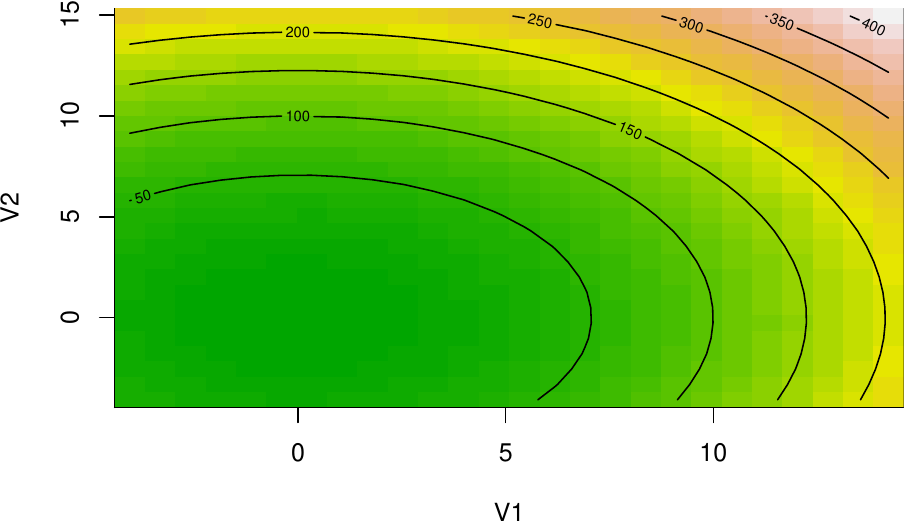} \end{center}

\hypertarget{rsm-and-spot}{%
\subsection{RSM and SPOT}\label{rsm-and-spot}}

We can use RSM for an interactive tuning approach. The starting point in
this example are the 100 design points that were generated during the
Kriging-based \texttt{SPOT} run in Example \texttt{res100K}. These 100
data points are used to build a response surface with
\texttt{buildRSM()}.

\begin{Shaded}
\begin{Highlighting}[]
\NormalTok{rsm100K }\OtherTok{\textless{}{-}} \FunctionTok{buildRSM}\NormalTok{(}\AttributeTok{x=}\NormalTok{res100K}\SpecialCharTok{$}\NormalTok{x,}
                    \AttributeTok{y=}\NormalTok{res100K}\SpecialCharTok{$}\NormalTok{y) }
\FunctionTok{summary}\NormalTok{(rsm100K}\SpecialCharTok{$}\NormalTok{rsmfit)}
\end{Highlighting}
\end{Shaded}

\begin{verbatim}
## 
## Call:
## rsm(formula = y ~ FO(x1, x2) + TWI(x1, x2) + PQ(x1, x2), data = codedData)
## 
##             Estimate Std. Error t value  Pr(>|t|)    
## (Intercept)  38.2446     7.5949  5.0355 2.293e-06 ***
## x1           31.2749     6.8422  4.5709 1.476e-05 ***
## x2          -24.8816     6.4151 -3.8786 0.0001947 ***
## x1:x2       -13.4514     8.0668 -1.6675 0.0987436 .  
## x1^2         -5.4131    10.9583 -0.4940 0.6224746    
## x2^2         -5.1288    12.1820 -0.4210 0.6747073    
## ---
## Signif. codes:  0 '***' 0.001 '**' 0.01 '*' 0.05 '.' 0.1 ' ' 1
## 
## Multiple R-squared:  0.4191, Adjusted R-squared:  0.3882 
## F-statistic: 13.57 on 5 and 94 DF,  p-value: 5.824e-10
## 
## Analysis of Variance Table
## 
## Response: y
##             Df  Sum Sq Mean Sq F value    Pr(>F)
## FO(x1, x2)   2 11552.2  5776.1 32.2138 2.213e-11
## TWI(x1, x2)  1   509.7   509.7  2.8426   0.09511
## PQ(x1, x2)   2   100.1    50.1  0.2792   0.75704
## Residuals   94 16854.8   179.3                  
## Lack of fit  4  1243.1   310.8  1.7916   0.13742
## Pure error  90 15611.6   173.5                  
## 
## Stationary point of response surface:
##        x1        x2 
## -9.378938  9.873582 
## 
## Stationary point in original units:
##        V1        V2 
## -342.9154  475.5640 
## 
## Eigenanalysis:
## eigen() decomposition
## $values
## [1]   1.456267 -11.998166
## 
## $vectors
##          [,1]       [,2]
## x1  0.6995942 -0.7145403
## x2 -0.7145403 -0.6995942
\end{verbatim}

Following the path of the steepest descent on the RSM meta model, we
obtain a new design point, which can be evaluated.

\begin{Shaded}
\begin{Highlighting}[]
\NormalTok{(xSteep }\OtherTok{\textless{}{-}} \FunctionTok{descentSpotRSM}\NormalTok{(rsm100K) )}
\end{Highlighting}
\end{Shaded}

\begin{verbatim}
## Path of steepest descent from ridge analysis:
\end{verbatim}

\begin{verbatim}
## $x
##        V1     V2
## 1  47.682 53.666
## 2  44.322 56.246
## 3  40.920 58.697
## 4  37.392 61.105
## 5  33.822 63.341
## 6  30.084 65.405
## 7  26.220 67.211
## 8  22.188 68.673
## 9  17.904 69.705
## 10 13.410 70.178
## 
## $y
##             [,1]
##  [1,] 34.2436148
##  [2,] 30.2840896
##  [3,] 26.3865774
##  [4,] 22.4370814
##  [5,] 18.5631730
##  [6,] 14.6604386
##  [7,] 10.7908208
##  [8,]  6.9425382
##  [9,]  3.0556599
## [10,] -0.8304278
\end{verbatim}

We have chosen the eighth point, i.e.,

\begin{Shaded}
\begin{Highlighting}[]
\NormalTok{xNew }\OtherTok{\textless{}{-}}\NormalTok{ xSteep}\SpecialCharTok{$}\NormalTok{x[}\DecValTok{8}\NormalTok{,]}
\end{Highlighting}
\end{Shaded}

Then we determine its function value.

\begin{Shaded}
\begin{Highlighting}[]
\NormalTok{(yNew }\OtherTok{\textless{}{-}} \FunctionTok{sann2spot}\NormalTok{(xNew))}
\end{Highlighting}
\end{Shaded}

\begin{verbatim}
##            [,1]
## [1,] 0.07183276
\end{verbatim}

Next, we can refine the \texttt{rsm} meta model by including this point
to the set of design points.

\begin{Shaded}
\begin{Highlighting}[]
\NormalTok{x101 }\OtherTok{\textless{}{-}} \FunctionTok{rbind}\NormalTok{(res100K}\SpecialCharTok{$}\NormalTok{x, xNew)}
\NormalTok{y101 }\OtherTok{\textless{}{-}} \FunctionTok{rbind}\NormalTok{(res100K}\SpecialCharTok{$}\NormalTok{y, yNew)}
\NormalTok{rsm101K }\OtherTok{\textless{}{-}} \FunctionTok{buildRSM}\NormalTok{(}\AttributeTok{x=}\NormalTok{x101,}
                    \AttributeTok{y=}\NormalTok{y101) }
\FunctionTok{summary}\NormalTok{(rsm101K}\SpecialCharTok{$}\NormalTok{rsmfit)}
\end{Highlighting}
\end{Shaded}

\begin{verbatim}
## 
## Call:
## rsm(formula = y ~ FO(x1, x2) + TWI(x1, x2) + PQ(x1, x2), data = codedData)
## 
##             Estimate Std. Error t value  Pr(>|t|)    
## (Intercept)  37.5519     7.4113  5.0668 1.987e-06 ***
## x1           31.8496     6.6977  4.7553 7.049e-06 ***
## x2          -25.8726     6.0128 -4.3029 4.092e-05 ***
## x1:x2       -12.3827     7.6887 -1.6105    0.1106    
## x1^2         -4.4403    10.7050 -0.4148    0.6792    
## x2^2         -5.5684    12.0934 -0.4604    0.6462    
## ---
## Signif. codes:  0 '***' 0.001 '**' 0.01 '*' 0.05 '.' 0.1 ' ' 1
## 
## Multiple R-squared:  0.4211, Adjusted R-squared:  0.3906 
## F-statistic: 13.82 on 5 and 95 DF,  p-value: 3.878e-10
## 
## Analysis of Variance Table
## 
## Response: y
##             Df  Sum Sq Mean Sq F value   Pr(>F)
## FO(x1, x2)   2 11709.4  5854.7 32.9261 1.37e-11
## TWI(x1, x2)  1   486.4   486.4  2.7352   0.1015
## PQ(x1, x2)   2    89.4    44.7  0.2515   0.7781
## Residuals   95 16892.3   177.8                 
## Lack of fit  5  1280.6   256.1  1.4766   0.2053
## Pure error  90 15611.6   173.5                 
## 
## Stationary point of response surface:
##        x1        x2 
## -12.40282  11.46718 
## 
## Stationary point in original units:
##        V1        V2 
## -469.9186  544.0889 
## 
## Eigenanalysis:
## eigen() decomposition
## $values
## [1]   1.212627 -11.221309
## 
## $vectors
##          [,1]      [,2]
## x1 -0.7384869 0.6742679
## x2  0.6742679 0.7384869
\end{verbatim}

The resulting plot is shown in the following Figure.

\begin{Shaded}
\begin{Highlighting}[]
\FunctionTok{plot}\NormalTok{(rsm101K)}
\end{Highlighting}
\end{Shaded}

\begin{center}\includegraphics[width=0.7\linewidth]{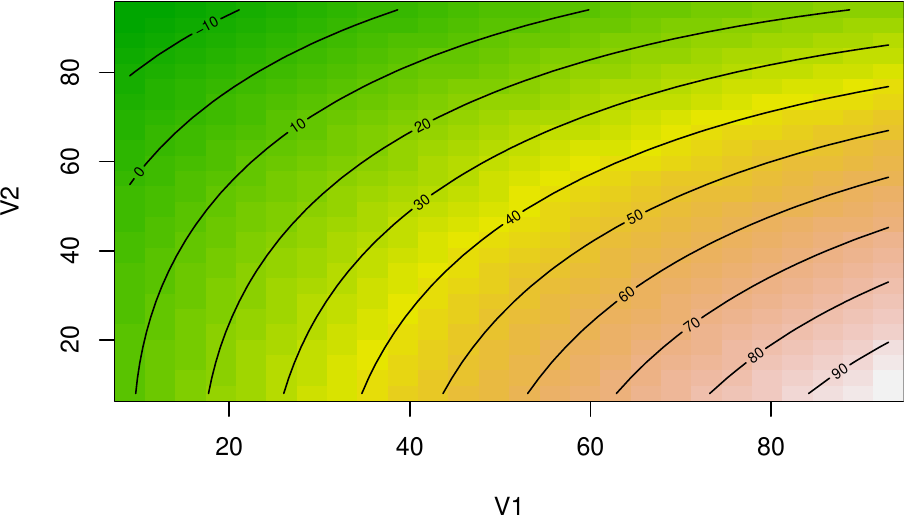} \end{center}

Now, we have two options for continuing the optimization. 1. Either, we
continue following the path of the steepest descent, i.e., we use
\texttt{SPOT}'s \texttt{descentSpotRSM()} function.

\begin{Shaded}
\begin{Highlighting}[]
\FunctionTok{descentSpotRSM}\NormalTok{(rsm101K)}
\end{Highlighting}
\end{Shaded}

\begin{verbatim}
## Path of steepest descent from ridge analysis:
\end{verbatim}

\begin{verbatim}
## $x
##        V1     V2
## 1  47.724 53.709
## 2  44.448 56.375
## 3  41.130 58.998
## 4  37.770 61.578
## 5  34.368 64.115
## 6  30.924 66.609
## 7  27.438 68.974
## 8  23.826 71.210
## 9  20.130 73.317
## 10 16.224 75.123
## 
## $y
##            [,1]
##  [1,] 33.449375
##  [2,] 29.395664
##  [3,] 25.358303
##  [4,] 21.336555
##  [5,] 17.329550
##  [6,] 13.336286
##  [7,]  9.402808
##  [8,]  5.461677
##  [9,]  1.539398
## [10,] -2.378995
\end{verbatim}

\begin{enumerate}
\def\labelenumi{\arabic{enumi}.}
\setcounter{enumi}{1}
\tightlist
\item
  Or, we can continue with \texttt{SPOT}. Following this second option,
  we have built an updated Kriging model using nine additional function
  evaluations.
\end{enumerate}

\begin{Shaded}
\begin{Highlighting}[]
\NormalTok{spotConfig}\SpecialCharTok{$}\NormalTok{model }\OtherTok{=}\NormalTok{ buildKriging}
\NormalTok{spotConfig}\SpecialCharTok{$}\NormalTok{optimizer }\OtherTok{=}\NormalTok{ optimLBFGSB}
\NormalTok{spotConfig}\SpecialCharTok{$}\NormalTok{modelControl }\OtherTok{=} \FunctionTok{list}\NormalTok{(}\AttributeTok{algTheta=}\NormalTok{optimLBFGSB)}
\NormalTok{spotConfig}\SpecialCharTok{$}\NormalTok{funEvals }\OtherTok{\textless{}{-}} \DecValTok{110}
\NormalTok{res110K }\OtherTok{\textless{}{-}} \FunctionTok{spotLoop}\NormalTok{(}\AttributeTok{x=}\NormalTok{x101,}
                    \AttributeTok{y=}\NormalTok{y101,}
                    \AttributeTok{fun=}\NormalTok{sann2spot,}
                    \AttributeTok{lower=}\NormalTok{lower,}
                    \AttributeTok{upper=}\NormalTok{upper,}
                    \AttributeTok{control=}\NormalTok{spotConfig)}
\end{Highlighting}
\end{Shaded}

Finally, we can plot the updated model. \texttt{buildRSM()} was used to
build a response surface. Data from the Kriging-based \texttt{SPOT} run
with 100 function evaluations, one additional point, which was
calculated using the steepest descent function
\texttt{descentSpotRSM()}, and one additional \texttt{SPOT} run with
nine additional design points, were used to generate this model.
Altogether, 110 design points were used to generate this model.

\begin{Shaded}
\begin{Highlighting}[]
\FunctionTok{plotModel}\NormalTok{(res110K}\SpecialCharTok{$}\NormalTok{modelFit)}
\end{Highlighting}
\end{Shaded}

\begin{center}\includegraphics[width=0.7\linewidth]{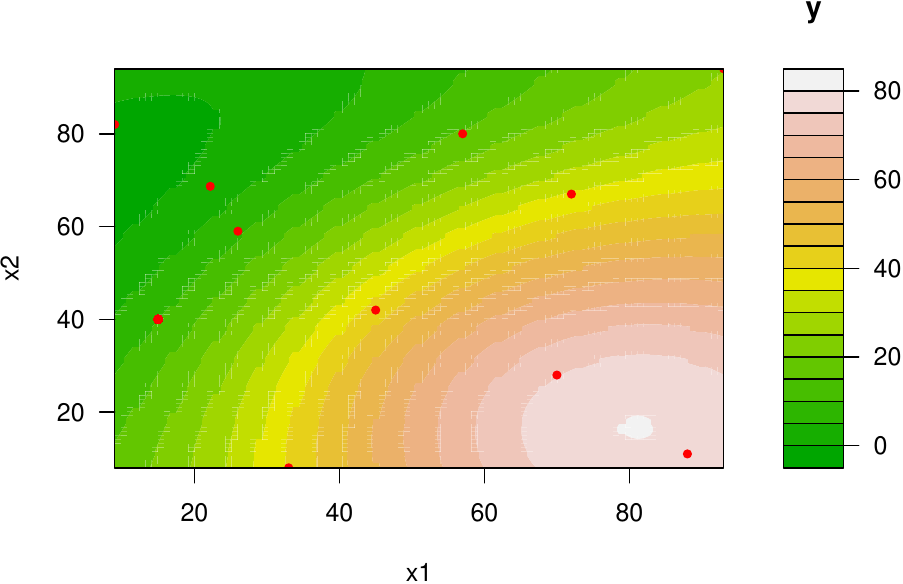} \end{center}

\hypertarget{sec:stats}{%
\section{Statistical Analysis}\label{sec:stats}}

This section describes basic approaches for the analysis of the results
from the previous example:

\begin{enumerate}
\def\labelenumi{\arabic{enumi}.}
\tightlist
\item
  tree-based analysis and
\item
  regression analysis.
\end{enumerate}

We will continue using the data from Example \texttt{resRf100}.

\hypertarget{tree-based-analysis}{%
\subsection{Tree-based Analysis}\label{tree-based-analysis}}

\texttt{SPOT}'s \texttt{buildRandomForest()} function is a wrapper
function for the \texttt{randomForest()} function from the
\texttt{randomForest} package. Since the \texttt{randomForest} package
has no default plot function, we switch to the \texttt{party} package.
This package provides the \texttt{ctree()} function, which can be
applied as follows:

\begin{Shaded}
\begin{Highlighting}[]
\NormalTok{tmaxtempz.df }\OtherTok{\textless{}{-}} \FunctionTok{data.frame}\NormalTok{(res100K}\SpecialCharTok{$}\NormalTok{x[,}\DecValTok{1}\NormalTok{], res100K}\SpecialCharTok{$}\NormalTok{x[,}\DecValTok{2}\NormalTok{], res100K}\SpecialCharTok{$}\NormalTok{y)}
\FunctionTok{names}\NormalTok{(tmaxtempz.df) }\OtherTok{\textless{}{-}} \FunctionTok{c}\NormalTok{(}\StringTok{"tmax"}\NormalTok{, }\StringTok{"temp"}\NormalTok{, }\StringTok{"y"}\NormalTok{)}
\NormalTok{tmaxtempz.tree }\OtherTok{\textless{}{-}}\NormalTok{ party}\SpecialCharTok{::}\FunctionTok{ctree}\NormalTok{(y }\SpecialCharTok{\textasciitilde{}}\NormalTok{ ., }\AttributeTok{data=}\NormalTok{tmaxtempz.df)}
\FunctionTok{plot}\NormalTok{(tmaxtempz.tree, }\AttributeTok{type=}\StringTok{"simple"}\NormalTok{)}
\end{Highlighting}
\end{Shaded}

\begin{figure}

{\centering \includegraphics[width=0.7\linewidth]{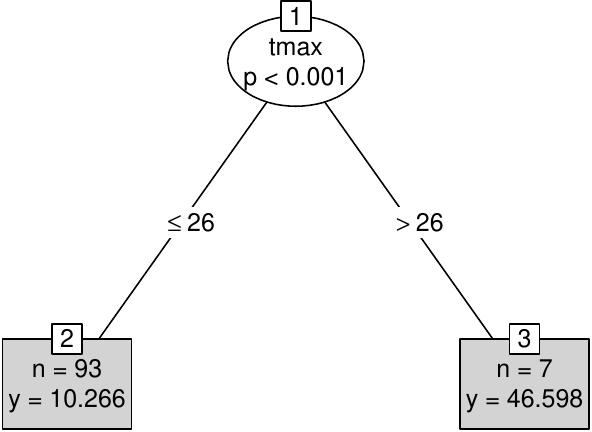} 

}

\caption{Tree based analysis with Kriging data. tmax and temp.}\label{fig:plotParty}
\end{figure}

\hypertarget{regression-analysis}{%
\subsection{Regression Analysis}\label{regression-analysis}}

\hypertarget{building-the-linear-model}{%
\subsubsection{Building the Linear
Model}\label{building-the-linear-model}}

Data from the \texttt{SPOT} run can be used for building linear models.
First, we extract the data from the result file. Then we can use the
standard \texttt{lm()} function for building the linear model.

\begin{Shaded}
\begin{Highlighting}[]
\NormalTok{xyz100K.df }\OtherTok{\textless{}{-}} \FunctionTok{data.frame}\NormalTok{(res100K}\SpecialCharTok{$}\NormalTok{x[,}\DecValTok{1}\NormalTok{], res100K}\SpecialCharTok{$}\NormalTok{x[,}\DecValTok{2}\NormalTok{], res100K}\SpecialCharTok{$}\NormalTok{y)}
\FunctionTok{names}\NormalTok{(xyz100K.df) }\OtherTok{\textless{}{-}} \FunctionTok{c}\NormalTok{(}\StringTok{"x"}\NormalTok{, }\StringTok{"y"}\NormalTok{, }\StringTok{"z"}\NormalTok{)}
\NormalTok{lm100K }\OtherTok{\textless{}{-}} \FunctionTok{lm}\NormalTok{(z }\SpecialCharTok{\textasciitilde{}}\NormalTok{ x}\SpecialCharTok{*}\NormalTok{y, }\AttributeTok{data=}\NormalTok{xyz100K.df)}
\FunctionTok{summary}\NormalTok{(lm100K)}
\end{Highlighting}
\end{Shaded}

\begin{verbatim}
## 
## Call:
## lm(formula = z ~ x * y, data = xyz100K.df)
## 
## Residuals:
##     Min      1Q  Median      3Q     Max 
## -12.266  -8.995  -5.397   4.465  53.929 
## 
## Coefficients:
##              Estimate Std. Error t value Pr(>|t|)    
## (Intercept)  6.553522  12.137952   0.540   0.5905    
## x            1.111358   0.226976   4.896 3.94e-06 ***
## y           -0.203703   0.290709  -0.701   0.4852    
## x:y         -0.007516   0.004425  -1.699   0.0926 .  
## ---
## Signif. codes:  0 '***' 0.001 '**' 0.01 '*' 0.05 '.' 0.1 ' ' 1
## 
## Residual standard error: 13.29 on 96 degrees of freedom
## Multiple R-squared:  0.4157, Adjusted R-squared:  0.3974 
## F-statistic: 22.77 on 3 and 96 DF,  p-value: 3.243e-11
\end{verbatim}

Diagnostic plots can be generated as follows:

\begin{Shaded}
\begin{Highlighting}[]
\FunctionTok{plot}\NormalTok{(lm100K)}
\end{Highlighting}
\end{Shaded}

\begin{center}\includegraphics[width=0.7\linewidth]{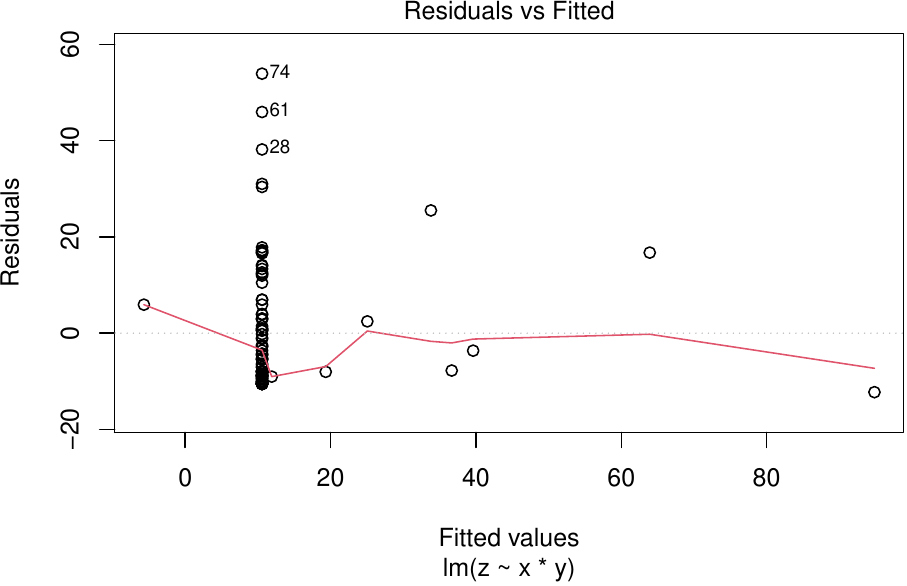} \end{center}

\begin{center}\includegraphics[width=0.7\linewidth]{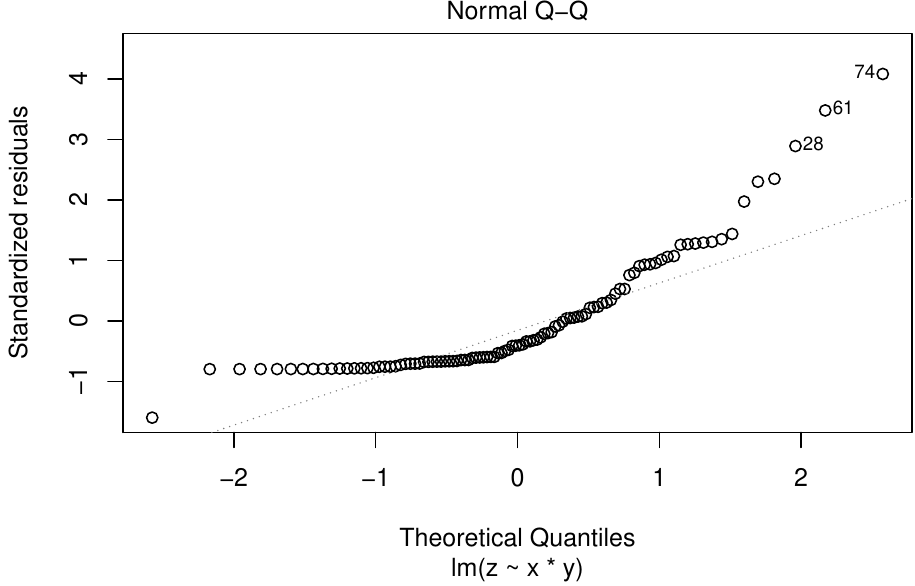} \end{center}

\begin{center}\includegraphics[width=0.7\linewidth]{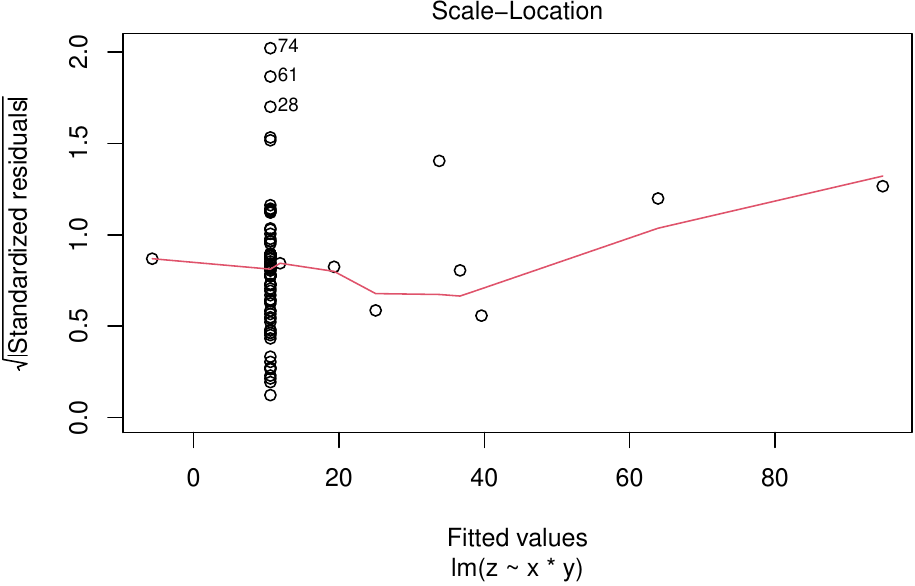} \end{center}

\begin{center}\includegraphics[width=0.7\linewidth]{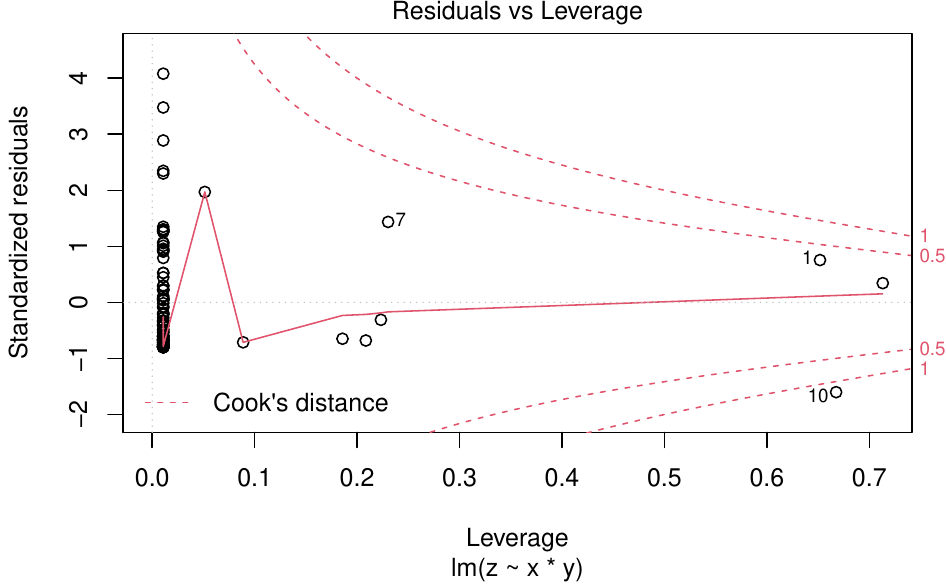} \end{center}

\hypertarget{estimating-variable-effects-based-on-linear-regression-models}{%
\subsubsection{Estimating Variable Effects Based on Linear Regression
Models}\label{estimating-variable-effects-based-on-linear-regression-models}}

\hypertarget{termplot}{%
\paragraph{termplot}\label{termplot}}

\texttt{R}'s \texttt{termplot()} function can be used to plot regression
terms against their predictors, optionally with standard errors and
partial residuals added.

\begin{Shaded}
\begin{Highlighting}[]
\FunctionTok{par}\NormalTok{(}\AttributeTok{mfrow=}\FunctionTok{c}\NormalTok{(}\DecValTok{1}\NormalTok{,}\DecValTok{2}\NormalTok{))}
\FunctionTok{termplot}\NormalTok{(lm100K, }\AttributeTok{partial =} \ConstantTok{TRUE}\NormalTok{, }\AttributeTok{smooth =}\NormalTok{ panel.smooth, }\AttributeTok{ask=}\ConstantTok{FALSE}\NormalTok{) }
\end{Highlighting}
\end{Shaded}

\begin{verbatim}
## Warning in termplot(lm100K, partial = TRUE, smooth = panel.smooth, ask = FALSE):
## 'model' appears to involve interactions: see the help page
\end{verbatim}

\begin{center}\includegraphics[width=0.7\linewidth]{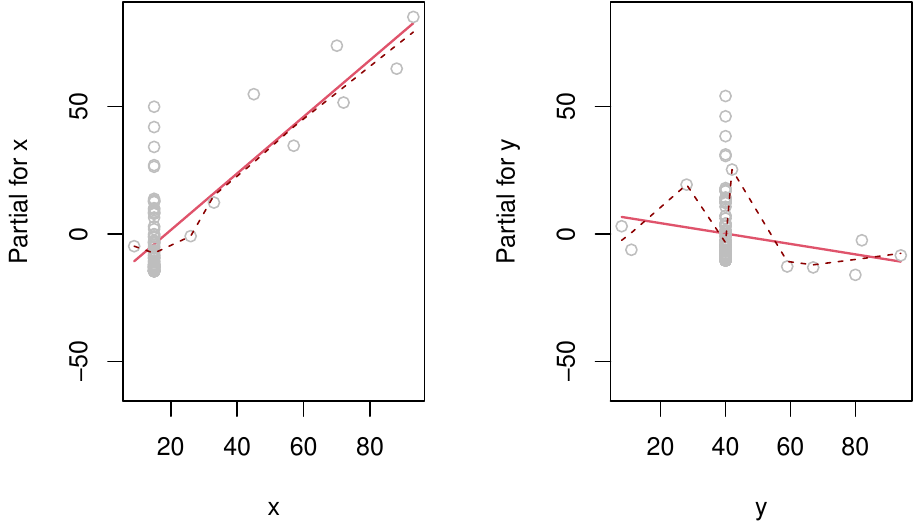} \end{center}

\begin{Shaded}
\begin{Highlighting}[]
\FunctionTok{par}\NormalTok{(}\AttributeTok{mfrom=}\FunctionTok{c}\NormalTok{(}\DecValTok{1}\NormalTok{,}\DecValTok{1}\NormalTok{))}
\end{Highlighting}
\end{Shaded}

\begin{verbatim}
## Warning in par(mfrom = c(1, 1)): "mfrom" ist kein Grafikparameter
\end{verbatim}

\hypertarget{avplots}{%
\paragraph{avPlots}\label{avplots}}

The \texttt{car} package provides the function \texttt{avPlots()}, which
can be used for visualization as follows:

\begin{Shaded}
\begin{Highlighting}[]
\FunctionTok{par}\NormalTok{(}\AttributeTok{mfrow=}\FunctionTok{c}\NormalTok{(}\DecValTok{1}\NormalTok{,}\DecValTok{3}\NormalTok{))}
\NormalTok{car}\SpecialCharTok{::}\FunctionTok{avPlots}\NormalTok{(lm100K,}\AttributeTok{ask=}\NormalTok{F)   }
\end{Highlighting}
\end{Shaded}

\begin{center}\includegraphics[width=0.7\linewidth]{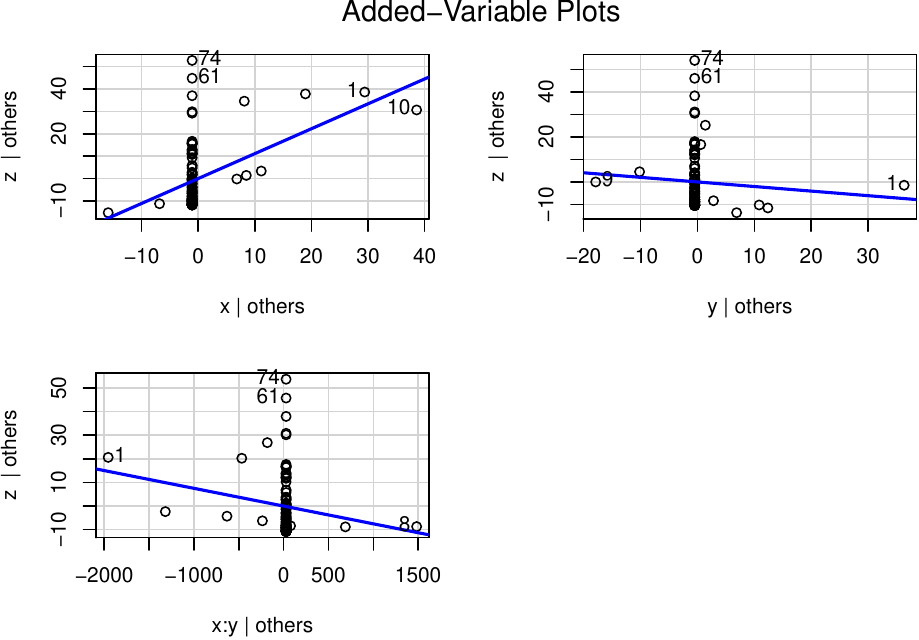} \end{center}

\begin{Shaded}
\begin{Highlighting}[]
\FunctionTok{par}\NormalTok{(}\AttributeTok{mfrow=}\FunctionTok{c}\NormalTok{(}\DecValTok{1}\NormalTok{,}\DecValTok{1}\NormalTok{))}
\end{Highlighting}
\end{Shaded}

\hypertarget{sensitivity-analysis}{%
\subsection{Sensitivity Analysis}\label{sensitivity-analysis}}

Sequential bifurcation implements an efficient procedure for screening
the most important variables, especially if the number of variables is
very high. Our example uses only two variables. It uses the res110K
Kriging model from above, which models two variables. First, we compile
a prediction function from the Kriging model \texttt{res110K}. This
function is passed to the \texttt{sequentialBifurcation} function.
Finally, a barplot is plotted to illustrate the results.

\begin{Shaded}
\begin{Highlighting}[]
\NormalTok{predictFunKriging }\OtherTok{\textless{}{-}} \ControlFlowTok{function}\NormalTok{(x)\{}
  \FunctionTok{predict}\NormalTok{(}\AttributeTok{object =}\NormalTok{ res110K}\SpecialCharTok{$}\NormalTok{modelFit,x)}
\NormalTok{\}}

\NormalTok{sens }\OtherTok{\textless{}{-}} \FunctionTok{sequentialBifurcation}\NormalTok{(predictFunKriging,}
\NormalTok{                       lower, upper,}
                      \AttributeTok{k=}\DecValTok{2}\NormalTok{, }\AttributeTok{interaction =} \ConstantTok{TRUE}\NormalTok{,}
                       \AttributeTok{verbosity =} \DecValTok{0}\NormalTok{)}
\NormalTok{ps }\OtherTok{\textless{}{-}} \FunctionTok{subgroups}\NormalTok{(sens)}
\NormalTok{colors }\OtherTok{\textless{}{-}}\NormalTok{ RColorBrewer}\SpecialCharTok{::}\FunctionTok{brewer.pal}\NormalTok{(}\DecValTok{12}\NormalTok{, }\StringTok{"Set3"}\NormalTok{) }
\FunctionTok{barplot}\NormalTok{(ps}\SpecialCharTok{$}\NormalTok{effect, }\AttributeTok{names.arg=}\NormalTok{ps}\SpecialCharTok{$}\NormalTok{group, }\AttributeTok{col=}\NormalTok{ colors)}
\end{Highlighting}
\end{Shaded}

\begin{center}\includegraphics[width=0.7\linewidth]{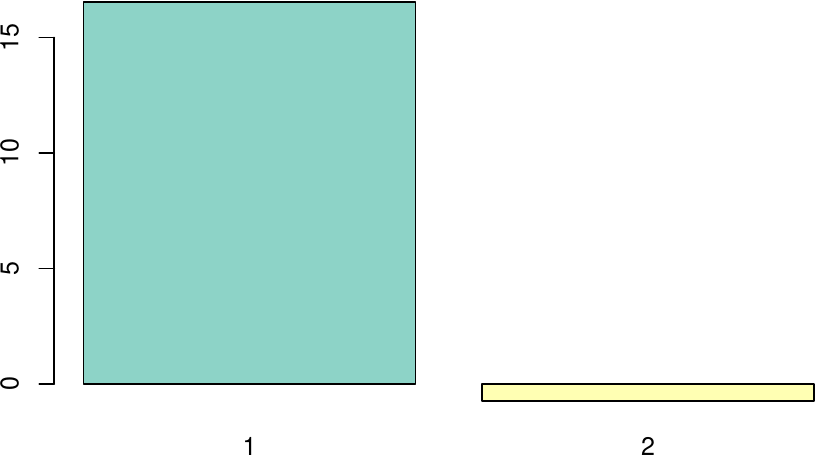} \end{center}

\hypertarget{sec:determ}{%
\section{Deterministic Problems and Surrogate Model Based
Optimization}\label{sec:determ}}

Previous sections discussed the tuning of non-deterministic, i.e.,
noisy, algorithms, which is the usual case when tuning evolutionary
algorithms. This section presents an application of \texttt{SPOT} in a
simple setting: We will describe how \texttt{SPOT} can be used for
optimizing deterministic problems directly, i.e., we apply \texttt{SPOT}
in the context of surrogate model based optimization.

To present a very simple example, \texttt{SPOT} will be used for
minimizing the sphere function. Level L2 from is omitted, and the tuning
algorithm for level L3 is applied directly to the real-world system on
level L1. So instead of tuning the stochastic \texttt{SANN} heuristic,
which in turn optimizes the sphere function, \texttt{SPOT} tries to find
the minimum of the deterministic sphere function directly. This example
illustrates necessary modifications of the \texttt{SPOT} configuration
in deterministic settings. Since no randomness occurs, repeats or other
mechanism to cope with noise are not necessary anymore. The interval
\([-5;5] \times [-5; 5]\) was chosen as the region of interest, i.e., we
are considering a two-dimensional optimization problem.

\begin{Shaded}
\begin{Highlighting}[]
\NormalTok{res }\OtherTok{\textless{}{-}} \FunctionTok{spot}\NormalTok{(}\AttributeTok{x=}\ConstantTok{NULL}\NormalTok{,funSphere,}\FunctionTok{c}\NormalTok{(}\SpecialCharTok{{-}}\DecValTok{5}\NormalTok{,}\SpecialCharTok{{-}}\DecValTok{5}\NormalTok{),}\FunctionTok{c}\NormalTok{(}\DecValTok{5}\NormalTok{,}\DecValTok{5}\NormalTok{), }\AttributeTok{control=}\FunctionTok{list}\NormalTok{(}\AttributeTok{optimizer=}\NormalTok{optimLBFGSB)) }
\end{Highlighting}
\end{Shaded}

We can extract the best solution with the following command.

\begin{Shaded}
\begin{Highlighting}[]
\NormalTok{res}\SpecialCharTok{$}\NormalTok{xbest}
\end{Highlighting}
\end{Shaded}

\begin{verbatim}
##              [,1]        [,2]
## [1,] -0.009598569 -0.01446771
\end{verbatim}

\begin{Shaded}
\begin{Highlighting}[]
\NormalTok{res}\SpecialCharTok{$}\NormalTok{ybest}
\end{Highlighting}
\end{Shaded}

\begin{verbatim}
##              [,1]
## [1,] 0.0003014473
\end{verbatim}

\hypertarget{sec:stack}{%
\section{Model Ensembles: Stacking}\label{sec:stack}}

\texttt{SPOT} provides several models that can be used as surrogates.
Sometimes it is not obvious, which surrogate should be chosen.
Ensemble-based models provide a well-established solution to this model
selection problem \cite{Bart16n}. Therefore, \texttt{SPOT} provides a
stacking approach, that combines several models in a sophisticated
manner. The stacking procedure is described in detail in
\cite{Bart16jcos}.

We will use the data from Example \texttt{plotTrained} to illustrate the
stacking approach.

\begin{Shaded}
\begin{Highlighting}[]
\NormalTok{fit.stack }\OtherTok{\textless{}{-}} \FunctionTok{buildEnsembleStack}\NormalTok{(x.test, y.test)}
\end{Highlighting}
\end{Shaded}

\begin{Shaded}
\begin{Highlighting}[]
\FunctionTok{plotModel}\NormalTok{(fit.stack)}
\end{Highlighting}
\end{Shaded}

\begin{center}\includegraphics[width=0.7\linewidth]{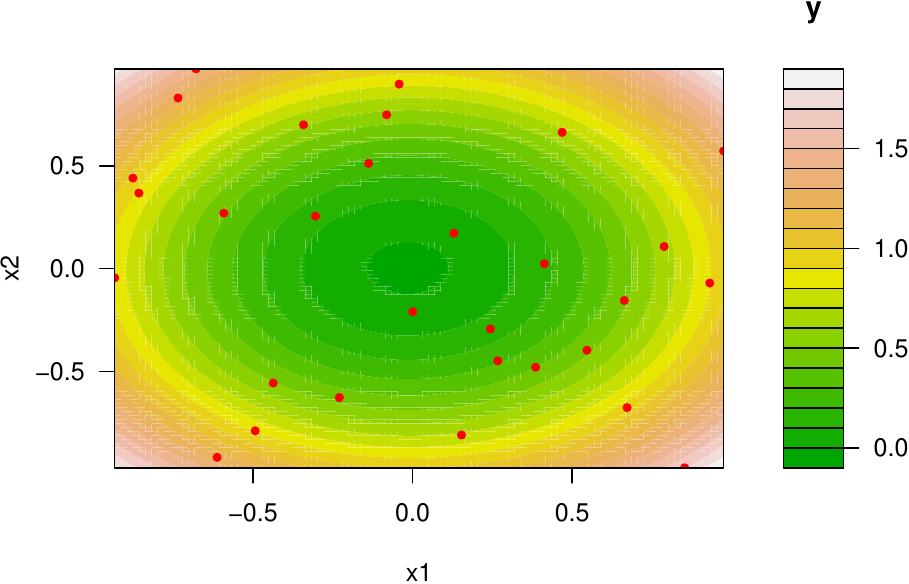} \end{center}

We can compare predicted and true values as follows:

\begin{Shaded}
\begin{Highlighting}[]
\NormalTok{xNew }\OtherTok{\textless{}{-}}  \FunctionTok{cbind}\NormalTok{(}\DecValTok{1}\NormalTok{,}\DecValTok{1}\NormalTok{,}\DecValTok{1}\NormalTok{)}
\FunctionTok{predict}\NormalTok{(fit.stack, xNew)}
\end{Highlighting}
\end{Shaded}

\begin{verbatim}
## $y
##        1 
## 2.857598
\end{verbatim}

\begin{Shaded}
\begin{Highlighting}[]
\FunctionTok{funSphere}\NormalTok{(xNew)}
\end{Highlighting}
\end{Shaded}

\begin{verbatim}
##      [,1]
## [1,]    3
\end{verbatim}

\hypertarget{sec:hybrid}{%
\section{Hybrid Approaches}\label{sec:hybrid}}

SPOT provides methods for hybrid optimization strategies. A surrogate
model can be used to perform a balanced explorative-exploitative search
during the first phase of the optimization. During the second phase of
the optimization, a local (direct, exploitative, aggressive) search is
performed to refine the results.

\hypertarget{sphere-function}{%
\subsection{Sphere Function}\label{sphere-function}}

\begin{Shaded}
\begin{Highlighting}[]
\FunctionTok{set.seed}\NormalTok{(}\DecValTok{1}\NormalTok{)}
\NormalTok{n }\OtherTok{\textless{}{-}} \DecValTok{2}
\NormalTok{low }\OtherTok{=} \SpecialCharTok{{-}}\DecValTok{10}
\NormalTok{up }\OtherTok{=} \DecValTok{10}
\NormalTok{a }\OtherTok{=} \FunctionTok{runif}\NormalTok{(n, low, }\DecValTok{0}\NormalTok{)}
\NormalTok{b }\OtherTok{=} \FunctionTok{runif}\NormalTok{(n, }\DecValTok{0}\NormalTok{, up)}
\NormalTok{x0 }\OtherTok{=}\NormalTok{ a }\SpecialCharTok{+} \FunctionTok{runif}\NormalTok{(n)}\SpecialCharTok{*}\NormalTok{(b}\SpecialCharTok{{-}}\NormalTok{a)}
\CommentTok{\#plot(a, type = "l", ylim=c(up,low))}
\CommentTok{\#lines(b)}
\CommentTok{\#lines(x0)}
\NormalTok{x0 }\OtherTok{=} \FunctionTok{matrix}\NormalTok{( x0, }\AttributeTok{nrow =} \DecValTok{1}\NormalTok{)}
\NormalTok{reps }\OtherTok{\textless{}{-}} \DecValTok{2}
\NormalTok{end }\OtherTok{\textless{}{-}} \DecValTok{20}\SpecialCharTok{*}\NormalTok{n }
\NormalTok{ninit }\OtherTok{\textless{}{-}} \DecValTok{2}\SpecialCharTok{*}\NormalTok{n}
\end{Highlighting}
\end{Shaded}

The first run uses all function evaluations for the surrogate model
optimization:

\begin{Shaded}
\begin{Highlighting}[]
\NormalTok{progSpot }\OtherTok{\textless{}{-}} \FunctionTok{matrix}\NormalTok{(}\ConstantTok{NA}\NormalTok{, }\AttributeTok{nrow =}\NormalTok{ reps, }\AttributeTok{ncol =} \DecValTok{2}\SpecialCharTok{*}\NormalTok{end)}
\ControlFlowTok{for}\NormalTok{(r }\ControlFlowTok{in} \DecValTok{1}\SpecialCharTok{:}\NormalTok{reps)\{}
  \FunctionTok{set.seed}\NormalTok{(r)}
\NormalTok{  x0 }\OtherTok{\textless{}{-}}\NormalTok{ a }\SpecialCharTok{+} \FunctionTok{runif}\NormalTok{(n)}\SpecialCharTok{*}\NormalTok{(b}\SpecialCharTok{{-}}\NormalTok{a)}
\NormalTok{  x0 }\OtherTok{=} \FunctionTok{matrix}\NormalTok{( x0, }\AttributeTok{nrow =} \DecValTok{1}\NormalTok{)}
\NormalTok{  sol }\OtherTok{\textless{}{-}} \FunctionTok{spot}\NormalTok{(}\AttributeTok{x=}\NormalTok{ x0, funSphere, a, b, }\AttributeTok{control=}\FunctionTok{list}\NormalTok{(}\AttributeTok{funEvals=}\DecValTok{2}\SpecialCharTok{*}\NormalTok{end,}
                                                   \AttributeTok{model =}\NormalTok{ buildKriging,}
                                                   \AttributeTok{optimizer=}\NormalTok{optimNLOPTR, }
                                                   \AttributeTok{directOptControl =} \FunctionTok{list}\NormalTok{(}\AttributeTok{funEvals=}\DecValTok{0}\NormalTok{),}
                                                   \AttributeTok{designControl =} \FunctionTok{list}\NormalTok{(}\AttributeTok{size =}\NormalTok{ ninit)))}
\NormalTok{  progSpot[r, ] }\OtherTok{\textless{}{-}} \FunctionTok{prepareBestObjectiveVal}\NormalTok{(sol}\SpecialCharTok{$}\NormalTok{y, }\DecValTok{2}\SpecialCharTok{*}\NormalTok{end)}
\NormalTok{\}}
\end{Highlighting}
\end{Shaded}

The next run splits the number of available objective function
evaluations equally between

\begin{itemize}
\tightlist
\item
  surrogate model based and
\item
  direct optimization:
\end{itemize}

\begin{Shaded}
\begin{Highlighting}[]
\NormalTok{progSpotHyb }\OtherTok{\textless{}{-}} \FunctionTok{matrix}\NormalTok{(}\ConstantTok{NA}\NormalTok{, }\AttributeTok{nrow =}\NormalTok{ reps, }\AttributeTok{ncol =} \DecValTok{2}\SpecialCharTok{*}\NormalTok{end)}
\ControlFlowTok{for}\NormalTok{(r }\ControlFlowTok{in} \DecValTok{1}\SpecialCharTok{:}\NormalTok{reps)\{}
  \FunctionTok{set.seed}\NormalTok{(r)}
\NormalTok{  x0 }\OtherTok{\textless{}{-}}\NormalTok{ a }\SpecialCharTok{+} \FunctionTok{runif}\NormalTok{(n)}\SpecialCharTok{*}\NormalTok{(b}\SpecialCharTok{{-}}\NormalTok{a)}
\NormalTok{  x0 }\OtherTok{=} \FunctionTok{matrix}\NormalTok{( x0, }\AttributeTok{nrow =} \DecValTok{1}\NormalTok{)}
\NormalTok{  solHyb }\OtherTok{\textless{}{-}} \FunctionTok{spot}\NormalTok{(}\AttributeTok{x=}\NormalTok{ x0, funSphere, a, b, }\AttributeTok{control=}\FunctionTok{list}\NormalTok{(}\AttributeTok{funEvals=}\NormalTok{end,}
                                                   \AttributeTok{model =}\NormalTok{ buildKriging,}
                                                   \AttributeTok{optimizer=}\NormalTok{optimNLOPTR, }
                                                   \AttributeTok{directOptControl =} \FunctionTok{list}\NormalTok{(}\AttributeTok{funEvals=}\NormalTok{end, }\AttributeTok{verbosity=}\DecValTok{1}\NormalTok{),}
                                                   \AttributeTok{designControl =} \FunctionTok{list}\NormalTok{(}\AttributeTok{size =}\NormalTok{ ninit),}
                                                   \AttributeTok{verbosity =} \DecValTok{0}\NormalTok{))}
\NormalTok{  progSpotHyb[r, ] }\OtherTok{\textless{}{-}} \FunctionTok{prepareBestObjectiveVal}\NormalTok{(solHyb}\SpecialCharTok{$}\NormalTok{y, }\DecValTok{2}\SpecialCharTok{*}\NormalTok{end)}
\NormalTok{\}}
\end{Highlighting}
\end{Shaded}

\begin{verbatim}
## [1] "optimNLOPTR starting point:"
##  num [1:2] -0.00716 -0.01011
## [1] "optimNLOPTR finished:"
## List of 6
##  $ x    : num [1:47, 1:2] -0.00716 -0.00716 -0.80819 3.54963 -5.16601 ...
##  $ y    : num [1:47, 1] 1.53e-04 1.53e-04 2.62 1.46e+01 2.87e+01 ...
##   ..- attr(*, "dimnames")=List of 2
##   .. ..$ : NULL
##   .. ..$ : NULL
##  $ xbest: num [1, 1:2] -0.00119 0.07418
##  $ ybest: num 0.0055
##  $ count: int 45
##  $ msg  : chr "NLOPT_MAXEVAL_REACHED: Optimization stopped because maxeval (above) was reached."
## [1] "optimNLOPTR starting point:"
##  num [1:2] 0.00479 0.01097
## [1] "optimNLOPTR finished:"
## List of 6
##  $ x    : num [1:47, 1:2] 0.00479 0.00479 -0.80819 3.54963 -5.16601 ...
##  $ y    : num [1:47, 1] 1.43e-04 1.43e-04 2.62 1.46e+01 2.87e+01 ...
##   ..- attr(*, "dimnames")=List of 2
##   .. ..$ : NULL
##   .. ..$ : NULL
##  $ xbest: num [1, 1:2] -0.00119 0.07418
##  $ ybest: num 0.0055
##  $ count: int 45
##  $ msg  : chr "NLOPT_MAXEVAL_REACHED: Optimization stopped because maxeval (above) was reached."
\end{verbatim}

\begin{Shaded}
\begin{Highlighting}[]
\FunctionTok{matplot}\NormalTok{(}\FunctionTok{t}\NormalTok{(progSpotHyb), }
        \AttributeTok{type=}\StringTok{"l"}\NormalTok{, }
        \AttributeTok{col=}\StringTok{"red"}\NormalTok{, }
        \AttributeTok{lty=}\DecValTok{1}\NormalTok{, }\AttributeTok{xlab=}\StringTok{"n: function evaluations"}\NormalTok{, }
        \AttributeTok{ylab=}\StringTok{"function value (log scale)"}\NormalTok{, }
        \AttributeTok{log=}\StringTok{"y"}\NormalTok{, }
        \AttributeTok{ylim =} \FunctionTok{c}\NormalTok{( }\FunctionTok{min}\NormalTok{(}\FunctionTok{c}\NormalTok{(progSpot, progSpotHyb)), }\FunctionTok{max}\NormalTok{(}\FunctionTok{c}\NormalTok{(progSpot, progSpotHyb))))}
\FunctionTok{abline}\NormalTok{(}\AttributeTok{v=}\NormalTok{ninit, }\AttributeTok{lty=}\DecValTok{2}\NormalTok{)}
\FunctionTok{matlines}\NormalTok{(}\FunctionTok{t}\NormalTok{(progSpot), }\AttributeTok{type=}\StringTok{"l"}\NormalTok{, }\AttributeTok{col=}\StringTok{"black"}\NormalTok{, }\AttributeTok{lty=}\DecValTok{2}\NormalTok{)}
\FunctionTok{legend}\NormalTok{(}\StringTok{"topright"}\NormalTok{, }\FunctionTok{c}\NormalTok{(}\StringTok{"SpotHyb"}\NormalTok{, }\StringTok{"Spot"}\NormalTok{), }\AttributeTok{col=}\FunctionTok{c}\NormalTok{(}\StringTok{"red"}\NormalTok{, }\StringTok{"black"}\NormalTok{), }\AttributeTok{lty=}\FunctionTok{c}\NormalTok{(}\DecValTok{1}\NormalTok{,}\DecValTok{2}\NormalTok{), }\AttributeTok{bty=}\StringTok{"n"}\NormalTok{)}
\end{Highlighting}
\end{Shaded}

\begin{center}\includegraphics[width=0.7\linewidth]{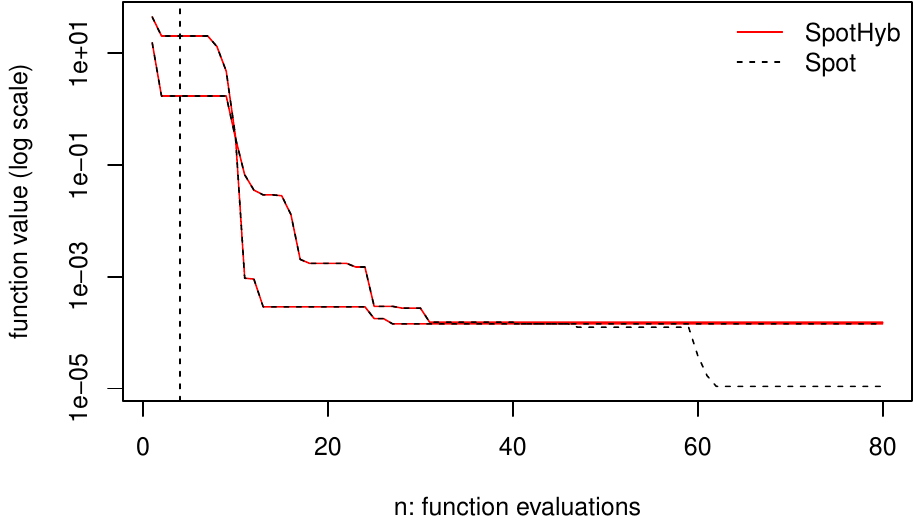} \end{center}

We can also access the x and y values from the complete run (surrogate
and direct):

\begin{Shaded}
\begin{Highlighting}[]
\NormalTok{yy }\OtherTok{\textless{}{-}}\NormalTok{ solHyb}\SpecialCharTok{$}\NormalTok{y}
\NormalTok{yx }\OtherTok{\textless{}{-}} \FunctionTok{funSphere}\NormalTok{(solHyb}\SpecialCharTok{$}\NormalTok{x)}
\NormalTok{yval }\OtherTok{=} \FunctionTok{cbind}\NormalTok{(yy,yx)}
\FunctionTok{head}\NormalTok{(yval)}
\end{Highlighting}
\end{Shaded}

\begin{verbatim}
##          [,1]     [,2]
## [1,] 44.62660 44.62660
## [2,] 20.36456 20.36456
## [3,] 24.18770 24.18770
## [4,] 40.41844 40.41844
## [5,] 68.04472 68.04472
## [6,] 20.37929 20.37929
\end{verbatim}

\hypertarget{sec:summary}{%
\section{Summary}\label{sec:summary}}

This report describes experimental methods for tuning algorithms. Using
a simple simulated annealing algorithm, it was demonstrated how
optimization algorithms can be tuned using the \texttt{SPOT}. Several
tools from the \texttt{SPOT} for automated and interactive tuning were
illustrated and the underling concepts of the \texttt{SPOT} approach
were explained. Central in the \texttt{SPOT} approach are techniques
such as exploratory fitness landscape analysis and response surface
methodology. Furthermore, we demonstrated how SPOT can be used as
optimizer and how a sophisticated ensemble approach is able to combine
several meta models via stacking.

\hypertarget{sec:appendix}{%
\section{Appendix}\label{sec:appendix}}

\hypertarget{sequential-parameter-optimization-examples-how-to-call-spot}{%
\subsection{Sequential Parameter Optimization Examples: How to Call
SPOT}\label{sequential-parameter-optimization-examples-how-to-call-spot}}

This section exemplifies how \texttt{spot} can be called.

\hypertarget{most-simple-example-kriging-lhs-predicted-mean-optimization-not-expected-improvement}{%
\subsubsection{Most simple example: Kriging + LHS + predicted mean
optimization (not expected
improvement)}\label{most-simple-example-kriging-lhs-predicted-mean-optimization-not-expected-improvement}}

\begin{Shaded}
\begin{Highlighting}[]
\NormalTok{res }\OtherTok{\textless{}{-}} \FunctionTok{spot}\NormalTok{(,funSphere,}\FunctionTok{c}\NormalTok{(}\SpecialCharTok{{-}}\DecValTok{2}\NormalTok{,}\SpecialCharTok{{-}}\DecValTok{3}\NormalTok{),}\FunctionTok{c}\NormalTok{(}\DecValTok{1}\NormalTok{,}\DecValTok{2}\NormalTok{),}\AttributeTok{control=}\FunctionTok{list}\NormalTok{(}\AttributeTok{funEvals=}\DecValTok{15}\NormalTok{))}
\NormalTok{res}\SpecialCharTok{$}\NormalTok{xbest}
\end{Highlighting}
\end{Shaded}

\begin{verbatim}
##            [,1]      [,2]
## [1,] -0.1086201 0.1184503
\end{verbatim}

\hypertarget{with-expected-improvement}{%
\subsubsection{With expected
improvement}\label{with-expected-improvement}}

\begin{Shaded}
\begin{Highlighting}[]
\NormalTok{res }\OtherTok{\textless{}{-}} \FunctionTok{spot}\NormalTok{(,funSphere,}\FunctionTok{c}\NormalTok{(}\SpecialCharTok{{-}}\DecValTok{2}\NormalTok{,}\SpecialCharTok{{-}}\DecValTok{3}\NormalTok{),}\FunctionTok{c}\NormalTok{(}\DecValTok{1}\NormalTok{,}\DecValTok{2}\NormalTok{),}
    \AttributeTok{control=}\FunctionTok{list}\NormalTok{(}\AttributeTok{funEvals=}\DecValTok{15}\NormalTok{,}\AttributeTok{modelControl=}\FunctionTok{list}\NormalTok{(}\AttributeTok{target=}\StringTok{"ei"}\NormalTok{)))}
\NormalTok{res}\SpecialCharTok{$}\NormalTok{xbest}
\end{Highlighting}
\end{Shaded}

\begin{verbatim}
##            [,1]      [,2]
## [1,] -0.1086201 0.1184503
\end{verbatim}

\hypertarget{with-additional-start-point}{%
\subsubsection{With additional start
point:}\label{with-additional-start-point}}

\begin{Shaded}
\begin{Highlighting}[]
\NormalTok{res }\OtherTok{\textless{}{-}} \FunctionTok{spot}\NormalTok{(}\FunctionTok{matrix}\NormalTok{(}\FunctionTok{c}\NormalTok{(}\FloatTok{0.05}\NormalTok{,}\FloatTok{0.1}\NormalTok{),}\DecValTok{1}\NormalTok{,}\DecValTok{2}\NormalTok{),funSphere,}\FunctionTok{c}\NormalTok{(}\SpecialCharTok{{-}}\DecValTok{2}\NormalTok{,}\SpecialCharTok{{-}}\DecValTok{3}\NormalTok{),}\FunctionTok{c}\NormalTok{(}\DecValTok{1}\NormalTok{,}\DecValTok{2}\NormalTok{))}
\NormalTok{res}\SpecialCharTok{$}\NormalTok{xbest}
\end{Highlighting}
\end{Shaded}

\begin{verbatim}
##             [,1]        [,2]
## [1,] -0.06104759 -0.05040567
\end{verbatim}

\hypertarget{larger-budget}{%
\subsubsection{Larger budget:}\label{larger-budget}}

\begin{Shaded}
\begin{Highlighting}[]
\NormalTok{res }\OtherTok{\textless{}{-}} \FunctionTok{spot}\NormalTok{(,funSphere,}\FunctionTok{c}\NormalTok{(}\SpecialCharTok{{-}}\DecValTok{2}\NormalTok{,}\SpecialCharTok{{-}}\DecValTok{3}\NormalTok{),}\FunctionTok{c}\NormalTok{(}\DecValTok{1}\NormalTok{,}\DecValTok{2}\NormalTok{),}
    \AttributeTok{control=}\FunctionTok{list}\NormalTok{(}\AttributeTok{funEvals=}\DecValTok{50}\NormalTok{))}
\NormalTok{res}\SpecialCharTok{$}\NormalTok{xbest}
\end{Highlighting}
\end{Shaded}

\begin{verbatim}
##            [,1]        [,2]
## [1,] 0.02088315 -0.03783177
\end{verbatim}

\hypertarget{use-local-optimization-instead-of-lhs}{%
\subsubsection{Use local optimization instead of
LHS}\label{use-local-optimization-instead-of-lhs}}

\begin{Shaded}
\begin{Highlighting}[]
\NormalTok{res }\OtherTok{\textless{}{-}} \FunctionTok{spot}\NormalTok{(,funSphere,}\FunctionTok{c}\NormalTok{(}\SpecialCharTok{{-}}\DecValTok{2}\NormalTok{,}\SpecialCharTok{{-}}\DecValTok{3}\NormalTok{),}\FunctionTok{c}\NormalTok{(}\DecValTok{1}\NormalTok{,}\DecValTok{2}\NormalTok{),}
   \AttributeTok{control=}\FunctionTok{list}\NormalTok{(}\AttributeTok{optimizer=}\NormalTok{optimLBFGSB))}
\NormalTok{ res}\SpecialCharTok{$}\NormalTok{xbest}
\end{Highlighting}
\end{Shaded}

\begin{verbatim}
##             [,1]        [,2]
## [1,] -0.01573485 -0.00886033
\end{verbatim}

\hypertarget{random-forest-instead-of-kriging}{%
\subsubsection{Random Forest instead of
Kriging}\label{random-forest-instead-of-kriging}}

\begin{Shaded}
\begin{Highlighting}[]
\NormalTok{res }\OtherTok{\textless{}{-}} \FunctionTok{spot}\NormalTok{(,funSphere,}\FunctionTok{c}\NormalTok{(}\SpecialCharTok{{-}}\DecValTok{2}\NormalTok{,}\SpecialCharTok{{-}}\DecValTok{3}\NormalTok{),}\FunctionTok{c}\NormalTok{(}\DecValTok{1}\NormalTok{,}\DecValTok{2}\NormalTok{),}
     \AttributeTok{control=}\FunctionTok{list}\NormalTok{(}\AttributeTok{model=}\NormalTok{buildRandomForest))}
\NormalTok{res}\SpecialCharTok{$}\NormalTok{xbest}
\end{Highlighting}
\end{Shaded}

\begin{verbatim}
##           [,1]      [,2]
## [1,] 0.1531584 0.3294388
\end{verbatim}

\hypertarget{lm-instead-of-kriging}{%
\subsubsection{LM instead of Kriging}\label{lm-instead-of-kriging}}

\begin{Shaded}
\begin{Highlighting}[]
\NormalTok{res }\OtherTok{\textless{}{-}} \FunctionTok{spot}\NormalTok{(,funSphere,}\FunctionTok{c}\NormalTok{(}\SpecialCharTok{{-}}\DecValTok{2}\NormalTok{,}\SpecialCharTok{{-}}\DecValTok{3}\NormalTok{),}\FunctionTok{c}\NormalTok{(}\DecValTok{1}\NormalTok{,}\DecValTok{2}\NormalTok{),}
     \AttributeTok{control=}\FunctionTok{list}\NormalTok{(}\AttributeTok{model=}\NormalTok{buildLM)) }\CommentTok{\#lm as surrogate}
\NormalTok{res}\SpecialCharTok{$}\NormalTok{xbest}
\end{Highlighting}
\end{Shaded}

\begin{verbatim}
##           [,1]      [,2]
## [1,] 0.1531584 0.3294388
\end{verbatim}

\hypertarget{lm-and-local-optimizer-which-for-this-simple-example-is-perfect}{%
\subsubsection{LM and local optimizer (which for this simple example is
perfect)}\label{lm-and-local-optimizer-which-for-this-simple-example-is-perfect}}

\begin{Shaded}
\begin{Highlighting}[]
\NormalTok{res }\OtherTok{\textless{}{-}} \FunctionTok{spot}\NormalTok{(,funSphere,}\FunctionTok{c}\NormalTok{(}\SpecialCharTok{{-}}\DecValTok{2}\NormalTok{,}\SpecialCharTok{{-}}\DecValTok{3}\NormalTok{),}\FunctionTok{c}\NormalTok{(}\DecValTok{1}\NormalTok{,}\DecValTok{2}\NormalTok{),}
   \AttributeTok{control=}\FunctionTok{list}\NormalTok{(}\AttributeTok{model=}\NormalTok{buildLM, }\AttributeTok{optimizer=}\NormalTok{optimLBFGSB))}
\NormalTok{res}\SpecialCharTok{$}\NormalTok{xbest}
\end{Highlighting}
\end{Shaded}

\begin{verbatim}
##             [,1]       [,2]
## [1,] -0.02651579 -0.1091904
\end{verbatim}

\hypertarget{lasso-and-local-optimizer-nloptr}{%
\subsubsection{Lasso and local optimizer
NLOPTR}\label{lasso-and-local-optimizer-nloptr}}

\begin{Shaded}
\begin{Highlighting}[]
\NormalTok{res }\OtherTok{\textless{}{-}} \FunctionTok{spot}\NormalTok{(,funSphere,}\FunctionTok{c}\NormalTok{(}\SpecialCharTok{{-}}\DecValTok{2}\NormalTok{,}\SpecialCharTok{{-}}\DecValTok{3}\NormalTok{),}\FunctionTok{c}\NormalTok{(}\DecValTok{1}\NormalTok{,}\DecValTok{2}\NormalTok{), }
   \AttributeTok{control=}\FunctionTok{list}\NormalTok{(}\AttributeTok{model=}\NormalTok{buildLasso, }\AttributeTok{optimizer =}\NormalTok{ optimNLOPTR))}
\end{Highlighting}
\end{Shaded}

\begin{verbatim}
## Warning: Option grouped=FALSE enforced in cv.glmnet, since < 3 observations per
## fold

## Warning: Option grouped=FALSE enforced in cv.glmnet, since < 3 observations per
## fold

## Warning: Option grouped=FALSE enforced in cv.glmnet, since < 3 observations per
## fold

## Warning: Option grouped=FALSE enforced in cv.glmnet, since < 3 observations per
## fold

## Warning: Option grouped=FALSE enforced in cv.glmnet, since < 3 observations per
## fold

## Warning: Option grouped=FALSE enforced in cv.glmnet, since < 3 observations per
## fold

## Warning: Option grouped=FALSE enforced in cv.glmnet, since < 3 observations per
## fold

## Warning: Option grouped=FALSE enforced in cv.glmnet, since < 3 observations per
## fold

## Warning: Option grouped=FALSE enforced in cv.glmnet, since < 3 observations per
## fold

## Warning: Option grouped=FALSE enforced in cv.glmnet, since < 3 observations per
## fold
\end{verbatim}

\begin{Shaded}
\begin{Highlighting}[]
\NormalTok{res}\SpecialCharTok{$}\NormalTok{xbest}
\end{Highlighting}
\end{Shaded}

\begin{verbatim}
##           [,1]      [,2]
## [1,] 0.1531584 0.3294388
\end{verbatim}

\hypertarget{kriging-and-local-optimizer-lbfgsb}{%
\subsubsection{Kriging and local optimizer
LBFGSB}\label{kriging-and-local-optimizer-lbfgsb}}

\begin{Shaded}
\begin{Highlighting}[]
\NormalTok{res }\OtherTok{\textless{}{-}} \FunctionTok{spot}\NormalTok{(,funSphere,}\FunctionTok{c}\NormalTok{(}\SpecialCharTok{{-}}\DecValTok{2}\NormalTok{,}\SpecialCharTok{{-}}\DecValTok{3}\NormalTok{),}\FunctionTok{c}\NormalTok{(}\DecValTok{1}\NormalTok{,}\DecValTok{2}\NormalTok{), }
   \AttributeTok{control=}\FunctionTok{list}\NormalTok{(}\AttributeTok{model=}\NormalTok{buildKriging, }\AttributeTok{optimizer =}\NormalTok{ optimLBFGSB))}
\NormalTok{res}\SpecialCharTok{$}\NormalTok{xbest}
\end{Highlighting}
\end{Shaded}

\begin{verbatim}
##             [,1]        [,2]
## [1,] -0.01573485 -0.00886033
\end{verbatim}

\hypertarget{kriging-and-local-optimizer-nloptr}{%
\subsubsection{Kriging and local optimizer
NLOPTR}\label{kriging-and-local-optimizer-nloptr}}

\begin{Shaded}
\begin{Highlighting}[]
\NormalTok{res }\OtherTok{\textless{}{-}} \FunctionTok{spot}\NormalTok{(,funSphere,}\FunctionTok{c}\NormalTok{(}\SpecialCharTok{{-}}\DecValTok{2}\NormalTok{,}\SpecialCharTok{{-}}\DecValTok{3}\NormalTok{),}\FunctionTok{c}\NormalTok{(}\DecValTok{1}\NormalTok{,}\DecValTok{2}\NormalTok{), }
     \AttributeTok{control=}\FunctionTok{list}\NormalTok{(}\AttributeTok{model=}\NormalTok{buildKriging, }\AttributeTok{optimizer =}\NormalTok{ optimNLOPTR))}
\NormalTok{res}\SpecialCharTok{$}\NormalTok{xbest}
\end{Highlighting}
\end{Shaded}

\begin{verbatim}
##             [,1]         [,2]
## [1,] -0.01440329 0.0006858711
\end{verbatim}

\hypertarget{or-a-different-kriging-model}{%
\subsubsection{Or a different Kriging
model:}\label{or-a-different-kriging-model}}

\begin{Shaded}
\begin{Highlighting}[]
\NormalTok{res }\OtherTok{\textless{}{-}} \FunctionTok{spot}\NormalTok{(,funSphere,}\FunctionTok{c}\NormalTok{(}\SpecialCharTok{{-}}\DecValTok{2}\NormalTok{,}\SpecialCharTok{{-}}\DecValTok{3}\NormalTok{),}\FunctionTok{c}\NormalTok{(}\DecValTok{1}\NormalTok{,}\DecValTok{2}\NormalTok{),}
 \AttributeTok{control=}\FunctionTok{list}\NormalTok{(}\AttributeTok{model=}\NormalTok{buildKrigingDACE, }\AttributeTok{optimizer=}\NormalTok{optimLBFGSB))}
\NormalTok{res}\SpecialCharTok{$}\NormalTok{xbest}
\end{Highlighting}
\end{Shaded}

\begin{verbatim}
##             [,1]         [,2]
## [1,] -0.03586733 -0.004407048
\end{verbatim}

\hypertarget{with-noise-this-takes-some-time}{%
\subsubsection{With noise: (this takes some
time)}\label{with-noise-this-takes-some-time}}

\begin{Shaded}
\begin{Highlighting}[]
\NormalTok{res1 }\OtherTok{\textless{}{-}} \FunctionTok{spot}\NormalTok{(,}\ControlFlowTok{function}\NormalTok{(x)}\FunctionTok{funSphere}\NormalTok{(x)}\SpecialCharTok{+}\FunctionTok{rnorm}\NormalTok{(}\FunctionTok{nrow}\NormalTok{(x)),}\FunctionTok{c}\NormalTok{(}\SpecialCharTok{{-}}\DecValTok{2}\NormalTok{,}\SpecialCharTok{{-}}\DecValTok{3}\NormalTok{),}\FunctionTok{c}\NormalTok{(}\DecValTok{1}\NormalTok{,}\DecValTok{2}\NormalTok{),}
        \AttributeTok{control=}\FunctionTok{list}\NormalTok{(}\AttributeTok{funEvals=}\DecValTok{100}\NormalTok{,}\AttributeTok{noise=}\ConstantTok{TRUE}\NormalTok{)) }\CommentTok{\#noisy objective}
\NormalTok{res2 }\OtherTok{\textless{}{-}} \FunctionTok{spot}\NormalTok{(,}\ControlFlowTok{function}\NormalTok{(x)}\FunctionTok{funSphere}\NormalTok{(x)}\SpecialCharTok{+}\FunctionTok{rnorm}\NormalTok{(}\FunctionTok{nrow}\NormalTok{(x)),}\FunctionTok{c}\NormalTok{(}\SpecialCharTok{{-}}\DecValTok{2}\NormalTok{,}\SpecialCharTok{{-}}\DecValTok{3}\NormalTok{),}\FunctionTok{c}\NormalTok{(}\DecValTok{1}\NormalTok{,}\DecValTok{2}\NormalTok{),}
        \AttributeTok{control=}\FunctionTok{list}\NormalTok{(}\AttributeTok{funEvals=}\DecValTok{100}\NormalTok{,}\AttributeTok{noise=}\ConstantTok{TRUE}\NormalTok{,}\AttributeTok{replicates=}\DecValTok{2}\NormalTok{,}
        \AttributeTok{designControl=}\FunctionTok{list}\NormalTok{(}\AttributeTok{replicates=}\DecValTok{2}\NormalTok{))) }\CommentTok{\#noise with replicated evaluations}
\NormalTok{res3 }\OtherTok{\textless{}{-}} \FunctionTok{spot}\NormalTok{(,}\ControlFlowTok{function}\NormalTok{(x)}\FunctionTok{funSphere}\NormalTok{(x)}\SpecialCharTok{+}\FunctionTok{rnorm}\NormalTok{(}\FunctionTok{nrow}\NormalTok{(x)),}\FunctionTok{c}\NormalTok{(}\SpecialCharTok{{-}}\DecValTok{2}\NormalTok{,}\SpecialCharTok{{-}}\DecValTok{3}\NormalTok{),}\FunctionTok{c}\NormalTok{(}\DecValTok{1}\NormalTok{,}\DecValTok{2}\NormalTok{),}
        \AttributeTok{control=}\FunctionTok{list}\NormalTok{(}\AttributeTok{funEvals=}\DecValTok{100}\NormalTok{,}\AttributeTok{noise=}\ConstantTok{TRUE}\NormalTok{,}\AttributeTok{replicates=}\DecValTok{2}\NormalTok{,}\AttributeTok{OCBA=}\ConstantTok{TRUE}\NormalTok{,}\AttributeTok{OCBABudget=}\DecValTok{1}\NormalTok{,}
        \AttributeTok{designControl=}\FunctionTok{list}\NormalTok{(}\AttributeTok{replicates=}\DecValTok{2}\NormalTok{))) }\CommentTok{\#and with OCBA}
\CommentTok{\# Check results with non{-}noisy function:}
\FunctionTok{funSphere}\NormalTok{(res1}\SpecialCharTok{$}\NormalTok{xbest)}
\end{Highlighting}
\end{Shaded}

\begin{verbatim}
##           [,1]
## [1,] 0.9811919
\end{verbatim}

\begin{Shaded}
\begin{Highlighting}[]
\FunctionTok{funSphere}\NormalTok{(res2}\SpecialCharTok{$}\NormalTok{xbest)}
\end{Highlighting}
\end{Shaded}

\begin{verbatim}
##           [,1]
## [1,] 0.0409259
\end{verbatim}

\begin{Shaded}
\begin{Highlighting}[]
\FunctionTok{funSphere}\NormalTok{(res3}\SpecialCharTok{$}\NormalTok{xbest)}
\end{Highlighting}
\end{Shaded}

\begin{verbatim}
##            [,1]
## [1,] 0.06636886
\end{verbatim}

\hypertarget{random-number-seed-handling}{%
\subsubsection{Random number seed
handling}\label{random-number-seed-handling}}

The following is for demonstration only, to be used for random number
seed handling in case of external noisy target functions.

\begin{Shaded}
\begin{Highlighting}[]
\NormalTok{res3 }\OtherTok{\textless{}{-}} \FunctionTok{spot}\NormalTok{(,}\ControlFlowTok{function}\NormalTok{(x,seed)\{}\FunctionTok{set.seed}\NormalTok{(seed);}\FunctionTok{funSphere}\NormalTok{(x)}\SpecialCharTok{+}\FunctionTok{rnorm}\NormalTok{(}\FunctionTok{nrow}\NormalTok{(x))\},}
     \FunctionTok{c}\NormalTok{(}\SpecialCharTok{{-}}\DecValTok{2}\NormalTok{,}\SpecialCharTok{{-}}\DecValTok{3}\NormalTok{),}\FunctionTok{c}\NormalTok{(}\DecValTok{1}\NormalTok{,}\DecValTok{2}\NormalTok{),}\AttributeTok{control=}\FunctionTok{list}\NormalTok{(}\AttributeTok{funEvals=}\DecValTok{30}\NormalTok{,}\AttributeTok{noise=}\ConstantTok{TRUE}\NormalTok{,}\AttributeTok{seedFun=}\DecValTok{1}\NormalTok{))}
\end{Highlighting}
\end{Shaded}

\hypertarget{handling-factor-variables}{%
\subsubsection{Handling factor
variables}\label{handling-factor-variables}}

Note: factors should be coded as integer values, i.e., 1,2,\ldots,n
First, we create a test function:

\begin{Shaded}
\begin{Highlighting}[]
\NormalTok{braninFunctionFactor }\OtherTok{\textless{}{-}} \ControlFlowTok{function}\NormalTok{ (x) \{}
\NormalTok{   y }\OtherTok{\textless{}{-}}\NormalTok{ (x[}\DecValTok{2}\NormalTok{]  }\SpecialCharTok{{-}} \FloatTok{5.1}\SpecialCharTok{/}\NormalTok{(}\DecValTok{4} \SpecialCharTok{*}\NormalTok{ pi}\SpecialCharTok{\^{}}\DecValTok{2}\NormalTok{) }\SpecialCharTok{*}\NormalTok{ (x[}\DecValTok{1}\NormalTok{] }\SpecialCharTok{\^{}}\DecValTok{2}\NormalTok{) }\SpecialCharTok{+} \DecValTok{5}\SpecialCharTok{/}\NormalTok{pi }\SpecialCharTok{*}\NormalTok{ x[}\DecValTok{1}\NormalTok{]  }\SpecialCharTok{{-}} \DecValTok{6}\NormalTok{)}\SpecialCharTok{\^{}}\DecValTok{2} \SpecialCharTok{+}
     \DecValTok{10} \SpecialCharTok{*}\NormalTok{ (}\DecValTok{1} \SpecialCharTok{{-}} \DecValTok{1}\SpecialCharTok{/}\NormalTok{(}\DecValTok{8} \SpecialCharTok{*}\NormalTok{ pi)) }\SpecialCharTok{*} \FunctionTok{cos}\NormalTok{(x[}\DecValTok{1}\NormalTok{] ) }\SpecialCharTok{+} \DecValTok{10}
   \ControlFlowTok{if}\NormalTok{(x[}\DecValTok{3}\NormalTok{]}\SpecialCharTok{==}\DecValTok{1}\NormalTok{)}
\NormalTok{     y }\OtherTok{\textless{}{-}}\NormalTok{ y }\SpecialCharTok{+}\DecValTok{1}
   \ControlFlowTok{else} \ControlFlowTok{if}\NormalTok{(x[}\DecValTok{3}\NormalTok{]}\SpecialCharTok{==}\DecValTok{2}\NormalTok{)}
\NormalTok{     y }\OtherTok{\textless{}{-}}\NormalTok{ y }\SpecialCharTok{{-}}\DecValTok{1}
   \FunctionTok{return}\NormalTok{(y)}
\NormalTok{\}}
\end{Highlighting}
\end{Shaded}

Vectorize the test function.

\begin{Shaded}
\begin{Highlighting}[]
\NormalTok{objFun }\OtherTok{\textless{}{-}} \ControlFlowTok{function}\NormalTok{(x)\{}\FunctionTok{apply}\NormalTok{(x,}\DecValTok{1}\NormalTok{,braninFunctionFactor)\}}
\end{Highlighting}
\end{Shaded}

Run \texttt{spot}.

\begin{Shaded}
\begin{Highlighting}[]
\FunctionTok{set.seed}\NormalTok{(}\DecValTok{1}\NormalTok{)}
\NormalTok{res }\OtherTok{\textless{}{-}} \FunctionTok{spot}\NormalTok{(}\AttributeTok{fun=}\NormalTok{objFun,}\AttributeTok{lower=}\FunctionTok{c}\NormalTok{(}\SpecialCharTok{{-}}\DecValTok{5}\NormalTok{,}\DecValTok{0}\NormalTok{,}\DecValTok{1}\NormalTok{),}\AttributeTok{upper=}\FunctionTok{c}\NormalTok{(}\DecValTok{10}\NormalTok{,}\DecValTok{15}\NormalTok{,}\DecValTok{3}\NormalTok{),}
            \AttributeTok{control=}\FunctionTok{list}\NormalTok{(}\AttributeTok{model=}\NormalTok{buildKriging,}
                         \AttributeTok{types=} \FunctionTok{c}\NormalTok{(}\StringTok{"numeric"}\NormalTok{,}\StringTok{"numeric"}\NormalTok{,}\StringTok{"factor"}\NormalTok{),}
                         \AttributeTok{optimizer=}\NormalTok{optimLHD))}
\NormalTok{ res}\SpecialCharTok{$}\NormalTok{xbest}
\end{Highlighting}
\end{Shaded}

\begin{verbatim}
##          [,1]     [,2] [,3]
## [1,] 2.619386 2.725642    2
\end{verbatim}

\begin{Shaded}
\begin{Highlighting}[]
\NormalTok{ res}\SpecialCharTok{$}\NormalTok{ybest}
\end{Highlighting}
\end{Shaded}

\begin{verbatim}
##           [,1]
## [1,] 0.6777176
\end{verbatim}

\hypertarget{high-dimensional-problem-runs-very-long}{%
\subsubsection{High dimensional problem (runs very
long)}\label{high-dimensional-problem-runs-very-long}}

First, we consider the default \texttt{spot} setting with
\texttt{buildKriging()}.

\begin{Shaded}
\begin{Highlighting}[]
\FunctionTok{set.seed}\NormalTok{(}\DecValTok{2}\NormalTok{)}
\NormalTok{res0 }\OtherTok{\textless{}{-}} \FunctionTok{spot}\NormalTok{(}\AttributeTok{x=}\ConstantTok{NULL}\NormalTok{, funSphere, }\FunctionTok{runif}\NormalTok{(}\DecValTok{30}\NormalTok{), }\DecValTok{1}\SpecialCharTok{+} \FunctionTok{runif}\NormalTok{(}\DecValTok{30}\NormalTok{),}\AttributeTok{control=}\FunctionTok{list}\NormalTok{(}\AttributeTok{funEvals=}\DecValTok{30}\NormalTok{))}
\NormalTok{res0}\SpecialCharTok{$}\NormalTok{ybest}
\end{Highlighting}
\end{Shaded}

\begin{verbatim}
##          [,1]
## [1,] 27.12115
\end{verbatim}

Then, we use the \texttt{buildGaussianProcess()} model.

\begin{Shaded}
\begin{Highlighting}[]
\NormalTok{res1 }\OtherTok{\textless{}{-}} \FunctionTok{spot}\NormalTok{(}\AttributeTok{x=}\ConstantTok{NULL}\NormalTok{, funSphere, }\FunctionTok{runif}\NormalTok{(}\DecValTok{30}\NormalTok{), }\DecValTok{1}\SpecialCharTok{+} \FunctionTok{runif}\NormalTok{(}\DecValTok{30}\NormalTok{),}\AttributeTok{control=}\FunctionTok{list}\NormalTok{(}\AttributeTok{funEvals=}\DecValTok{30}\NormalTok{, }
\AttributeTok{model =}\NormalTok{ buildGaussianProcess))}
\NormalTok{res1}\SpecialCharTok{$}\NormalTok{ybest}
\end{Highlighting}
\end{Shaded}

\begin{verbatim}
##          [,1]
## [1,] 23.95562
\end{verbatim}

\begin{Shaded}
\begin{Highlighting}[]
\NormalTok{resb }\OtherTok{\textless{}{-}}\NormalTok{ microbenchmark}\SpecialCharTok{::}\FunctionTok{microbenchmark}\NormalTok{(}
  \FunctionTok{spot}\NormalTok{(}\AttributeTok{x=}\ConstantTok{NULL}\NormalTok{, funSphere, }\FunctionTok{runif}\NormalTok{(}\DecValTok{30}\NormalTok{), }\DecValTok{1}\SpecialCharTok{+} \FunctionTok{runif}\NormalTok{(}\DecValTok{30}\NormalTok{),}\AttributeTok{control=}\FunctionTok{list}\NormalTok{(}\AttributeTok{funEvals=}\DecValTok{30}\NormalTok{)),}
  \FunctionTok{spot}\NormalTok{(}\AttributeTok{x=}\ConstantTok{NULL}\NormalTok{, funSphere, }\FunctionTok{runif}\NormalTok{(}\DecValTok{30}\NormalTok{), }\DecValTok{1}\SpecialCharTok{+} \FunctionTok{runif}\NormalTok{(}\DecValTok{30}\NormalTok{),}\AttributeTok{control=}\FunctionTok{list}\NormalTok{(}\AttributeTok{funEvals=}\DecValTok{30}\NormalTok{, }\AttributeTok{model =}\NormalTok{ buildGaussianProcess)),}
  \AttributeTok{times =}\NormalTok{ 5L)}
\FunctionTok{print}\NormalTok{(resb)}
\end{Highlighting}
\end{Shaded}

\begin{verbatim}
## Unit: seconds
##                                                                                                                   expr
##                                     spot(x = NULL, funSphere, runif(30), 1 + runif(30), control = list(funEvals = 30))
##  spot(x = NULL, funSphere, runif(30), 1 + runif(30), control = list(funEvals = 30,      model = buildGaussianProcess))
##       min        lq      mean    median        uq       max neval cld
##  65.92593 67.633514 69.581661 70.302811 71.834703 72.211345     5   b
##   2.03337  2.056508  2.072568  2.077195  2.090665  2.105101     5  a
\end{verbatim}

\begin{Shaded}
\begin{Highlighting}[]
\FunctionTok{boxplot}\NormalTok{(resb)}
\end{Highlighting}
\end{Shaded}

\includegraphics{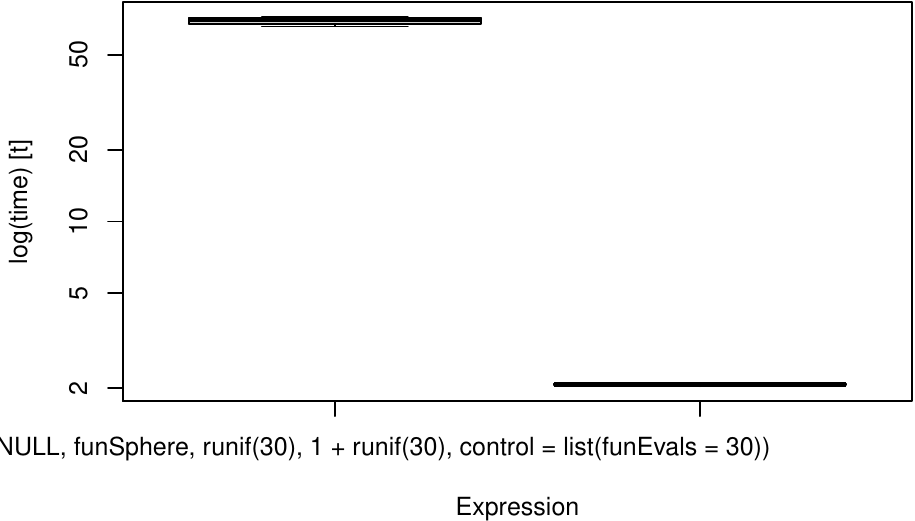}

\bibliographystyle{unsrt}
\bibliography{bartzAll.bib}

\begin{thebibliography}{10}

\bibitem{Bart06a}
Thomas Bartz-Beielstein.
\newblock {\em {Experimental Research in Evolutionary Computation---The New
  Experimentalism}}.
\newblock Natural Computing Series. Springer, Berlin, Heidelberg, New York,
  2006.

\bibitem{rcor21a}
{R Core Team}.
\newblock {\em R: A Language and Environment for Statistical Computing}.
\newblock R Foundation for Statistical Computing, Vienna, Austria, 2021.

\bibitem{Kirk83a}
S~Kirkpatrick, C~D Gelatt, and M~P Vecchi.
\newblock {Optimization by simulated annealing}.
\newblock {\em science}, 220(4598):671--680, 1983.

\bibitem{Eibe03a}
A~E Eiben and J~E Smith.
\newblock {\em {Introduction to Evolutionary Computing}}.
\newblock Springer, Berlin, Heidelberg, New York, 2003.

\bibitem{Bart16n}
Thomas Bartz-Beielstein and Martin Zaefferer.
\newblock Model-based methods for continuous and discrete global optimization.
\newblock {\em Applied Soft Computing}, 55:154 -- 167, 2017.

\bibitem{Brei01a}
L~Breiman.
\newblock {Random Forests}.
\newblock {\em Machine Learning}, 45(1):5--32, 2001.

\bibitem{Forr08a}
Alexander Forrester, Andr{\'a}s S{\'o}bester, and Andy Keane.
\newblock {\em {Engineering Design via Surrogate Modelling}}.
\newblock Wiley, 2008.

\bibitem{Box87a}
G~E~P Box and N~R Draper.
\newblock {\em {Empirical Model Building and Response Surfaces}}.
\newblock Wiley, New York NY, 1987.

\bibitem{Lent12a}
Russell~V Lenth.
\newblock {Response-Surface Methods in R Using rsm (Updated to version 2.00)}.
\newblock Technical report, 2012.

\bibitem{Bart16jcos}
Thomas Bartz-Beielstein.
\newblock {Stacked Generalization of Surrogate Models - A Practical Approach}.
\newblock Technical Report 5/2016, TH K{\"o}ln, K{\"o}ln, 2016.
\newblock https://cos.bibl.th-koeln.de/frontdoor/index/index/docId/375.

\end{thebibliography}

\end{document}